\documentclass{aa}  %[longauth]
\pdfoutput=1
\usepackage{graphicx, multirow, tablefootnote, longtable}
\usepackage{txfonts, listings}
\usepackage{natbib}
\usepackage{xcolor, tikz}
\usepackage[percent]{overpic}

\bibpunct{(}{)}{;}{a}{}{,}
\begin{document}

   \title{Photo-astrometric distances, extinctions, and astrophysical parameters for {\it Gaia} DR2 stars brighter than $G=18$}

   \author{F. Anders\inst{1,2,3}, A. Khalatyan\inst{2}, C. Chiappini\inst{2,3}, A. B. Queiroz\inst{2,3}, B. X. Santiago\inst{4,3}, C. Jordi\inst{1}, L. Girardi\inst{5}, \\ A. G. A. Brown\inst{6}, G. Matijevi\v{c}\inst{2}, G. Monari\inst{2}, T. Cantat-Gaudin\inst{1}, M. Weiler\inst{1}, S. Khan\inst{7}, A. Miglio\inst{7}, I. Carrillo\inst{2}, \\ M. Romero-G\'{o}mez\inst{1}, I. Minchev\inst{2}, R. S. de Jong\inst{2}, T. Antoja\inst{1}, P. Ramos\inst{1}, M. Steinmetz\inst{2}, H. Enke\inst{2}}

   \authorrunning{F. Anders et al.}

    \institute{Institut de Ci\`encies del Cosmos, Universitat de Barcelona (IEEC-UB), Carrer Mart\'i i Franqu\`es 1, 08028 Barcelona, Spain\\
              \email{fanders@icc.ub.edu}
	\and{Leibniz-Institut f\"ur Astrophysik Potsdam (AIP), An der Sternwarte 16, 14482 Potsdam, Germany}
	\and{Laborat\'orio Interinstitucional de e-Astronomia - LIneA, Rua Gal. Jos\'e Cristino 77, Rio de Janeiro, RJ - 20921-400, Brazil}
     \and{Instituto de F\'\i sica, Universidade Federal do Rio Grande do Sul, Caixa Postal 15051, Porto Alegre, RS - 91501-970, Brazil}
     \and{Osservatorio Astronomico di Padova, INAF, Vicolo dell'Osservatorio 5, 35122 Padova, Italy}
     \and{Leiden Observatory, P.O. Box 9513, 2300 RA, Leiden, The Netherlands}
     \and{School of Physics and Astronomy, University of Birmingham, Edgbaston, Birmingham, B 15 2TT, United Kingdom}
}
   \date{Received 25.04.2019; accepted 27.06.2019}

  \abstract
  {Combining the precise parallaxes and optical photometry delivered by {\it Gaia}'s second data release ({\it Gaia} DR2) with the photometric catalogues of Pan-STARRS1, 2MASS, and AllWISE, we derived Bayesian stellar parameters, distances, and extinctions for 265 million of the 285 million objects brighter than $G=18$. Because of the wide wavelength range used, our results substantially improve the accuracy and precision of previous extinction and effective temperature estimates. After cleaning our results for both unreliable input and output data, we retain 137 million stars, for which we achieve a median precision of 5\% in distance, 0.20 mag in $V$-band extinction, and 245 K in effective temperature for $G\leq14$, degrading towards fainter magnitudes (12\%, 0.20 mag, and 245 K at $G=16$; 16\%, 0.23 mag, and 260 K at $G=17$, respectively). We find a very good agreement with the asteroseismic surface gravities and distances of 7000 stars in the {\it Kepler}, {\it K2}-C3, and {\it K2}-C6 fields, with stellar parameters from the APOGEE survey, and with distances to star clusters. Our results are available through the ADQL query interface of the {\it Gaia} mirror at the Leibniz-Institut f\"{u}r Astrophysik Potsdam ({\tt gaia.aip.de}) and as binary tables at {\tt data.aip.de}. As a first application, we provide distance- and extinction-corrected colour-magnitude diagrams, extinction maps as a function of distance, and extensive density maps. These demonstrate the potential of our value-added dataset for mapping the three-dimensional structure of our Galaxy. In particular, we see a clear manifestation of the Galactic bar in the stellar density distributions, an observation that can almost be considered direct imaging of the Galactic bar. %Finally, our results also serve to assess targeting strategies of the future 4-metre Multi-Object Spectroscopic Telescope (4MOST).
  }

   \maketitle

%________________________________________________________________

\section{Introduction}

Galactic astrophysics is currently in a similar phase as geography was in the 15th century: large parts of the Earth were unknown to contemporary scientists, only crude maps of most of the known parts of the Earth existed, and even the orbit of our planet was still under debate. Nowadays, major parts of the Milky Way are still hidden by thick layers of dust, but we are beginning to discover and to map our Galaxy in a much more accurate fashion by virtue of dedicated large photometric, astrometric, and spectroscopic surveys. 

In this context, the astrometric European Space Agency mission {\it Gaia} \citep{GaiaCollaboration2016} represents a major leap in our understanding of the Milky Way's stellar content: its measurement precision as well as the absolute number counts surpass previous astrometric datasets by several orders of magnitude. The recent {\it Gaia} Data Release 2 ({\it Gaia} DR2; \citealt{GaiaCollaboration2018}), covered the first 22 months of observations (from a currently predicted total of approximately ten years) with positions and photometry for $1.7\cdot10^9$ sources \citep{Evans2018}, proper motions and parallaxes for $1.3\cdot10^9$ sources \citep{Lindegren2018}, astrophysical parameters for $\simeq 10^8$ stars \citep{Andrae2018}, and radial velocities for $7\cdot10^6$ of them \citep{Sartoretti2018, Katz2019}.

\begin{table*}
\centering
\caption{Summary of calibrations and data curation applied for fiducial {\tt StarHorse} run. Details are provided in Sect. \ref{data}.}
\begin{tabular}{c|c|l|l}
Parameter & Parameter regime & Calibration choice & Reference \\
\hline
 &         $G < 14$    &   ${\tt parallax} +0.05$ mas  & \citet{Lindegren2018a, Zinn2019}  \\
 $\varpi^{\rm cal}$  &    $14 < G < 16.5$  &  ${\tt parallax} +(0.1676-0.0084\cdot {\tt phot\_g\_mean\_mag})$ mas       &  \citet{Lindegren2018a}, linear interpolation  \\
                           &        $ G > 16.5$  &   ${\tt parallax}+0.029$ mas  & \citet{Lindegren2018}  \\
\hline
 &    $G < 11$    &   $1.2 \cdot {\tt parallax\_error}$  & \\
 $\sigma_{\varpi}^{\rm cal}$  &    $11 < G < 15$  &   $(0.22\cdot {\tt phot\_g\_mean\_mag}-1.22)\cdot {\tt parallax\_error}$      &  Fit to \citet{Lindegren2018a} data  \\
                           &        $ G > 15$  &   $({\rm e}^{-({\tt phot\_g\_mean\_mag}-15)}+1.08) \cdot {\tt parallax\_error}$  &    \\
\hline
 &    $G < 6$    &   {\tt phot\_g\_mean\_mag} + $0.0271 \cdot (6 - {\tt phot\_g\_mean\_mag})$  & \\
 $G$  &    $6 < G < 16$  & ${\tt phot\_g\_mean\_mag} - 0.0032 \cdot ({\tt phot\_g\_mean\_mag}-6)$ &  \citet{MaizApellaniz2018}  \\
                           &        $ G > 16$  &   ${\tt phot\_g\_mean\_mag} - 0.032 $  &  \\
\hline
 \multirow{2}{*}{$G_{\rm BP}$}  &    $G < 10.87$    &   using bright $G_{\rm BP}$ filter curve   & \multirow{2}{*}{\citet{MaizApellaniz2018}} \\
 &    $G > 10.87$  & using faint $G_{\rm BP}$ filter curve  &   \\
\hline
 $ g_{\rm PS1}$   &   &   ${\tt g\_mean\_psf\_mag} - 0.020 $  &   \\
 $ r_{\rm PS1}$   &   &   ${\tt r\_mean\_psf\_mag} - 0.033 $  &   \\
 $ i_{\rm PS1}$   &  $G>14$ &   ${\tt i\_mean\_psf\_mag} - 0.024 $  & \citet{Scolnic2015}  \\
 $ z_{\rm PS1}$   &   &   ${\tt z\_mean\_psf\_mag} - 0.028 $  &   \\
 $ y_{\rm PS1}$   &   &   ${\tt y\_mean\_psf\_mag} - 0.011 $  &   \\
\hline
\multirow{2}{*}{$\sigma_{\rm mag}$}  & {\it Gaia}, 2MASS, WISE  & $\max\{{\tt \sigma_{\rm mag, source}}, 0.03 {\rm mag}\} $  &    \\
  & Pan-STARRS1  & $\max\{{\tt \sigma_{\rm mag, source}}, 0.04 {\rm mag}\} $  &    \\
\end{tabular}
\label{calibtable}
\end{table*}

The {\it Gaia} DR2 dataset thus represents a treasure trove for many branches of Galactic astrophysics. Various advances have since been achieved in the field of Galactic dynamics \citep[e.g.][]{GaiaCollaboration2018b, GaiaCollaboration2018c, Antoja2018a, Kawata2018a, Quillen2018, Ramos2018, Laporte2019, Monari2019, Trick2019}, star clusters and associations \citep[e.g.][]{GaiaCollaboration2018e, Cantat-Gaudin2018, Cantat-Gaudin2018a, Cantat-Gaudin2018b, Castro-Ginard2018, Soubiran2018, Zari2018, Baumgardt2019, Bossini2019, deBoer2019, Meingast2019}, the Galactic star-formation history \citep{Helmi2018, Mor2019}, hyper-velocity stars \citep[e.g.][]{Bromley2018, Scholz2018, Shen2018, Boubert2018, Boubert2019, Erkal2019}, among others. Apart from stellar science, the precise {\it Gaia} DR2 photometry, in combination with the high quality of the stellar parallax measurements, can also be used to map the distribution of dust in the Galaxy. The availability of precise individual distance and extinction determinations (mainly from high-resolution spectroscopic surveys, and also recently from {\it Gaia}) has led to a significant improvement of interstellar dust maps within the past years and months (e.g. \citealt{Lallement2014, Green2015, Capitanio2017, RezaeiKh.2017, RezaeiKh.2018, Lallement2018, Lallement2019, Yan2019}, \citealt{Leike2019, Chen2019}). 

In addition to the main {\it Gaia} DR2 data products (parallaxes, proper motions, radial velocities, and photometry), the {\it Gaia} DR2 data allowed for the immediate computation of quantities relevant for Galactic stellar population studies. These are the Bayesian geometric distance estimates computed by \citet{Bailer-Jones2018} and the first stellar parameters and extinction estimates from the {\it Gaia} {\it Apsis} pipeline \citep{Andrae2018}. The latter authors deliberately used only {\it Gaia} DR2 data products to infer line-of-sight extinctions as well as effective temperatures, radii, and luminosities. This proved to be a difficult exercise since the three broad {\it Gaia} passbands contain little information to discriminate between effective temperature and interstellar extinction. As a result, the {\it Apsis} $T_{\rm eff}$ estimates were obtained under the assumption of zero extinction (thus suffering from systematics in the Galactic plane) and the uncertainties in individual $G$-band extinction and $E(G_{\rm BP}-G_{\rm RP})$ colour excess estimates are so large that these values should only be used in ensemble studies \citep{Andrae2018, GaiaCollaboration2018}. 

The lack of more precise extinction estimates prevented the use of {\it Gaia} data for stellar population studies in a larger volume outside the low-extinction regime \citep{GaiaCollaboration2018e, Antoja2018a}. Many of the new Galactic archaeology results derived from {\it Gaia} DR2 still concentrate on a small portion of the {\it Gaia} data. This is partly due to the necessity of full phase-space information (\citealt{GaiaCollaboration2018c, Antoja2018a}), but also partly due to extinction uncertainties hampering the direct inference of desired quantities \citep{GaiaCollaboration2018e, Helmi2018, Romero-Gomez2019, Mor2019}.  

In this spirit, the aim of this paper is to enlarge the volume in which we can make use of the {\it Gaia} DR2 data by providing more accurate and precise extinctions and stellar parameters (most importantly $T_{\rm eff}$, but also estimates of surface gravity, metallicity, and mass), and more accurate distances for distant giant stars. Although the data quality degrades notably around a magnitude of $G\sim 16.5$, we provide useful information for  considerable fraction of stars down to $G=18$. To this end, we use the {\tt python} code {\tt StarHorse}, originally designed to determine stellar parameters and distances for spectroscopic surveys \citep{Santiago2016, Queiroz2018}.\footnote{In particular, \citet{Queiroz2018} released distances and extinctions for around 1 million stars observed by APOGEE DR14 \citep{Abolfathi2018}, RAVE DR5 \citep{Kunder2017}, GES DR3 \citep{Gilmore2012a}, and GALAH DR1 \citep{Martell2017}.} Of the 285 million objects with $G\leq18$ contained in {\it Gaia} DR2, our code delivered results for $\sim266$ million stars. Applying a number of conservative quality criteria on the input and output data, we achieve a sample cleaned on the basis of data quality flags (see Sect. \ref{flags}) of around 137 million stars with reliable stellar parameters, distances, and extinctions. 

The paper is structured as follows: Section \ref{data} presents the input data used in the parameter estimation. The following Sect. \ref{starhorse} describes the basics of our code, focussing on updates with respect to its previous applications to spectroscopic stellar surveys. Section \ref{flags} in particular explains how we flagged the {\tt StarHorse} results for {\it Gaia} DR2. Since we decided to provide results for all objects that our code converged for, any user of our value-added catalogue should pay particular attention to this subsection. We present some first astrophysical results in Sect. \ref{results}, mainly focussing on extinction-corrected colour-magnitude diagrams, stellar density maps, extinction maps, and the emergence of the Galactic bar. We discuss the precision and accuracy of the {\tt StarHorse} parameters in Sect. \ref{uncertainties}, providing comparisons to open clusters and stellar parameters obtained from high-resolution spectroscopy. We also compare to previous results obtained from {\it Gaia} DR2 in Sec. \ref{dpac_comp}. We conclude the paper with a summary and a brief outlook on possible applications of {\tt StarHorse} or similar codes to future {\it Gaia} data releases.

\begin{table*}
\centering
\caption{Statistics of some of the currently available astro-spectro-photometric distances and extinctions based on {\it Gaia} data, in comparison to the results obtained in this paper. The last three columns refer to the median precision in relative distance, $V$-band extinction, and effective temperature, respectively. For the definition of the {\tt StarHorse} flags we refer to Sect. \ref{flags}.}
\begin{tabular}{l|c|c|r|c|c|c}
Reference & Survey(s) & mag limits & \# objects & $\sigma_d / d$ & $\sigma_{A_V}$ & $\sigma_{T_{\rm eff}}$ \\
\hline
\citet{Queiroz2018}  &  {\it Gaia} DR1 + spectroscopy & &  1.5M & 15 \% & 0.07 mag & -- \\
\citet{Mints2018a}  &  {\it Gaia} DR1 + spectroscopy & & 3.8M & 15 \% & -- & -- \\
\citet{Sanders2018}  &  {\it Gaia} DR2 + spectroscopy & & 3.1M & 3 \% & 0.01 mag & 40 K \\
Santiago et al. (in prep.)  &  {\it Gaia} DR2 + spectroscopy & & 2M &  5 \% & 0.07 mag & 40 K \\
\hline
\citet{Bailer-Jones2018}  &  {\it Gaia} DR2 & $G\lesssim21$ & 1330M & 25 \% & -- & -- \\
\citet{McMillan2018}  &  {\it Gaia} DR2 & $G\lesssim13$ & 7M & 6 \%  & -- & -- \\
\citet{Andrae2018}  &  {\it Gaia} DR2 & $G\leq17$ & 80M & -- & 0.46 mag & 324 K \\
\hline
This work  &  {\it Gaia} DR2 + photometry & $G<18$ & 285M &  &  \\
\qquad {\color{black} {\tt StarHorse} converged} & &  & 265,637,087  & 28 \% & 0.25 mag & 310 K \\
\qquad {\color{blue} $\varpi^{\rm cal}/\sigma_{\varpi}^{\rm cal}>5$} & & & {\color{blue} 103,108,516} & {\color{blue} 9 \%} & {\color{blue} 0.20 mag} & {\color{blue} 265 K}  \\
\qquad {\color{cyan} {\tt SH\_GAIAFLAG=}"000"} & & & {\color{cyan} 232,974,244} & {\color{cyan} 26 \%} & {\color{cyan} 0.24 mag} & {\color{cyan} 305 K}  \\
\qquad {\color{orange} {\tt SH\_OUTFLAG=}"00000"} & & & {\color{orange} 151,506,183} & {\color{orange} 13 \%} & {\color{orange} 0.22 mag} & {\color{orange} 250 K}  \\
\qquad {\color{red} both flags} & & & {\color{red} 136,606,128}& {\color{red} 13 \%}& {\color{red} 0.22 mag} & {\color{red} 250 K}   \\
\hline
\qquad All passbands available & & & 60,520,497 & 23 \% & 0.20 mag & 270 K\\
\qquad \qquad {\color{red} both flags}                         & & & {\color{red} 34,447,306} & {\color{red} 12 \%} & {\color{red} 0.18 mag} & {\color{red} 220 K}\\
\qquad {\it Gaia} DR2+2MASS+AllWISE & & & 72,754,432 & 13 \% & 0.23 mag & 255 K\\
\qquad \qquad {\color{red} both flags clean}                    & & & {\color{red} 52,148,742} & {\color{red} 9 \%} & {\color{red} 0.21 mag} & {\color{red} 230 K}\\
\qquad {\it Gaia} DR2+2MASS & & & 58,295,744 & 37 \% & 0.32 mag & 390 K\\
\qquad  \qquad {\color{red} both flags clean}           & & & {\color{red} 27,616,169} & {\color{red} 17 \%} & {\color{red} 0.28 mag} & {\color{red} 300 K}\\
\qquad {\it Gaia} DR2 only & & & 12,486,568 & 44 \% & 0.40 mag & 1000 K\\
\qquad  \qquad {\color{red} both flags clean}          & & & {\color{red} 2,003,978} & {\color{red} 18 \%} & {\color{red} 0.35 mag} & {\color{red} 390 K}\\
\hline
\qquad $G \leq 14$ & & & 16,143,700  & 5 \% & 0.20 mag & 250 K \\
\qquad  \qquad {\color{red} both flags clean}            & & & {\color{red} 14,432,712}  & {\color{red} 5 \%} & {\color{red} 0.20 mag} & {\color{red} 245 K} \\
\qquad $14 < G \leq 16$ & & & 57,368,469 & 12 \% & 0.20 mag & 250 K \\
\qquad   \qquad {\color{red} both flags clean}                & & & {\color{red} 49,171,794} & {\color{red} 12 \%} & {\color{red} 0.20 mag} & {\color{red} 245 K} \\
\qquad $16 < G \leq 17$ & & & 72,801,366 & 24 \% & 0.24 mag & 300 K \\
\qquad  \qquad {\color{red} both flags clean}                 & & & {\color{red} 43,398,790} & {\color{red} 16 \%} & {\color{red} 0.23 mag} & {\color{red} 260 K} \\
\qquad $17 < G \leq 18$ & & & 119,323,552 & 50 \% & 0.29 mag & 380 K \\
\qquad  \qquad {\color{red} both flags clean}                 & & & {\color{red} 29,602,832} & {\color{red} 14 \%} & {\color{red} 0.24 mag} & {\color{red} 230 K} \\
\end{tabular}
\label{summarytable}
\end{table*}

\section{Data}\label{data}

The {\it Gaia} satellite is measuring positions, parallaxes, proper motions and photometry for well over $10^9$ sources down to $G \simeq 20.7$, and obtaining physical parameters and radial velocities for millions of brighter stars. Particularly important for our purposes are the parallaxes, whose precision varies from $< 0.1$ mas for $G \leq 17$ to $\simeq 0.7$ mas for $G=20$ \citep{Lindegren2018}. Initial tests showed that reliable {\tt StarHorse} results (that represent an improvement with respect to purely photometric distances) can be obtained up to $G\sim18$. We therefore downloaded {\it Gaia} DR2 data for all stars with measured parallaxes up to that magnitude.

It is well known that the parallaxes delivered by {\it Gaia} DR2 are not entirely free from systematics \citep{GaiaCollaboration2018, Lindegren2018, Stassun2018, Zinn2019, Khan2019}\footnote{For a short and comprehensive review, see \citet{Lindegren2018a}, accessible at \url{https://www.cosmos.esa.int/web/gaia/dr2-known-issues}}. In particular, \citet{Arenou2018} have shown that the parallax zero-point is subject to a sub-100$\mu$as offset depending on position, and possibly magnitude, parallax, and/or colour. Since our distance inference depends critically on the accuracy of the input parallaxes, but the positional dependence is too complex to calibrate out at the moment, we opted for the following first-order calibrations detailed in Table \ref{calibtable}: in the bright regime ($G<14$), we apply a correction of +0.05 mas similar to the global offset found by \citet{Zinn2019} and \citet{Khan2019} from asteroseismic and spectroscopic observations in the {\it Kepler} field. It should be noted, however, that \citet{Khan2019}, in agreement with the quasar comparison shown in \citet{Arenou2018}, find different offsets for the {\it Kepler}-2 fields C3 and C6, indicating that also in the bright regime the parallax zero-point depends on sky position. In the faint regime ($G>16.5$), we use the +0.029 mas correction derived by \citet{Lindegren2018} from AllWISE quasars. For intermediate $G$ magnitudes, the parallax correction is linearly interpolated between these two values. 

\citet{Lindegren2018, Arenou2018}, and others have demonstrated that, similar to the {\it Gaia} DR2 parallaxes, also the parallax uncertainties are prone to moderate systematics, in the sense that they are typically slightly underestimated. For this work (see Table \ref{calibtable}) we follow a modified version of the recalibration advertised by \citet{Lindegren2018a}: in the faint regime ($G>15$), the external-to-internal uncertainty ratio exponentially drops to 1.08, while at the bright end ($G<12$) this factor is set to 1.2. In the intermediate regime, we again opt for linear interpolation, a choice that is supported by the data presented by \citet[][slide 15]{Lindegren2018a}. 

We note that this re-scaling of the parallax errors takes into account the systematic term $\sigma_s$ (which roughly accounts for the variations of the parallax zero-point over the sky, with magnitude, colour etc.; equation 2 in \citealt{Lindegren2018a}) only approximately. By choosing the recalibration detailed in Table \ref{calibtable} we have effectively accounted for $\sigma_s$ in the faint regime. In the bright regime, our recalibrated parallax uncertainties are slightly lower than in the Lindegren model (for bright stars our minimum parallax error is 0.018, below the systematic floor of $\sigma_s=0.021$ proposed by Lindegren). While this will be corrected in future runs, we have verified that the results change very little when correctly including the $\sigma_s$ term.

Apart from the parallaxes, we also make use of the three-band {\it Gaia} DR2 photometry ($G, G_{\rm BP}, G_{\rm RP}$). While these are of an unprecedented precision, several recent works \citep{Weiler2018, MaizApellaniz2018, Casagrande2018a} have shown by comparison with absolute spectrophotometry that the $G$ band suffers from a magnitude-dependent offset, and that the nominal passbands need to be slightly corrected. Therefore, in order to compare the {\it Gaia} DR2 $G$ magnitudes to the synthetic {\it Gaia} DR2 photometry from stellar models, we have applied the $G$ magnitude corrections, as well as the new passband definitions, given by \citet{MaizApellaniz2018}.

Furthermore, we supplement the {\it Gaia} data with additional Pan-STARRS1 $grizy$ \citep{Scolnic2015}, 2MASS $JHK_s$, and AllWISE $W1W2$ photometry, using the cross-matches provided by the {\it Gaia} team \citep{GaiaCollaboration2018, Marrese2019}. After initial tests, we only used Pan-STARRS1 photometry for stars with magnitudes fainter than $G=$14 that do not suffer from saturation problems. For all passbands, missing photometric uncertainties were substituted by fiducial maximum uncertainties of 0.3 mag. We also introduced an error floor of 0.04 mag. For {\it Gaia}, 2MASS, and AllWISE, we use an uncertainty floor of 0.03 mag, which can be considered a minimum value for the accuracy of the synthetic photometry used by our method. We verified that this choice does not impact our results.

\section{{\tt StarHorse} runs}\label{starhorse}

\subsection{The code}

The advent of massive multiplex spectroscopic stellar surveys has led to the development of a growing number of codes that aim to determine precise distances and extinctions to vast numbers of field stars (for example, \citealt{Breddels2010, Zwitter2010, Burnett2010, Binney2014a, Santiago2016, Wang2016a, Mints2018a, Das2019, Leung2019}).

The {\tt StarHorse} code \citep{Queiroz2018} is a Bayesian parameter estimation code that compares a number of observed quantities (be it photometric magnitudes, spectroscopically derived stellar parameters, or parallaxes) to stellar evolutionary models. In a nutshell, it finds the posterior probability over a grid of stellar models, distances, and extinctions, given the set of observations plus a number of priors. The priors include the stellar initial mass function (in our case \citealt{Chabrier2003}), density laws for the main components of the Milky Way (thin disc, thick disc, bulge, and halo), as well as broad metallicity and age priors for those components. We refer to \citet{Queiroz2018} for more details. In this work we also used a broad top-hat prior on extinction ($-0.3\leq A_V\leq 4.0$) for stars with low parallax signal-to-noise ratios ($\varpi^{\rm cal}/\sigma_{\varpi}^{\rm cal}<5$), ensuring the convergence of the code. This should be kept in mind when interpreting our results for highly extincted stars in the inner Galaxy. The impact of our choice of the priors on the results for the inner regions of the Galaxy are studied in more detail in Queiroz et al. (in prep.).

The first version of the code was developed by \citet{Santiago2016} in the context of the RAVE survey \citep{Steinmetz2006} and the SDSS-III \citep{Eisenstein2011} spectroscopic surveys SEGUE \citep{Yanny2009} and APOGEE \citep{Majewski2017}.                                                                                                                                                           In \citet{Queiroz2018} the code was ported to {\tt python 2.7} and made more flexible in the choice of input, priors, etc. With respect to that publication, we have implemented some important changes that were necessary to apply {\tt StarHorse} to the huge {\it Gaia} DR2 dataset. 

\begin{figure}\centering
 	\includegraphics[width=0.49\textwidth]{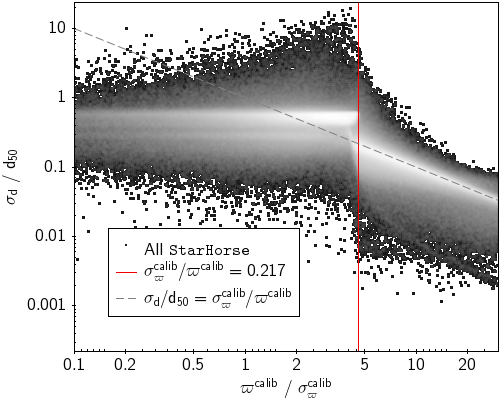}
 	\includegraphics[width=0.49\textwidth]{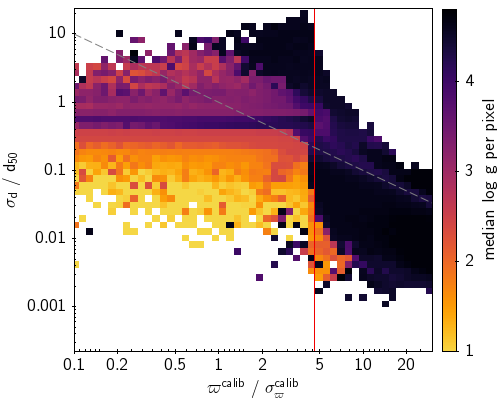}
 	\caption{Dependence of {\tt StarHorse} posterior distance uncertainty on the (recalibrated) {\it Gaia} DR2 parallax uncertainty. Top: Density plot. Bottom: coloured by median $\log g$ in each pixel. The grey dashed line indicates unity, the red vertical line indicates the approximate value below which the inverse parallax PDF becomes seriously biased and noisy (see  \citealt{Bailer-Jones2015a}).}
 	\label{sigdist_piepi}
\end{figure}

\subsection{Code updates and improvements}

With respect to \citet{Queiroz2018}, a few updates to the {\tt StarHorse} code have been carried out. Most importantly, we now take better account of dust extinction when comparing synthetic and observed photometry, an update that was necessary due to the use of the broad-band optical {\it Gaia} passbands.

{\it Dust-attenuated synthetic photometry:} As explained in  \citet{Holtzman1995, Sirianni2005}, or \citet{Girardi2008}, dust-attenuated photometry of very broad photometric passbands (such as the {\it Gaia} DR2 ones) should take into account that the passband extinction coefficient $A_{i}/A_V$ for a star varies as a function of its source spectrum $F_{\lambda}$ (most importantly its $T_{\rm eff}$) as well as extinction $A_V$\footnote{For simplicity we call our extinction parameter $A_V$, although it refers to $A_{5420\AA}$, as advertised by \citet{Schlafly2016}.} itself:
$$ \frac{A_i}{A_V} = \frac{2.5}{A_V} \cdot \log_{10} \dfrac{\int F_{\lambda} \cdot T^i_{\lambda} {\rm d}\lambda}{\int F_{\lambda} \cdot T^i_{\lambda} \cdot 10^{-0.4 a_{\lambda}\cdot A_V}{\rm d}\lambda}.$$
Here, $T^i_{\lambda}$ is the transmission curve, and $a_{\lambda}$ is the extinction law. Therefore, one has to compute the coefficients $A_{i}/A_V$ for each stellar model and each extinction value considered. In most of the recent literature concerning stellar distances, this effect is not taken into account, because for narrow-band and infra-red passbands, the extinction coefficient is roughly constant. For the {\it Gaia} passbands, however, this is not the case any more \citep{Jordi2010}. In the new version of {\tt StarHorse} we therefore use the Kurucz grid of synthetic stellar spectra \citep{Kurucz1993}\footnote{Provided by the Spanish Virtual Observatory's Theoretical Spectra web server (\url{http://svo2.cab.inta-csic.es/theory/newov2/index.php}).} to compute a grid of bolometric corrections as a function of $T_{\rm eff}$ and $A_V$ for each passband, and for our default extinction law \citep{Schlafly2016}. 

{\it Additional output:} While \citet{Queiroz2018} used spectroscopically determined stellar parameters as input and therefore only reported distances and extinctions (and in the case of high-resolution spectroscopy also masses and ages; e.g. \citealt{Anders2018a}), the absence of spectroscopically determined effective temperatures, gravities, and metallicities in the case of {\it Gaia}+photometry data led to the decision to also report the posterior values of $T_{\rm eff}, \log g,$ and [M/H]. Since the photometric estimates for $\log g$, [M/H], and stellar mass are of significantly lower precision, we regard these as {\it secondary} output parameters, in contrast to the {\it primary} output parameters $d, A_V$, and $T_{\rm eff}$. The secondary parameters were mainly obtained to test the targeting strategy of the 4MOST low-resolution disc and bulge survey \citep[4MIDABLE-LR;][]{Chiappini2019}, and the functionality of the 4MOST simulator (4FS; see \citealt{deJong2019}). %Due to the diverse precision and accuracy of the output parameters, we denominate the posterior distances, extinctions, and effective temperatures as {\it primary} output parameters, and the stellar masses, metallicities, and surface gravities as {\it secondary} output parameters.  
Furthermore, in addition to the $V$-band extinction values $A_V$, we also provide median extinction values in the {\it Gaia} DR2 passbands $G, G_{BP},$ and $G_{BP}$, as well as extinction-corrected absolute magnitude $M_{G_0}$, and dereddened colour $(G_{BP}-G_{BP})_0$.

{\it Computational updates:} Since \citet{Queiroz2018}, the {\tt StarHorse} code was migrated {\tt python 2.7} to {\tt python 3.6} and runs on the {\tt newton} cluster at the Leibniz-Institut f\"ur Astrophysik Potsdam (AIP). Due to several improvements in the data handling, the runtime was reduced by a factor of 6 as compared to the previous version used in \citet{Queiroz2018}. 

\subsection{{\tt StarHorse setup}} \label{setup}

\begin{figure}\centering
    \begin{tikzpicture}
       \node[anchor=south west,inner sep=0] (image) at (0,0) {\includegraphics[width=0.5\textwidth]{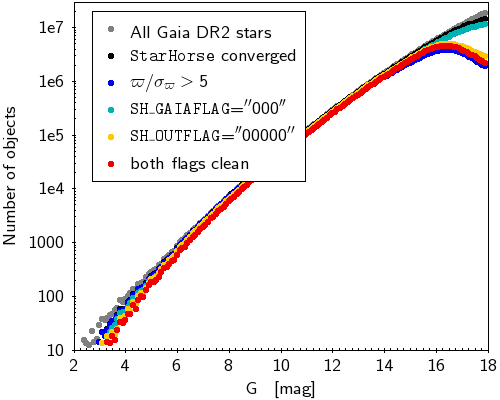}};
       \begin{scope}[x={(image.south east)},y={(image.north west)}]
       \node[anchor=south west,inner sep=0] (image) at (0.52,0.13) {\includegraphics[width=0.225\textwidth]{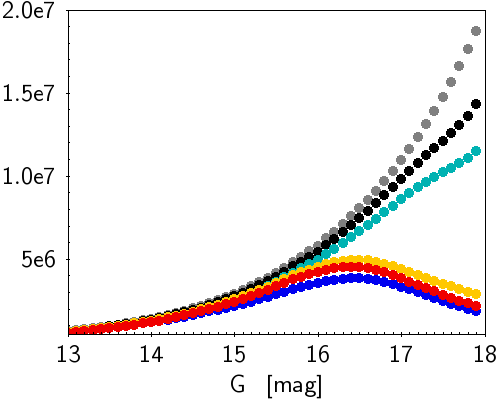}};
       \end{scope}
    \end{tikzpicture}
 	\caption{{\it Gaia} DR2 $G$ magnitude histogram, illustrating the magnitude coverage of the different {\tt StarHorse} sub-samples defined in Table \ref{summarytable}. Inset: zoom into the magnitude range $13<G<18$ with linear $y$ axis, illustrating the degrading parallax quality around $G\sim16.5$.}
 	\label{gmaghisto}
 \end{figure}
 
We then ran {\tt StarHorse} code \citep{Santiago2016, Queiroz2018}. In this work we used a grid of PARSEC 1.2S stellar models \citep{Bressan2012, Chen2014, Tang2014} in the 2MASS, Pan-STARRS1, {\it Gaia} DR2 rederived \citep{MaizApellaniz2018}, and WISE photometric systems available on the CMD webpage maintained by L. Girardi\footnote{\url{http://stev.oapd.inaf.it/cgi-bin/cmd_3.0}}. For $G\geq14$, we use a model grid equally spaced by 0.1 dex in log age as well as in metallicity [M/H]. Due to the higher precision of the {\it Gaia} DR2 parallaxes for $G<14$, we used a finer grid with 0.05 dex spacing in the bright regime. 

\begin{figure*}\centering
 	\includegraphics[width=0.99\textwidth]{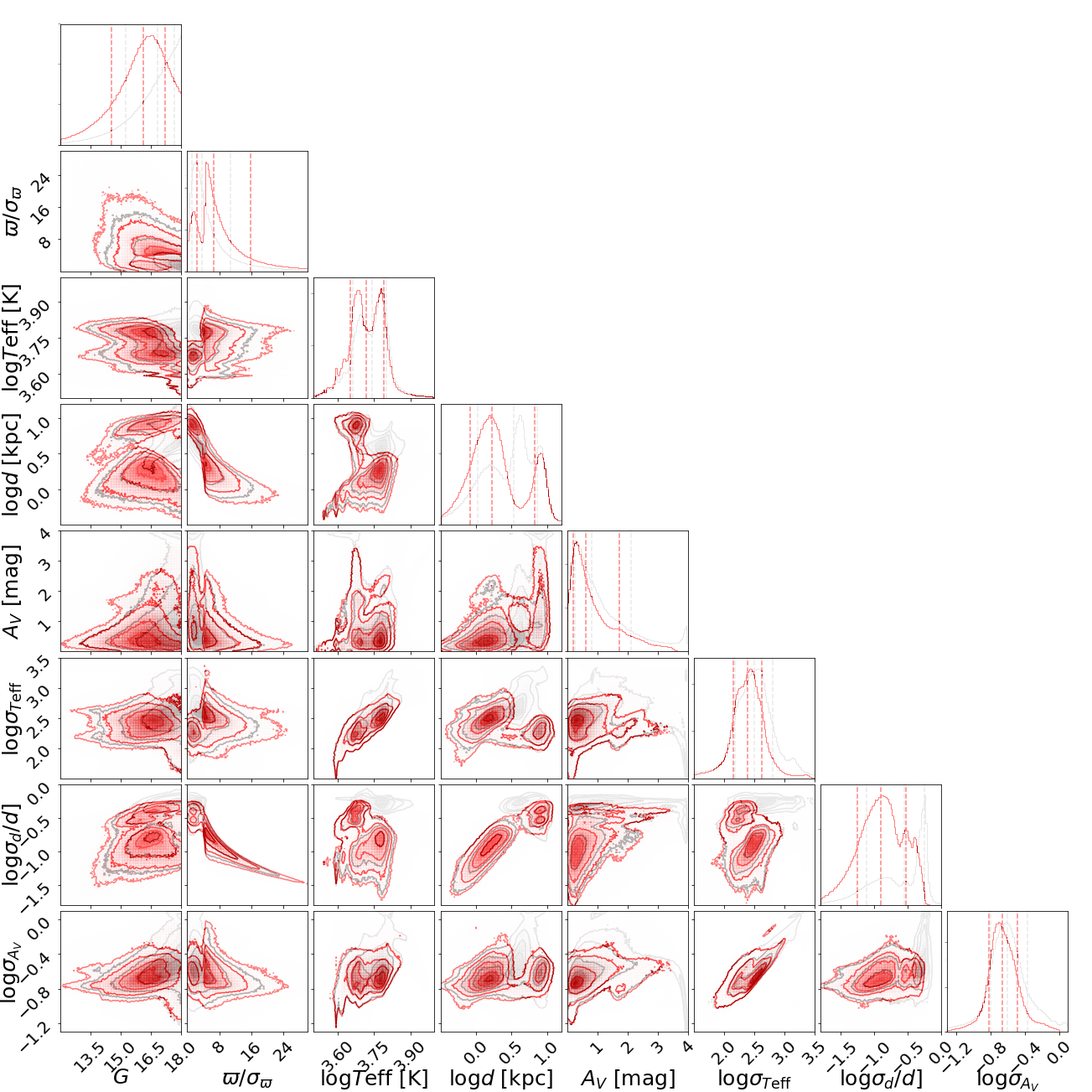}
 	\caption{{\tt corner} plot showing the correlations and distributions of {\tt StarHorse} median primary posterior output values $T_{\rm eff}, d,$ and $A_V$, their corresponding uncertainties, and the $G$ magnitude and parallax precision ($\varpi/\sigma_{\varpi}$). The grey contours show the distribution of the full sample, while the red contours show the distribution of all sources with {\tt SH\_OUTPUTFLAG}$==$"00000" and {\tt SH\_GAIAFLAG}$==$"000".}
 	\label{bigcorner}
 \end{figure*}

For computational reasons, depending on the parallax quality we used different ways to construct the range of possible distance values: for stars with well-determined parallaxes ($\varpi^{\rm cal}/\sigma_{\varpi}^{\rm cal} > 5$), we required the distances to lie within $\{ 1/(\varpi^{\rm cal}+4\cdot \sigma_{\varpi}^{\rm cal}), 1/(\varpi^{\rm cal}-4\cdot \sigma_{\varpi}^{\rm cal})\}$. For stars with less precisely measured parallaxes, we used their $G$ magnitudes to constrain the distance range for each possible stellar model (for details, see \citealt{Queiroz2018}).

For the case of {\it Gaia} DR2 run (i.e. in absence of spectroscopic data), the code took 1 second per star to run on the coarse grid ($G>14$, 270M stars), and 20 seconds per star on the fine grid ($G\leq14$, 16M stars). In total, the computational cost for this {\tt StarHorse} run thus was $\sim164,000$ CPU hours (19 years on a single CPU).
The global statistics for our output results are summarised in Table \ref{summarytable} and discussed in detail in Sect. \ref{results}.

\begin{figure*}\centering
 	\includegraphics[width=0.49\textwidth]{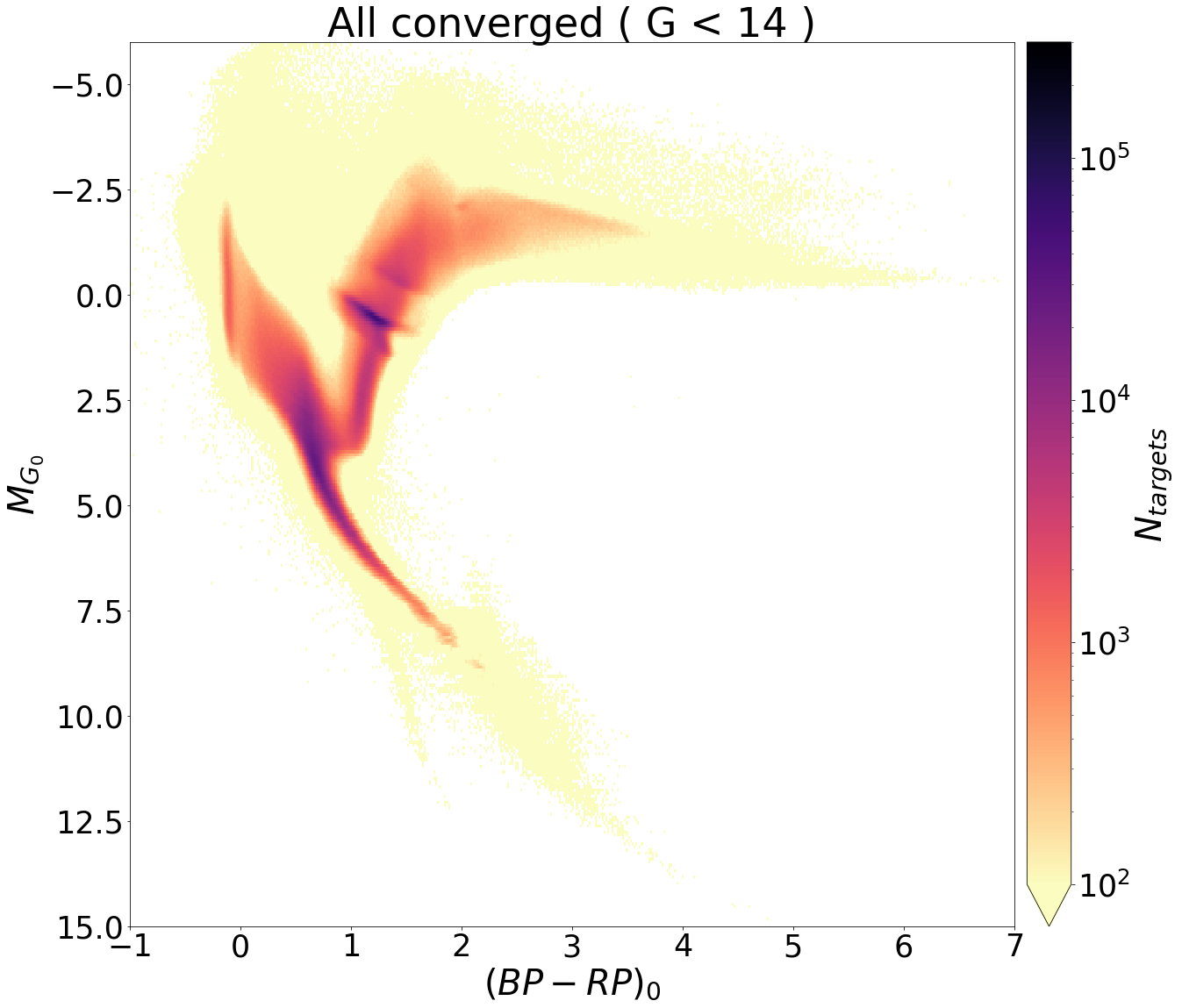}
 	\includegraphics[width=0.49\textwidth]{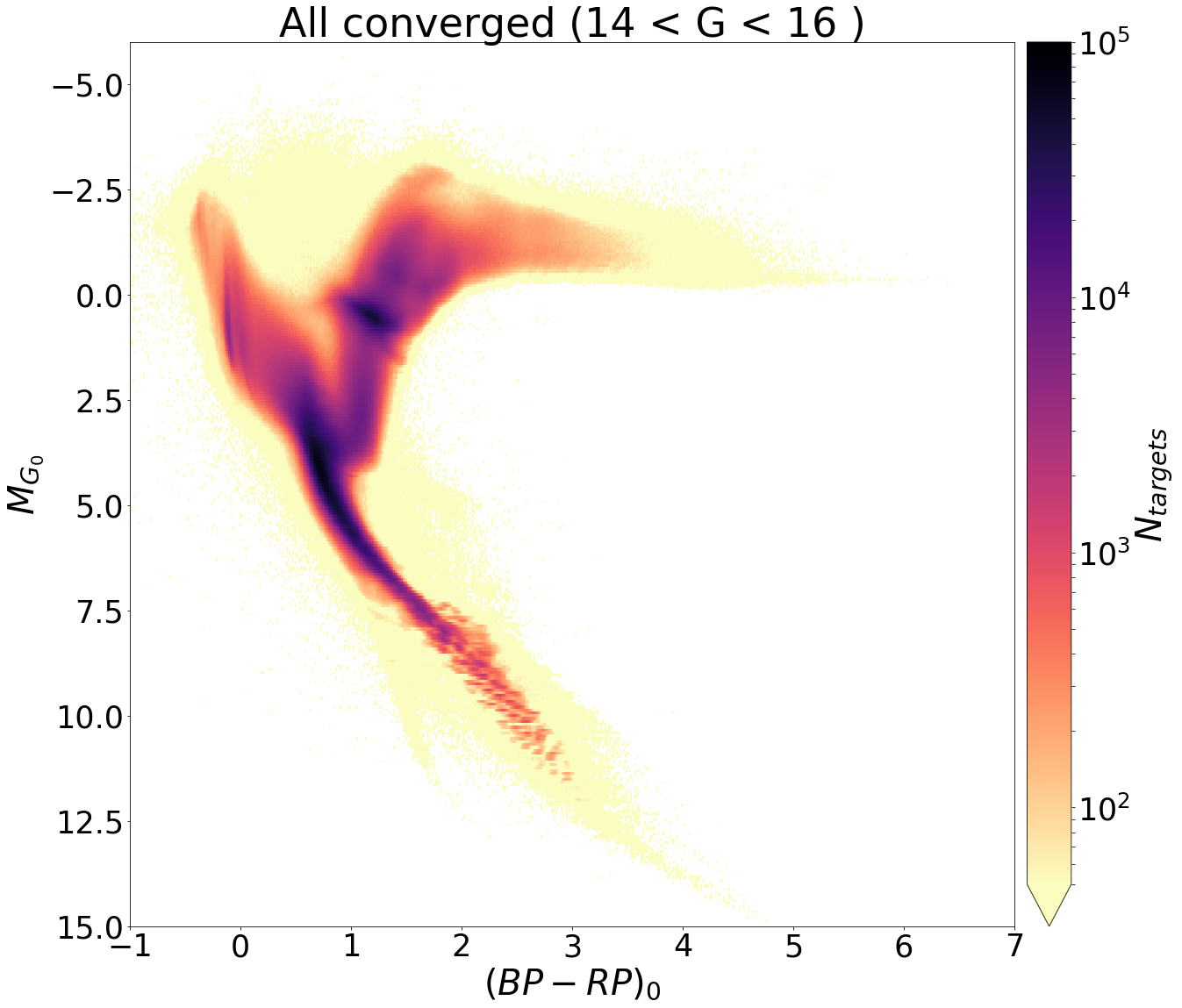}
 	\includegraphics[width=0.49\textwidth]{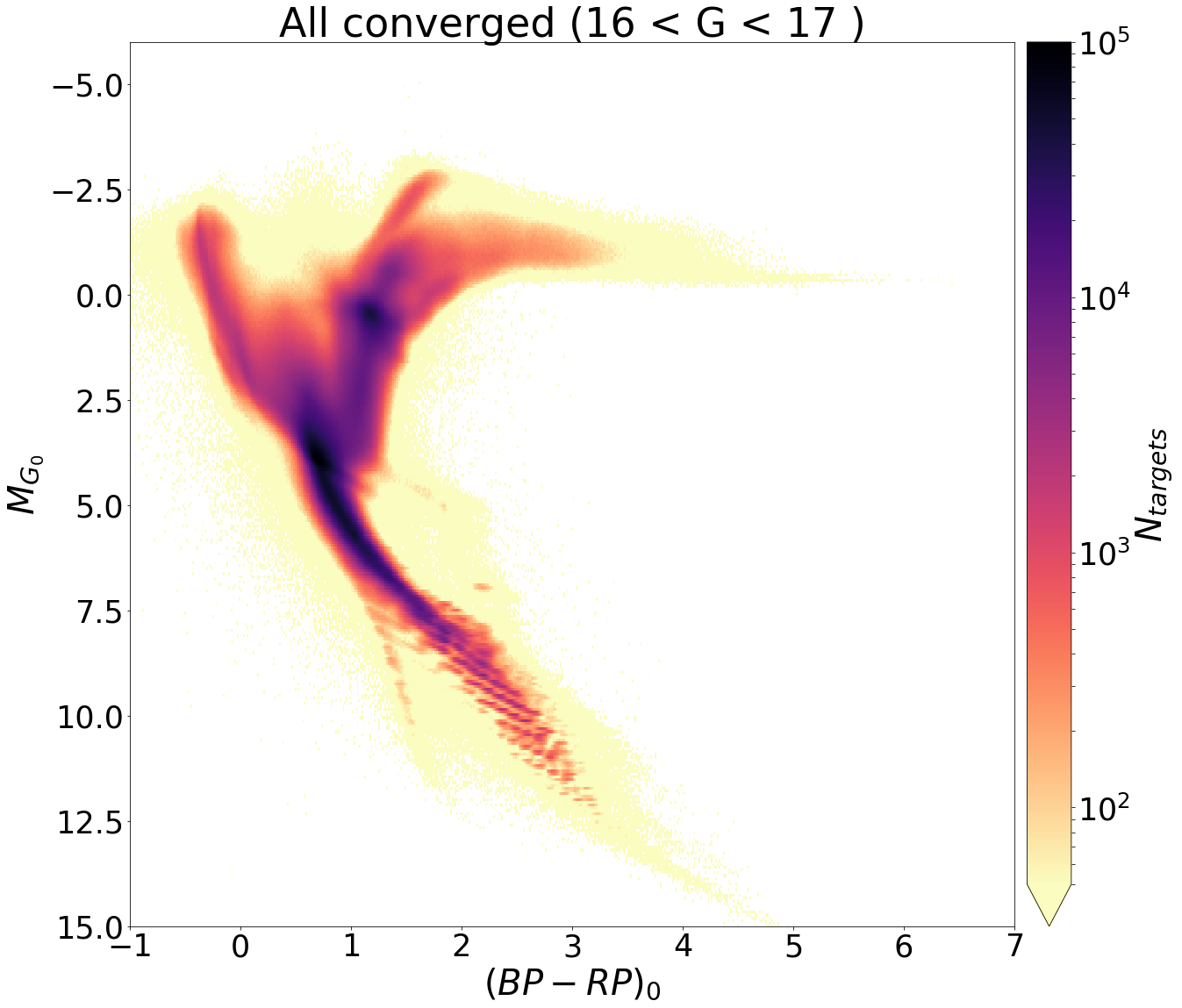}
 	\includegraphics[width=0.49\textwidth]{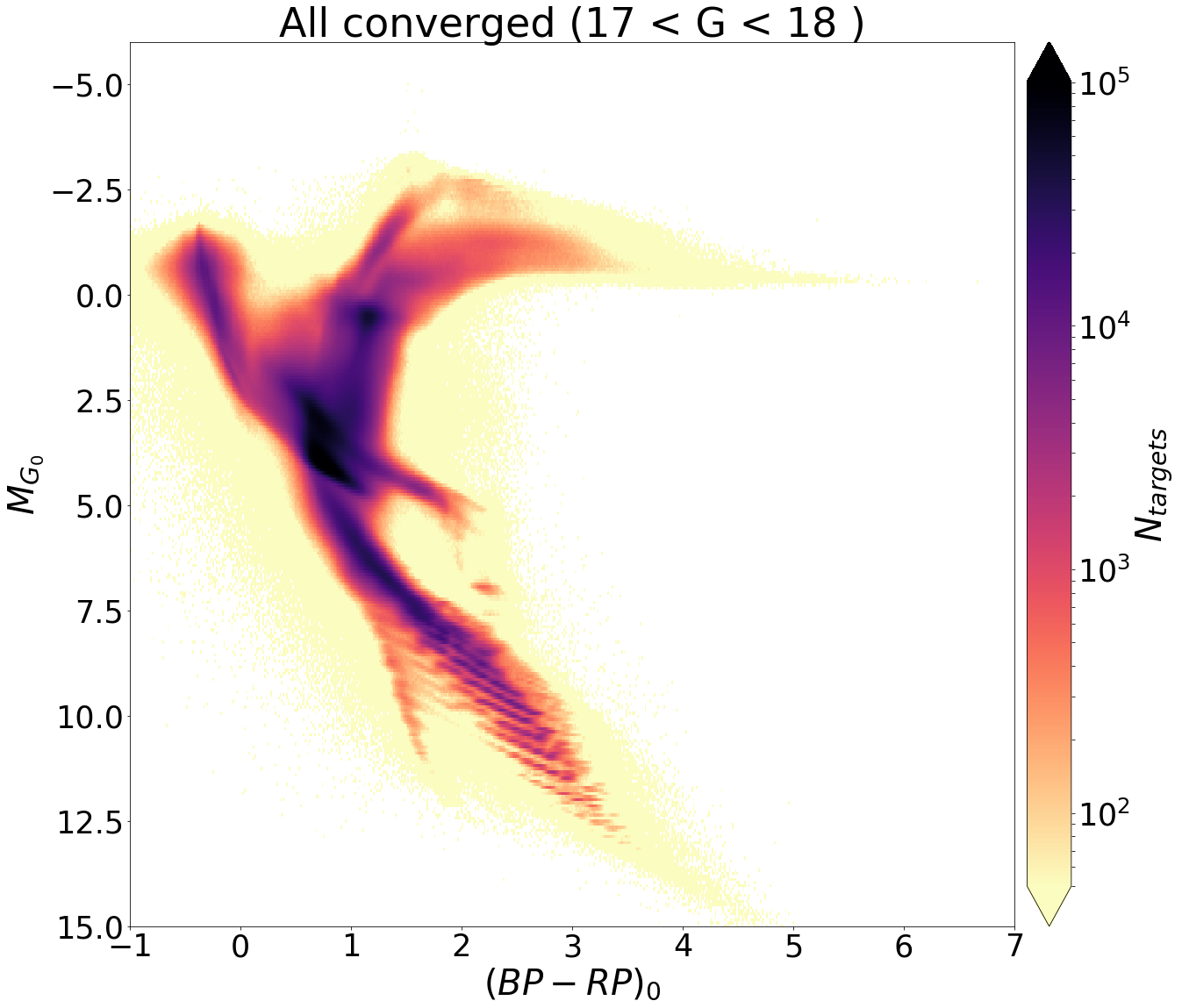}
 	\caption{{\tt StarHorse} posterior {\it Gaia} DR2 colour-magnitude diagrams for all converged stars in four magnitude bins, showing the degrading data quality from $G<14$ to $G>17$, making the use of the {\tt SH\_OUTFLAG} mandatory especially the faint regime (see Fig. \ref{cmds1}).
    }
 	\label{cmds2}
 \end{figure*}

\begin{figure*}\centering
 	\includegraphics[width=0.49\textwidth]{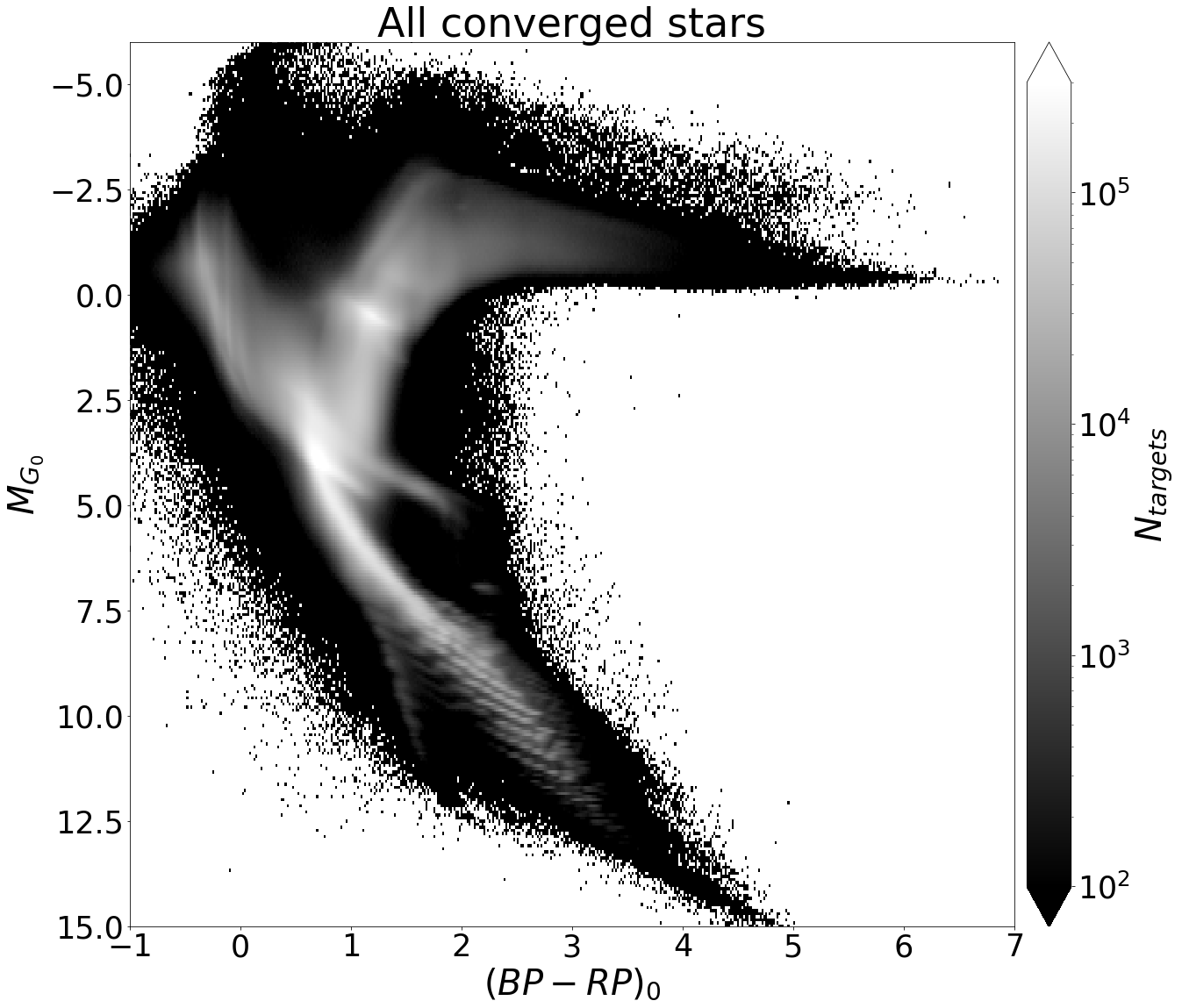}
 	\includegraphics[width=0.49\textwidth]{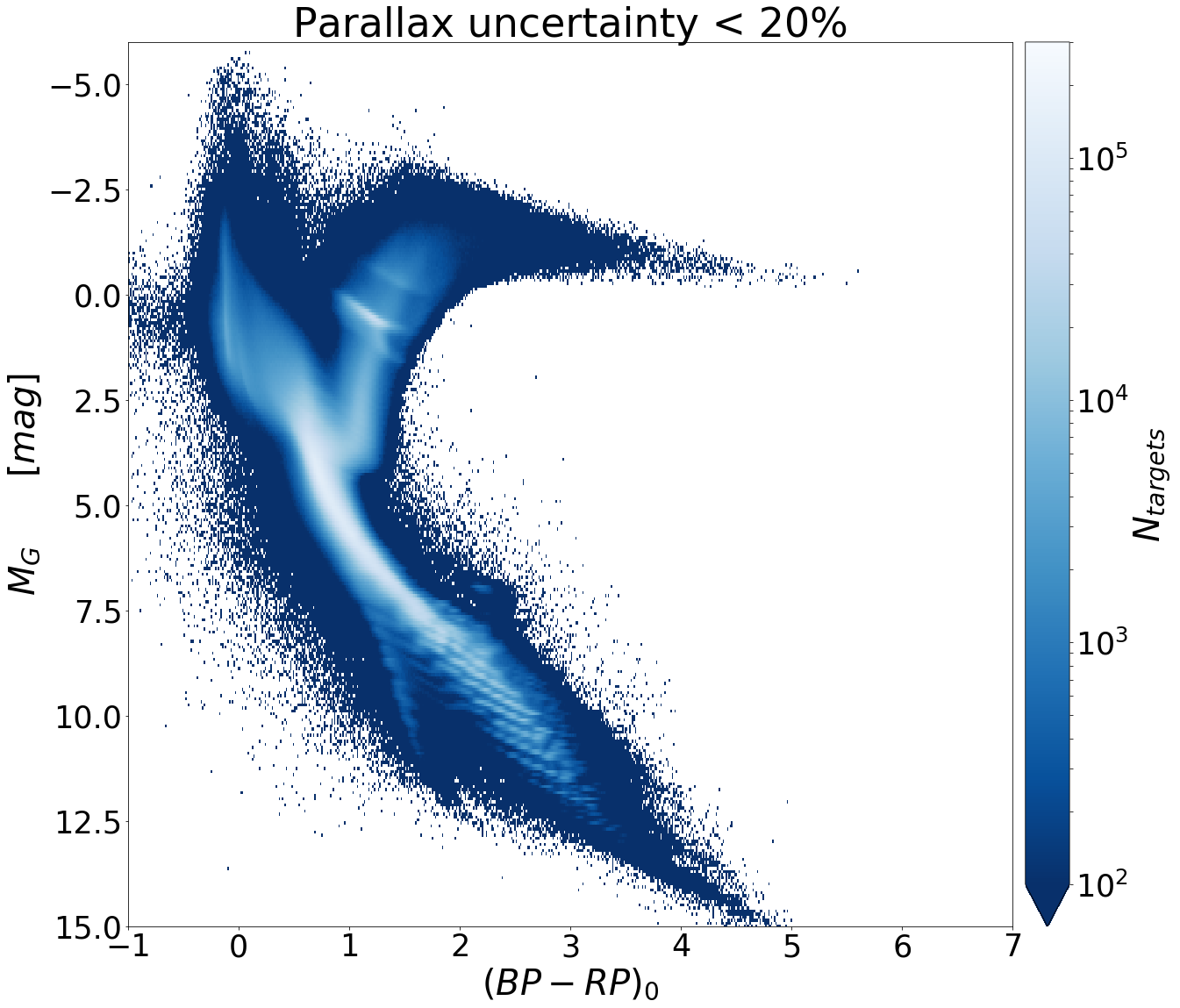}
 	\includegraphics[width=0.49\textwidth]{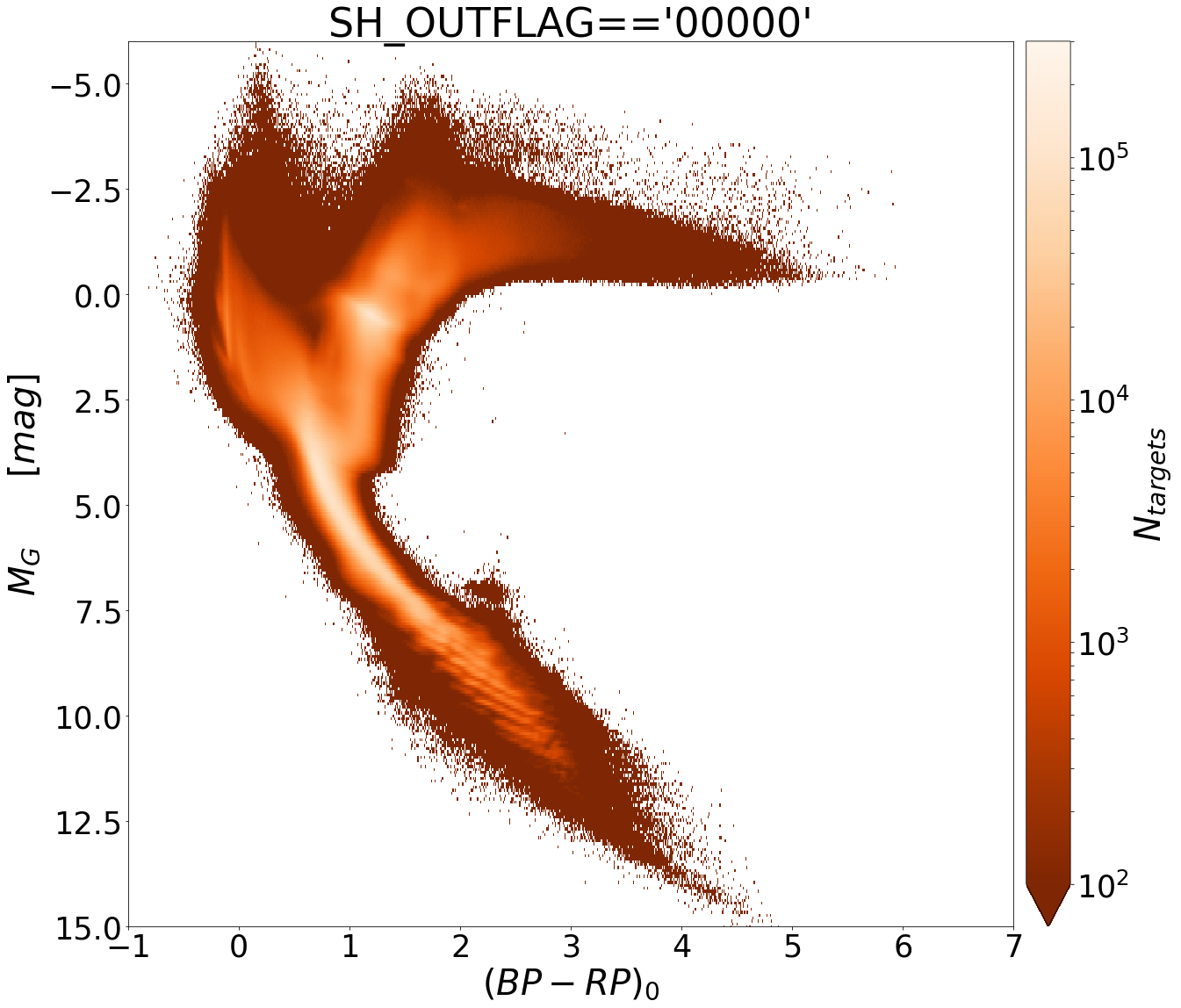}
 	\includegraphics[width=0.49\textwidth]{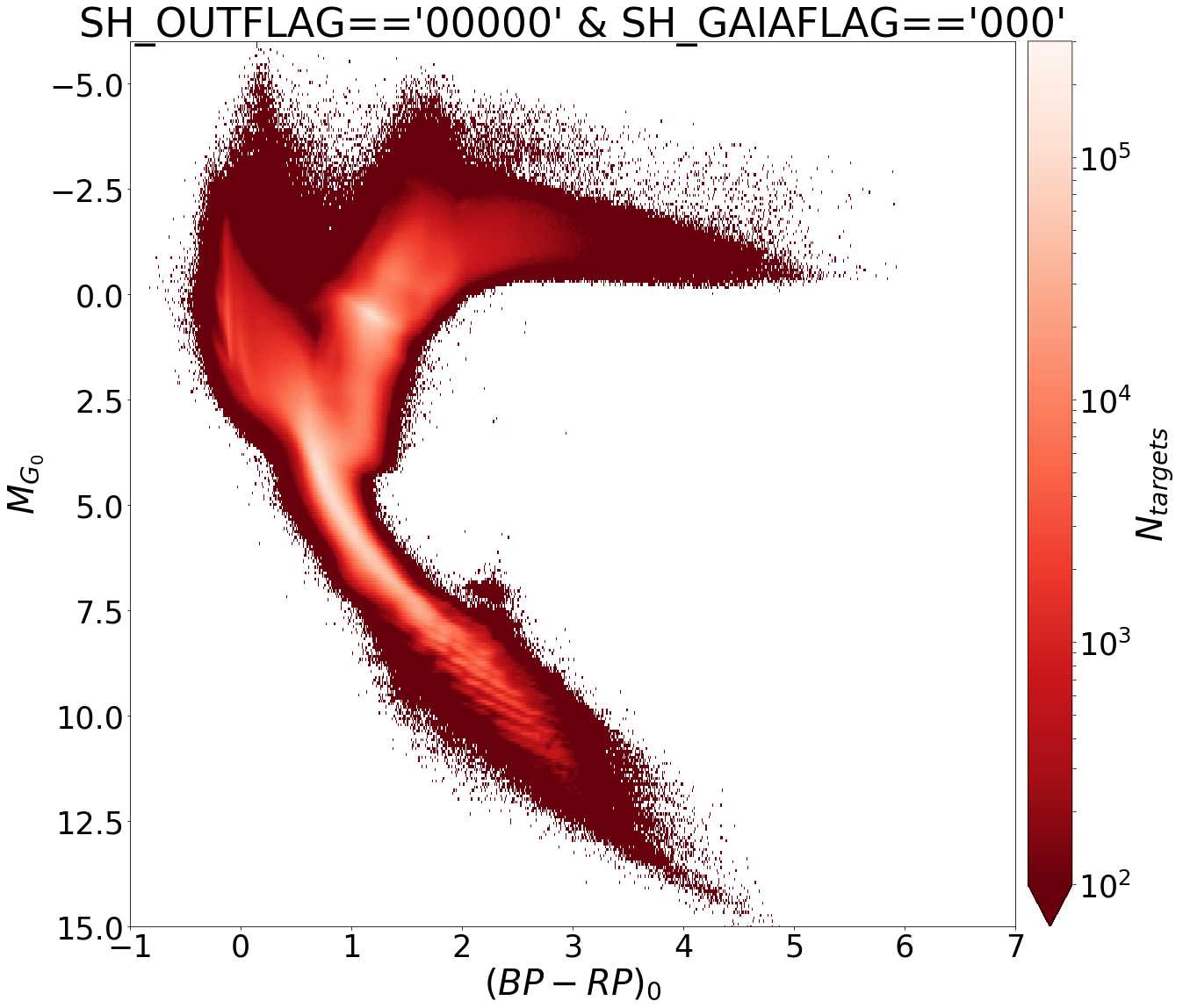}
 	\caption{{\tt StarHorse} {\it Gaia} DR2 colour-magnitude diagrams, colour-coded as in Table \ref{summarytable}. Top left: CMD resulting from all stars for which the code converged (266 million stars). Top right: emphasising sources with (recalibrated) parallaxes better than 20\% (103 million stars). Bottom row: emphasising the effect of cleaning the results by means of the {\tt StarHorse} flags (see discussion in Sec. \ref{flags}) Bottom left: cleaning only by {\tt SH\_OUTFLAG} (152 million stars). Bottom right: cleaning by both {\tt SH\_OUTFLAG} and {\tt SH\_GAIAFLAG} (137 million stars).
    }
 	\label{cmds1}
 \end{figure*}

\begin{figure*}\centering
 	\includegraphics[width=0.33\textwidth]{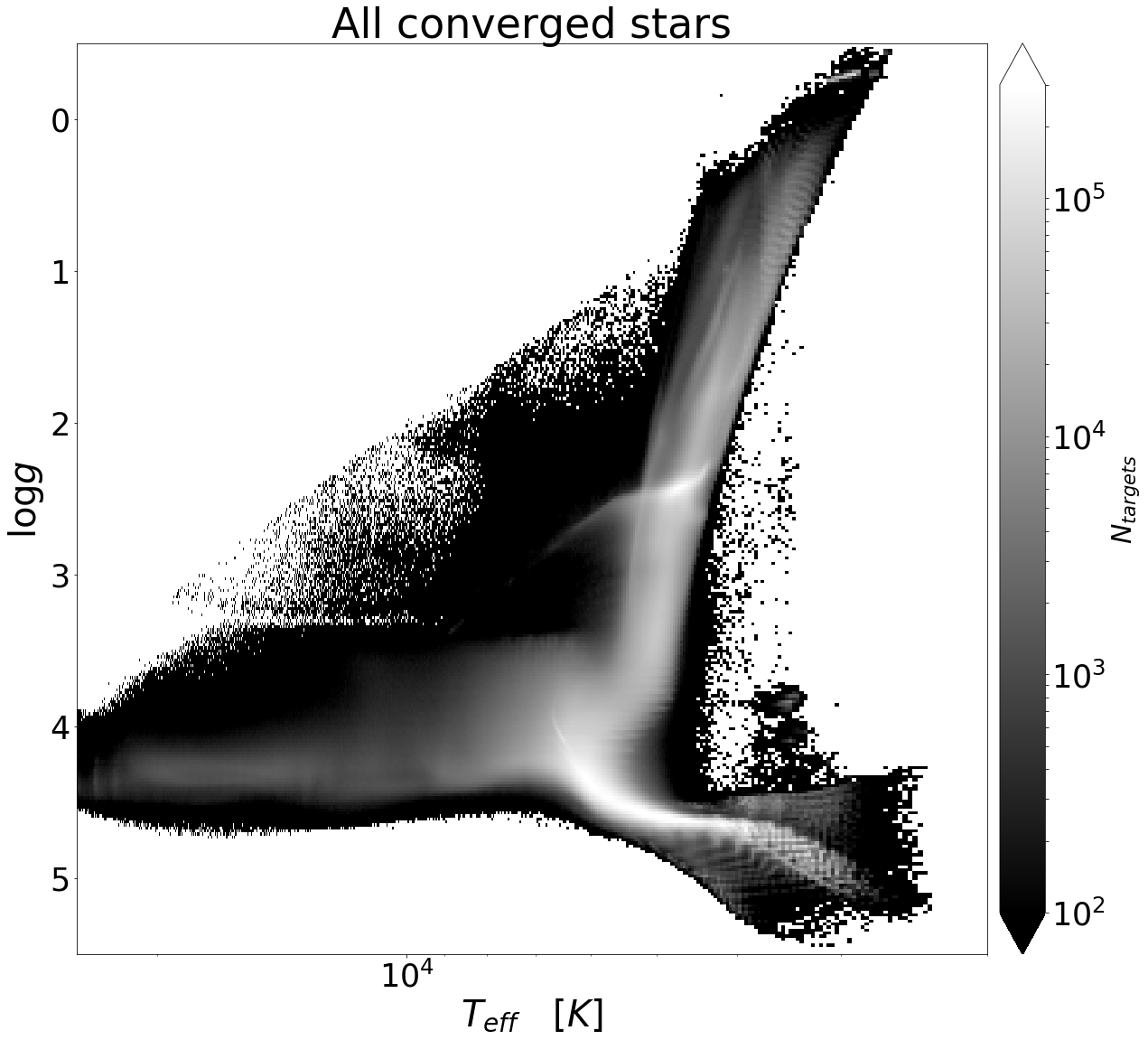}
 	\includegraphics[width=0.33\textwidth]{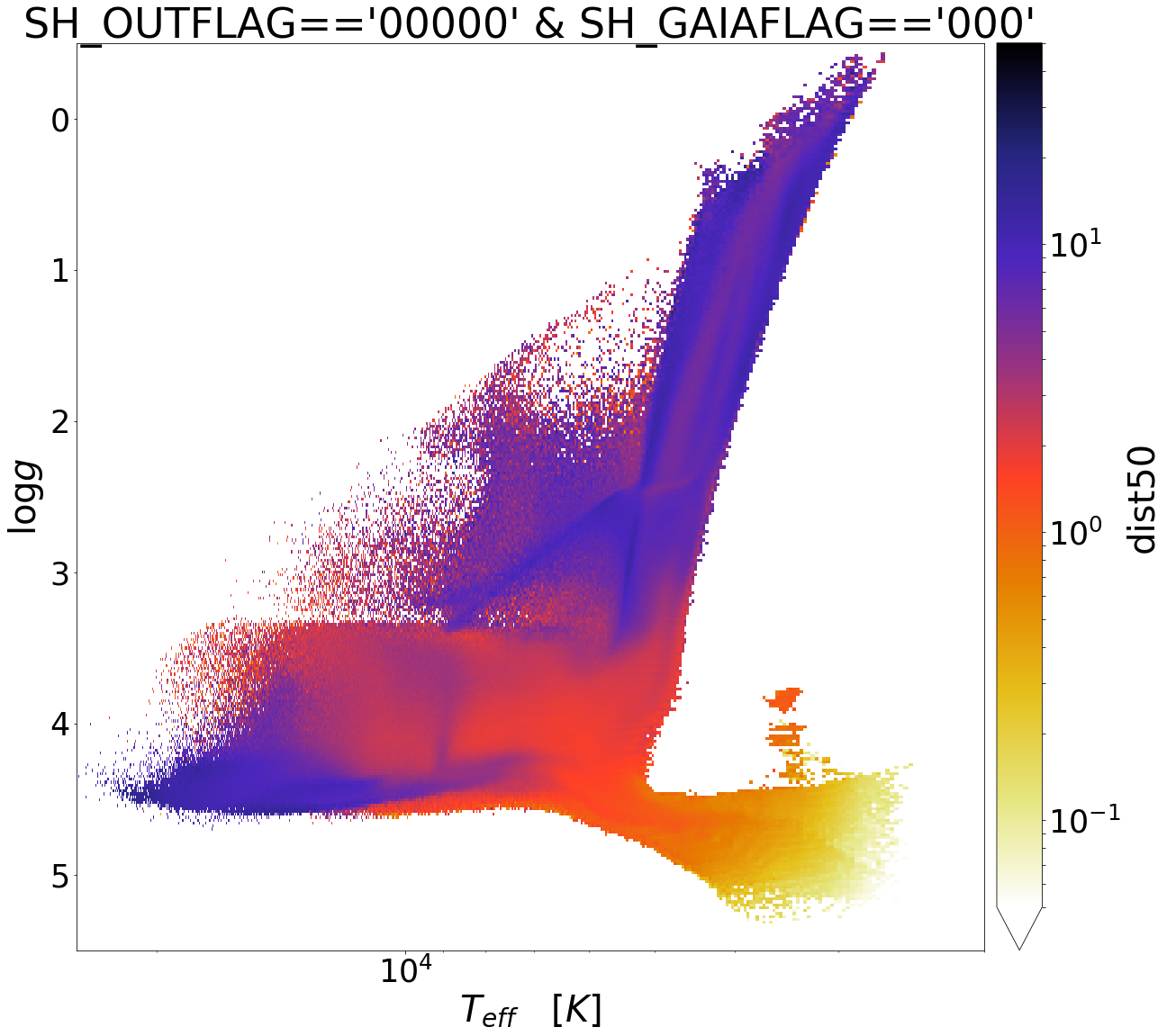}
 	\includegraphics[width=0.33\textwidth]{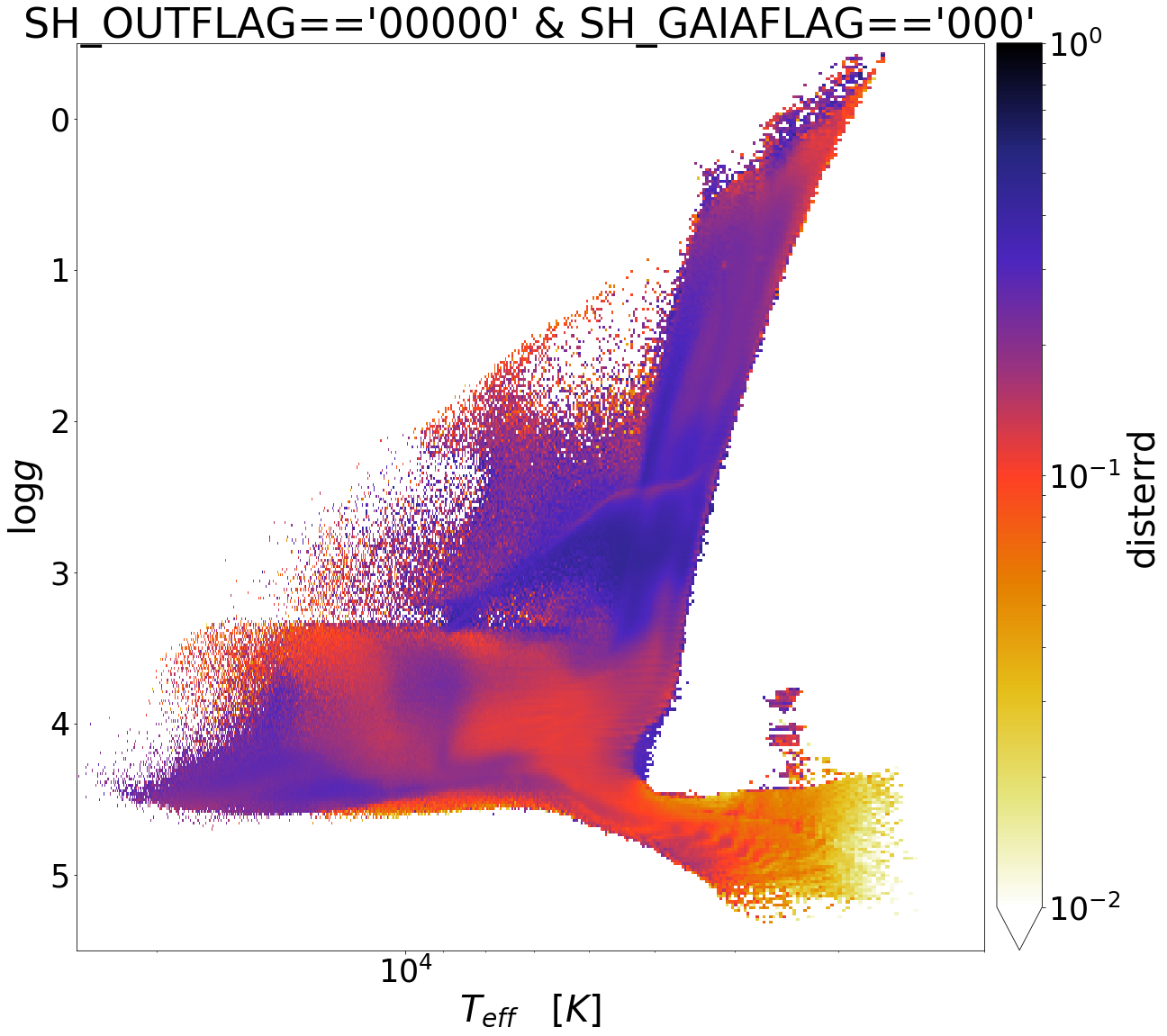}\\
 	\includegraphics[width=0.33\textwidth]{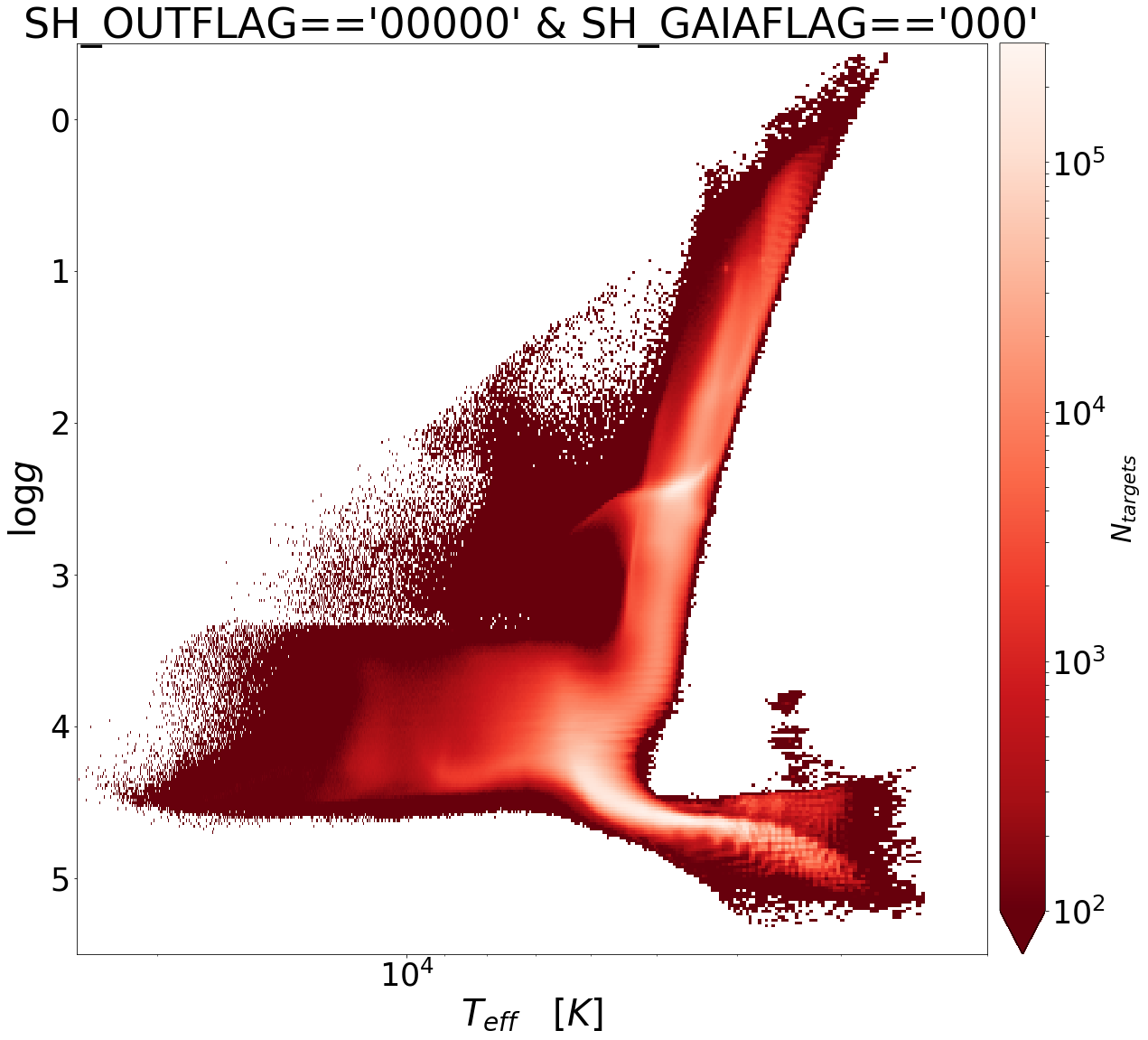}
 	\includegraphics[width=0.33\textwidth]{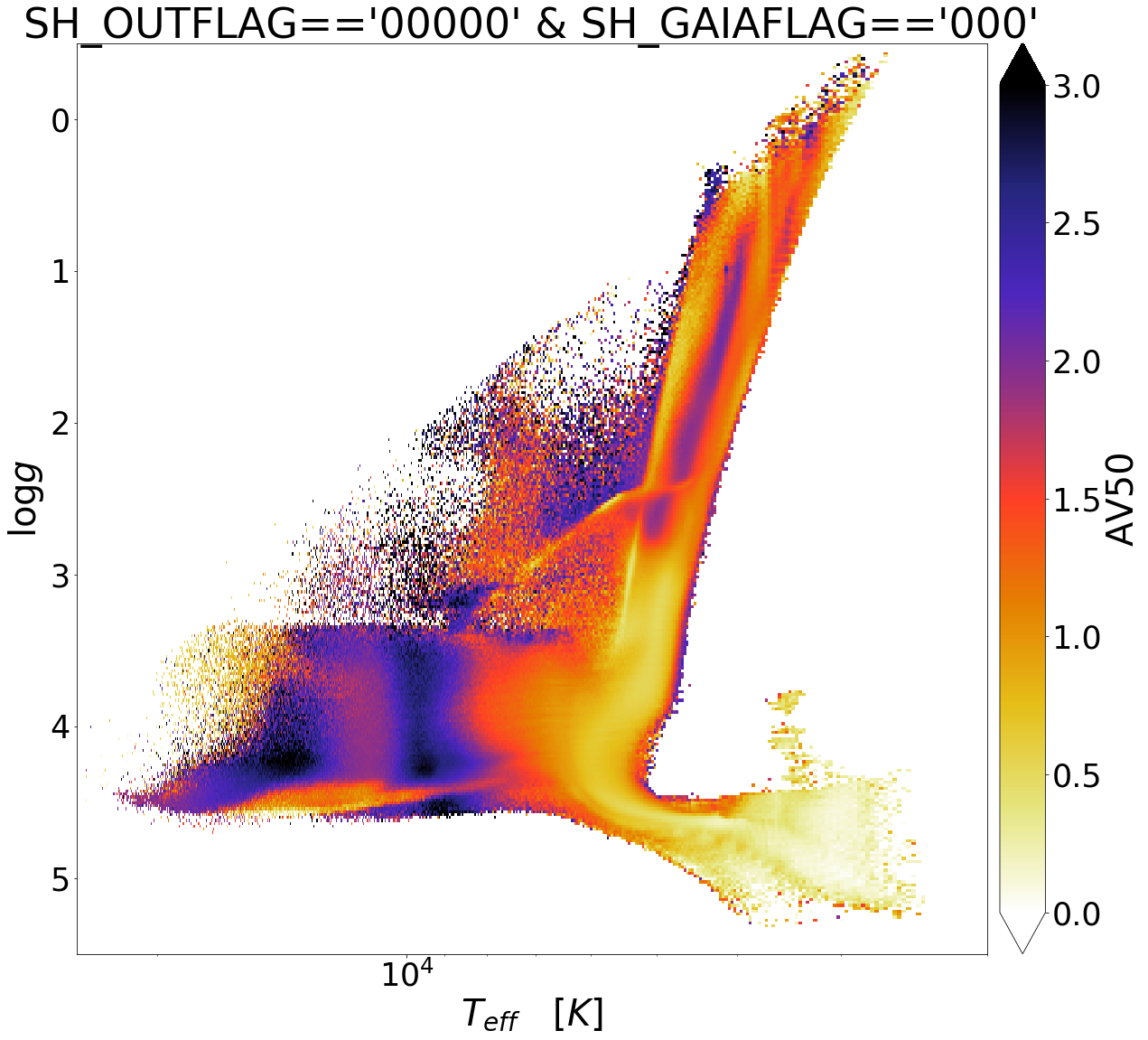}
 	\includegraphics[width=0.33\textwidth]{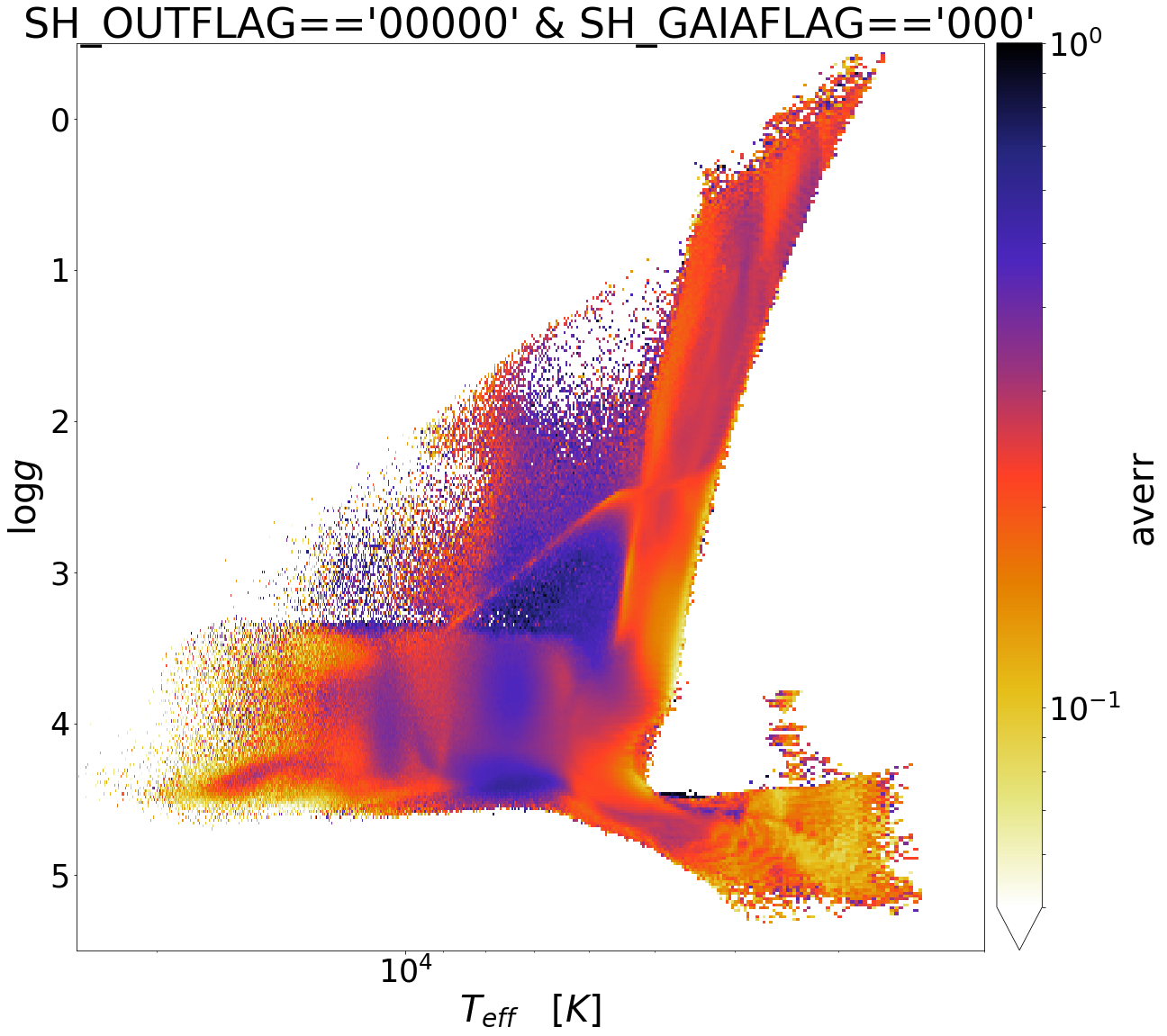}\\
    \caption{{\tt StarHorse}-derived {\it Kiel} diagrams. Top left: overall density plot before applying quality cuts. Top middle: colour-coded by median distance. Top right: colour-coded by median distance uncertainty. Lower left: flag-cleaned sample coloured by density. Lower middle: colour-coded by median $A_V$. Lower right: colour-coded by median $A_V$ uncertainty. 
    }
 	\label{kieldiagrams1}
\end{figure*}

\subsection{Input and output flags}\label{flags}

Along with the output of our code (median statistics of the marginal posterior in distance, extinction, and stellar parameters), we provide a set of flags to help the user decide which subset of the data to use for their particular science case. These flags correspond to the following columns.

\subsubsection{\tt SH\_GAIAFLAG}
This flag describes the overall astrometric and photometric quality of the {\it Gaia} DR2 data for each star in a three-digit flag (similar to the {\it Gaia} DR2-native {\tt priam\_flag}\footnote{\url{https://gea.esac.esa.int/archive/documentation/GDR2/Data_analysis/chap_cu8par/sec_cu8par_data/ssec_cu8par_data_flags.html}}). Balancing simplicity and the recommendations of \citet{Lindegren2018} and \citet{Lindegren2018a}, we limit this flag to the following three digits:
\begin{enumerate}
\item {\it Renormalised unit weight error flag:} \citet{Lindegren2018a} recently showed that instead of following the astrometric quality requirements used by \citet{GaiaCollaboration2018, Lindegren2018}, and \citet{Arenou2018}, similar or better cleaning of spurious {\it Gaia} DR2 astrometry can be obtained by requiring a maximum value for the so-called renormalised unit weight error ({\tt ruwe}). We therefore defined the first digit as follows: 

\quad IF {\tt ruwe}$<1.4$ THEN 0 ELSE 1
%\item IF {\tt astrometric\_excess\_noise} $< 1.0$ THEN 0 ELSE 1
\item {\it Colour excess factor flag:} \citet{Evans2018} and \citet{Arenou2018} recommend the use of the {\tt phot\_bp\_rp\_excess\_factor} to flag spurious {\it Gaia} DR2 photometry. We follow their recommendation and define the second digit as:

\quad IF $G_{\rm BP}-G_{\rm RP}$ IS NULL THEN 2 ELIF $1.0+0.015\cdot (G_{\rm BP}-G_{\rm RP})^2 < {\tt phot\_bp\_rp\_excess\_factor} < 1.3+0.060\cdot (G_{\rm BP}-G_{\rm RP})^2$ THEN 0 ELSE 1
%\item IF {\tt visibility\_periods\_used} $> 8.$ THEN 0 ELSE 1
%\item {\tt duplicate\_flag}
\item {\it Variability flag:} The third digit equals the {\it Gaia} DR2-native {\tt phot\_variable\_flag}.
\end{enumerate}
\subsubsection{\tt SH\_PHOTOFLAG}
The human-readable {\tt SH\_PHOTOFLAG} input flag details which combination of photometric data ({\it Gaia}, Pan-STARRS1, 2MASS, WISE) was used as input for {\tt StarHorse}. For example, if photometry in all passbands was available for a star, the {\tt SH\_PHOTOFLAG} entry reads {\tt GBPRPgrizyJHKsW1W2}. If only {\it Gaia} DR2 $G$ and Pan-STARRS1 $izy$ magnitudes were available, the flag reads {\tt Gizy}. In addition, in the rare case that no uncertainty for a particular photometric band was available from the original catalogue and the fiducial uncertainty of 0.3 mag was used instead (see Sect. \ref{data}), we added a '{\tt \#}' to the corresponding passband. For example, if a star has complete {\it Gaia} DR2, Pan-STARRS1, and WISE photometry, but the $r$ and $W2$ magnitudes come without uncertainties, then the {\tt SH\_PHOTOFLAG} entry would be {\tt GBPRPgr\#izyW1W2\#}.

\subsubsection{\tt SH\_PARALLAXFLAG}
The {\tt SH\_PARALLAXFLAG} input flag informs about the precision of the {\it Gaia} DR2 parallaxes (accounting for zero-point shift and uncertainty corrections; see Table \ref{calibtable}). For ${\varpi}^{\rm cal}/\sigma_{\varpi}^{\rm cal}>5$ (parallaxes better than 20\%), the flag reads {\tt gtr5}, else {\tt leq5}. As explained in Sec. \ref{setup}, this has consequences for the construction of the posterior PDF in the {\tt StarHorse} code: if the parallax is precise, then the range of possible distances is computed directly from the parallax itself (allowing for 4$\sigma$ deviations). On the other hand, if only uncertain parallaxes are available, the range of possible distance moduli is constructed based on the measured $G$ magnitude. We verified that this choice does not produce different results for stars near the decision boundary (the rupture in Fig. \ref{sigdist_piepi} does not occur at the decision boundary ${\varpi}^{\rm cal}/\sigma_{\varpi}^{\rm cal}=5$, but at $\simeq1./0.22\simeq4.55$, which is where the standard deviation of the inverse parallax PDF increases sharply; see \citealt{Bailer-Jones2015a}, Sect. 4.1).

\subsubsection{\tt SH\_OUTFLAG}\label{outflag}
The {\tt StarHorse} output flag, similar to {\tt SH\_GAIAFLAG}, consists of several digits that inform about the fidelity of the {\tt StarHorse} output parameters.

\begin{figure*}\centering
 	\includegraphics[width=0.49\textwidth]{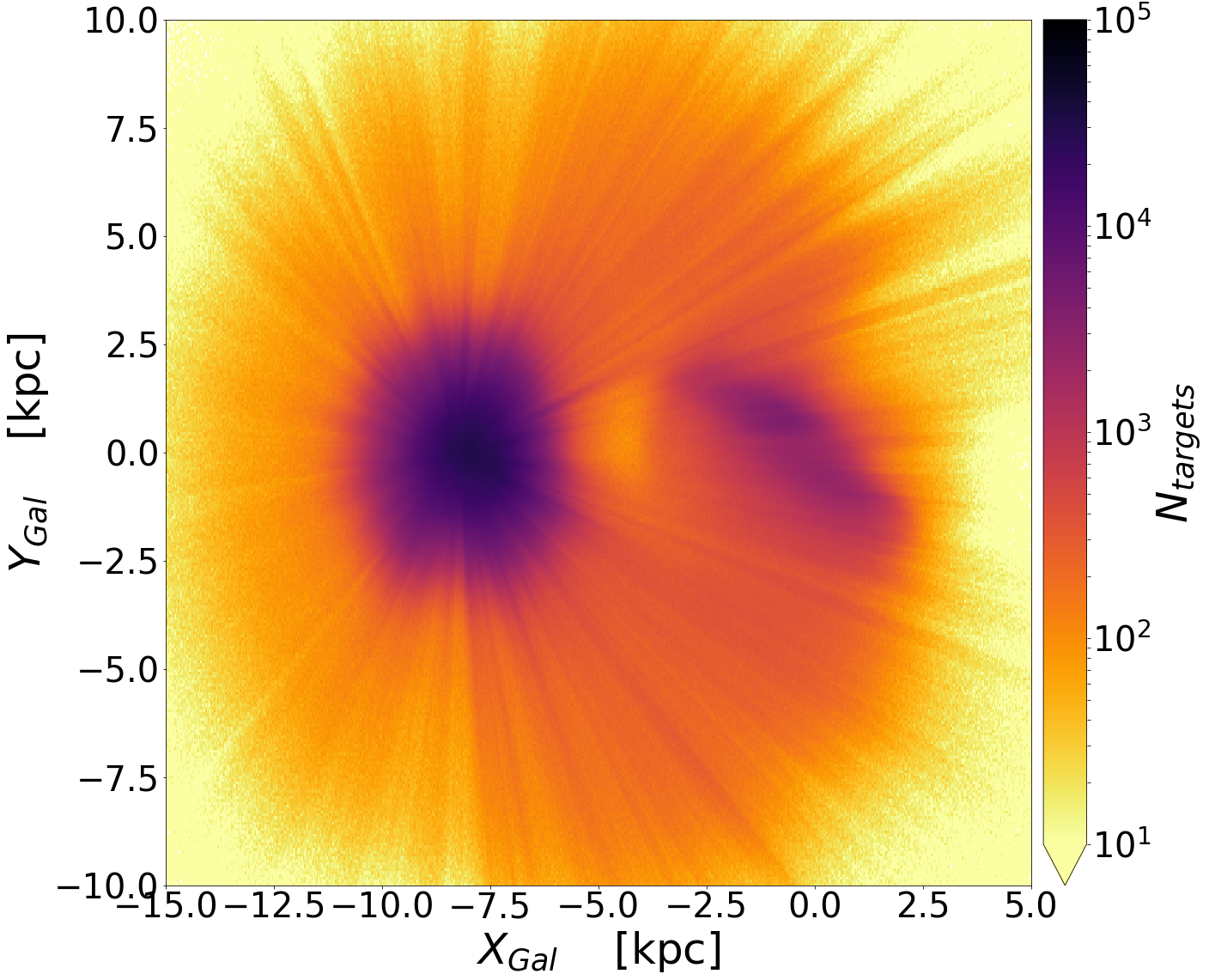}
 	\includegraphics[width=0.49\textwidth]{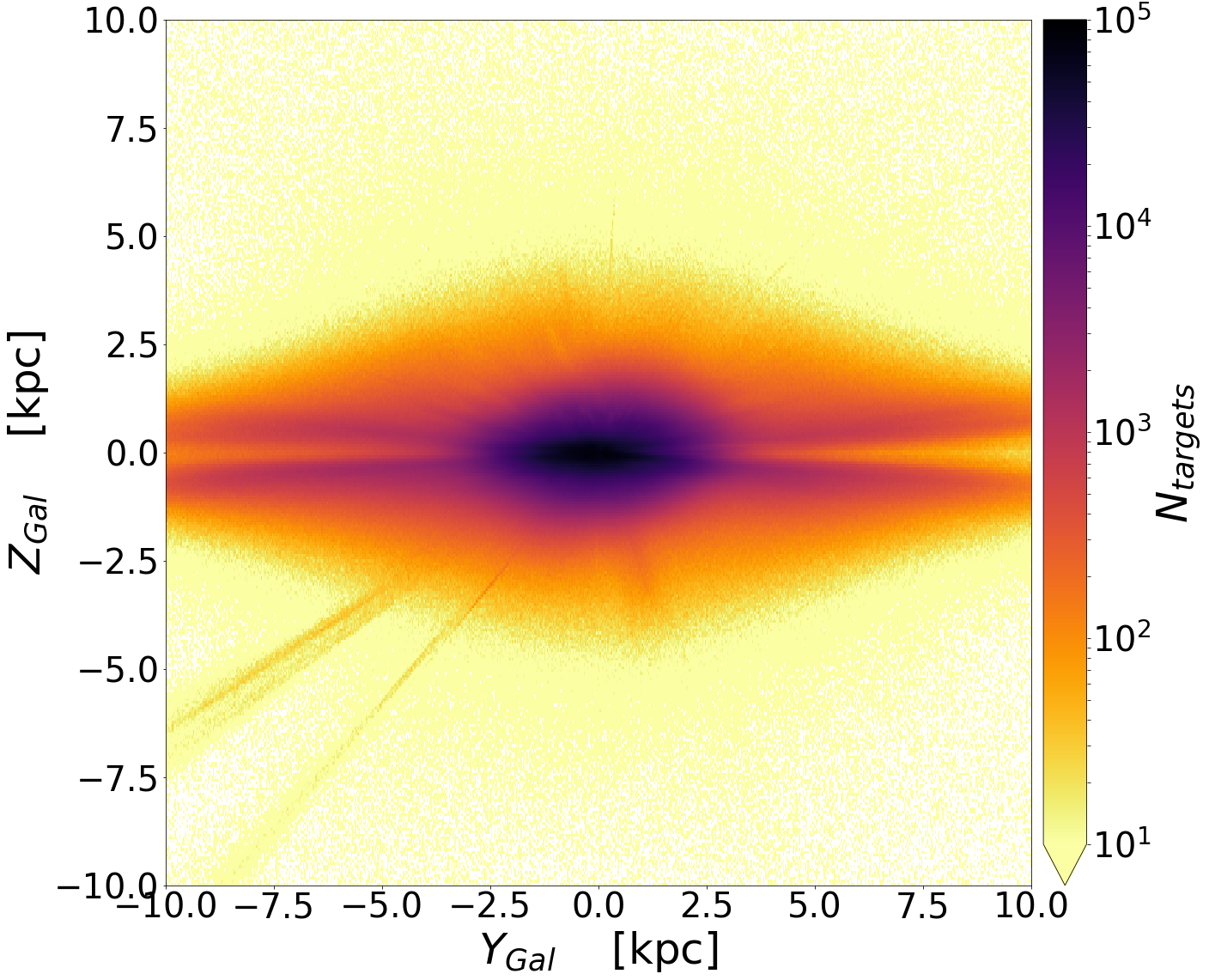}
 	\includegraphics[width=0.49\textwidth]{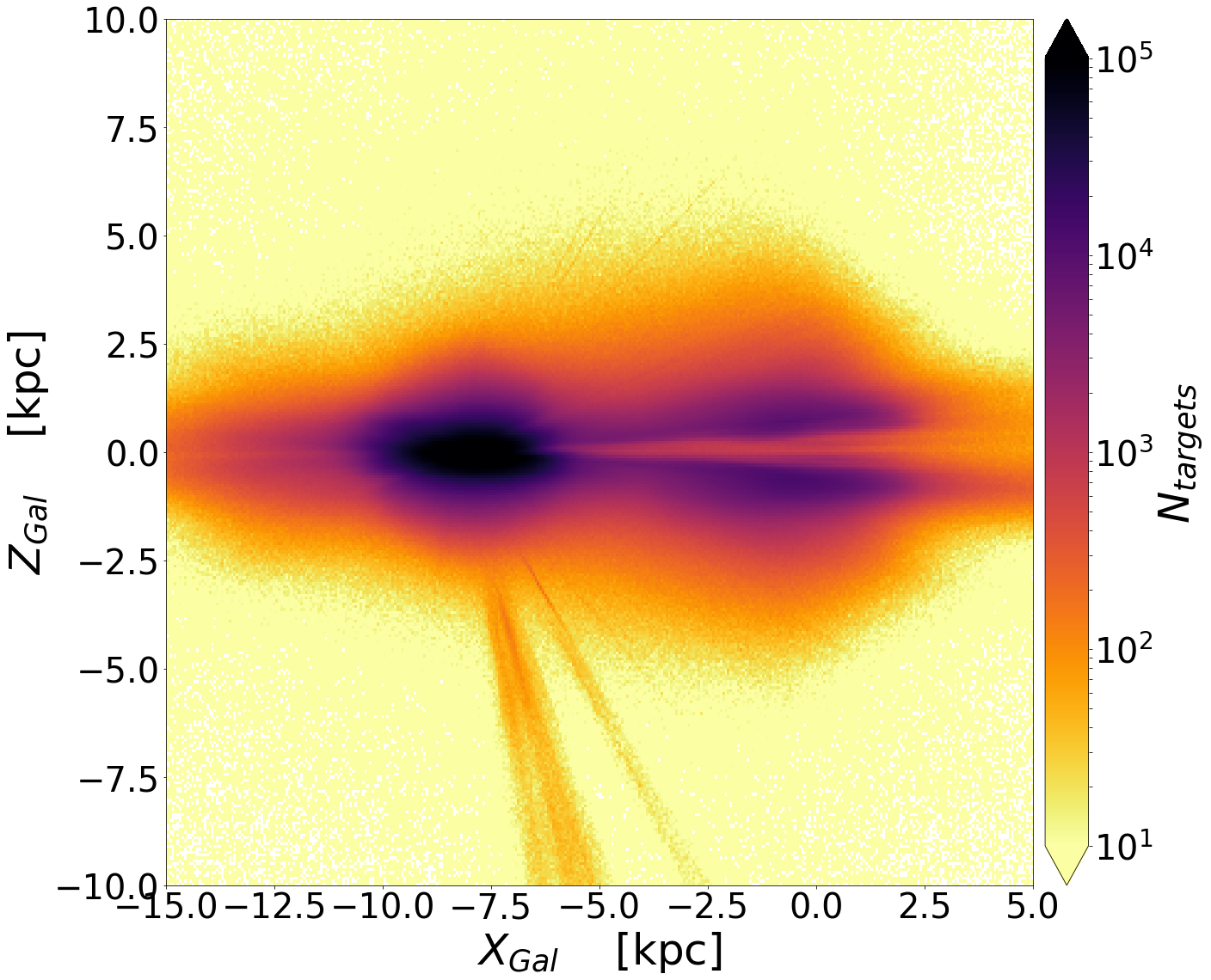}
 	\includegraphics[width=0.49\textwidth]{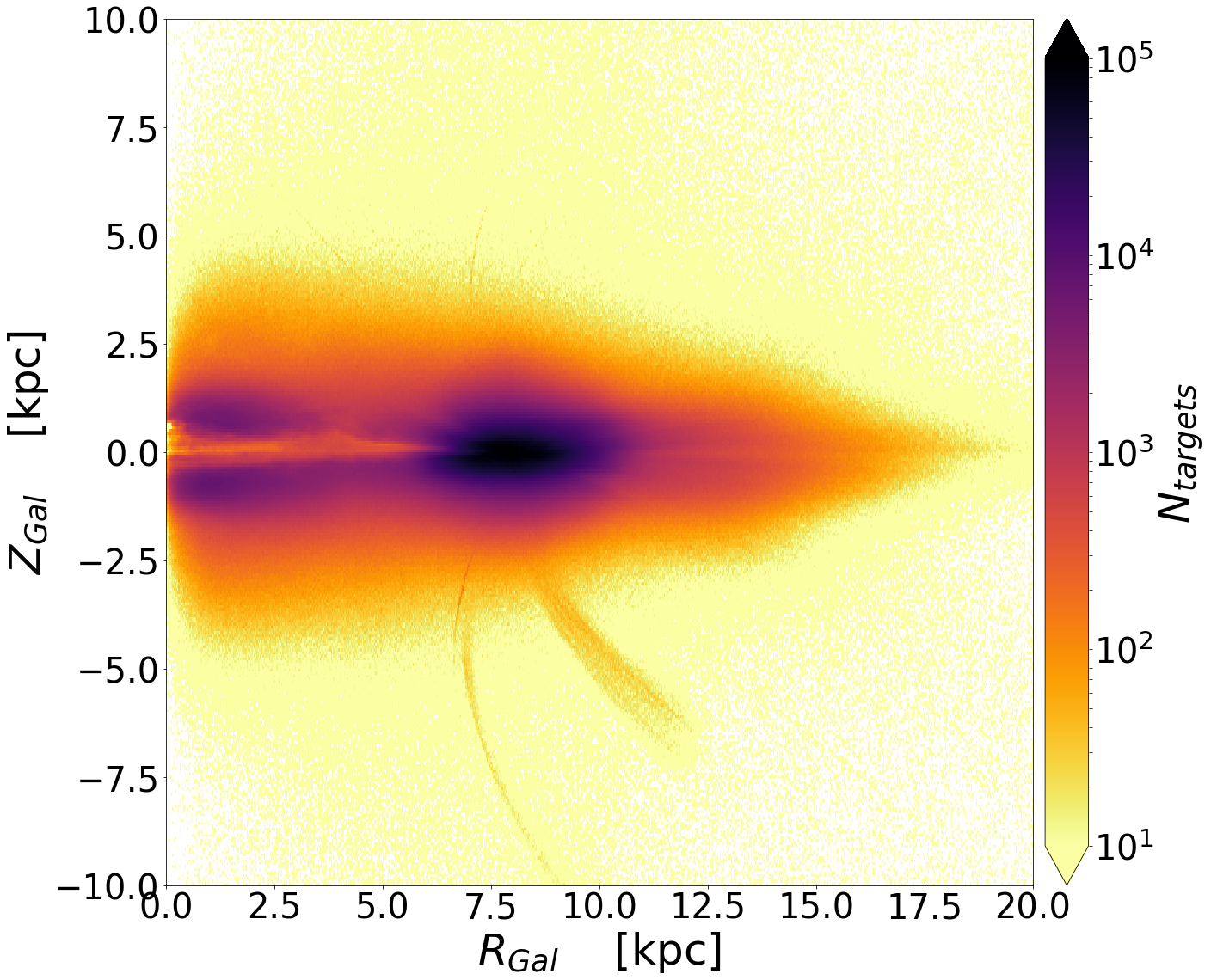}
 	\caption{Left: {\tt StarHorse} density maps for the {\tt SH\_GAIAFLAG}$=="000", ${\tt SH\_OUTFLAG}$="00000"$ sample in Galactocentric co-ordinates. Top left: $XY$ map. Top right: $YZ$ map. Bottom left: $XZ$ map. Bottom right: $RZ$ map. These density maps demonstrate that {\it Gaia} DR2 already allows to probe stellar populations in the Galactic bulge and beyond.}
 	\label{rzmaps}
\end{figure*}

\begin{figure}\centering
 	\includegraphics[width=0.5\textwidth]{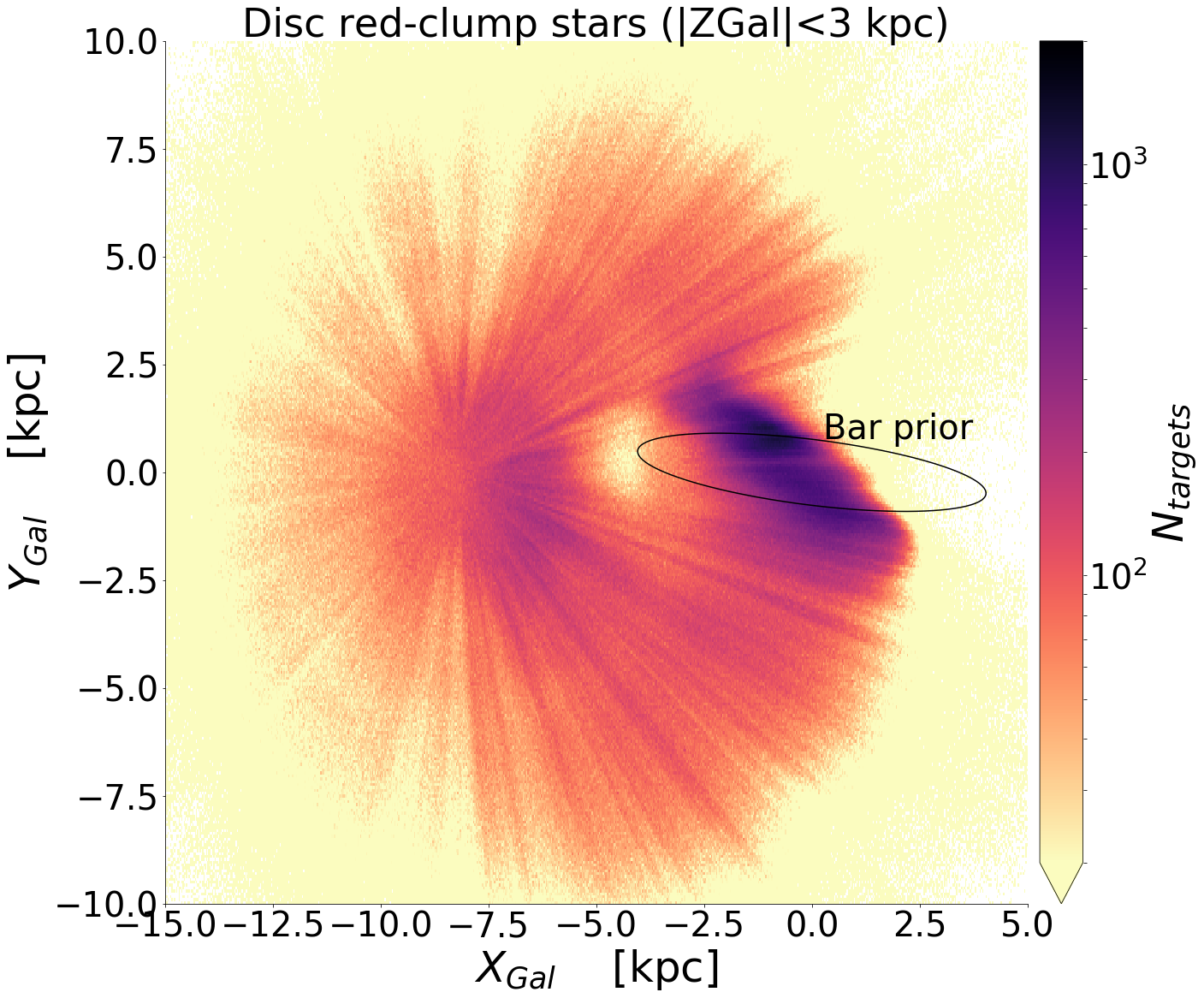}
 	\caption{XY density map, selecting only flag-cleaned red-clump stars less than 3 kpc away from the Galactic midplane. 10,807,155 stars are contained in this figure. The ellipse indicates the shape of the bar/bulge density prior adopted for this work (\citealt{Robin2012a} model B; see \citealt{Queiroz2018} for details).}
 	\label{bar_rc_stars}
 \end{figure}

\begin{figure}\centering
 	\includegraphics[width=0.49\textwidth, trim={0 3.9cm 0 4.9cm},clip=True]{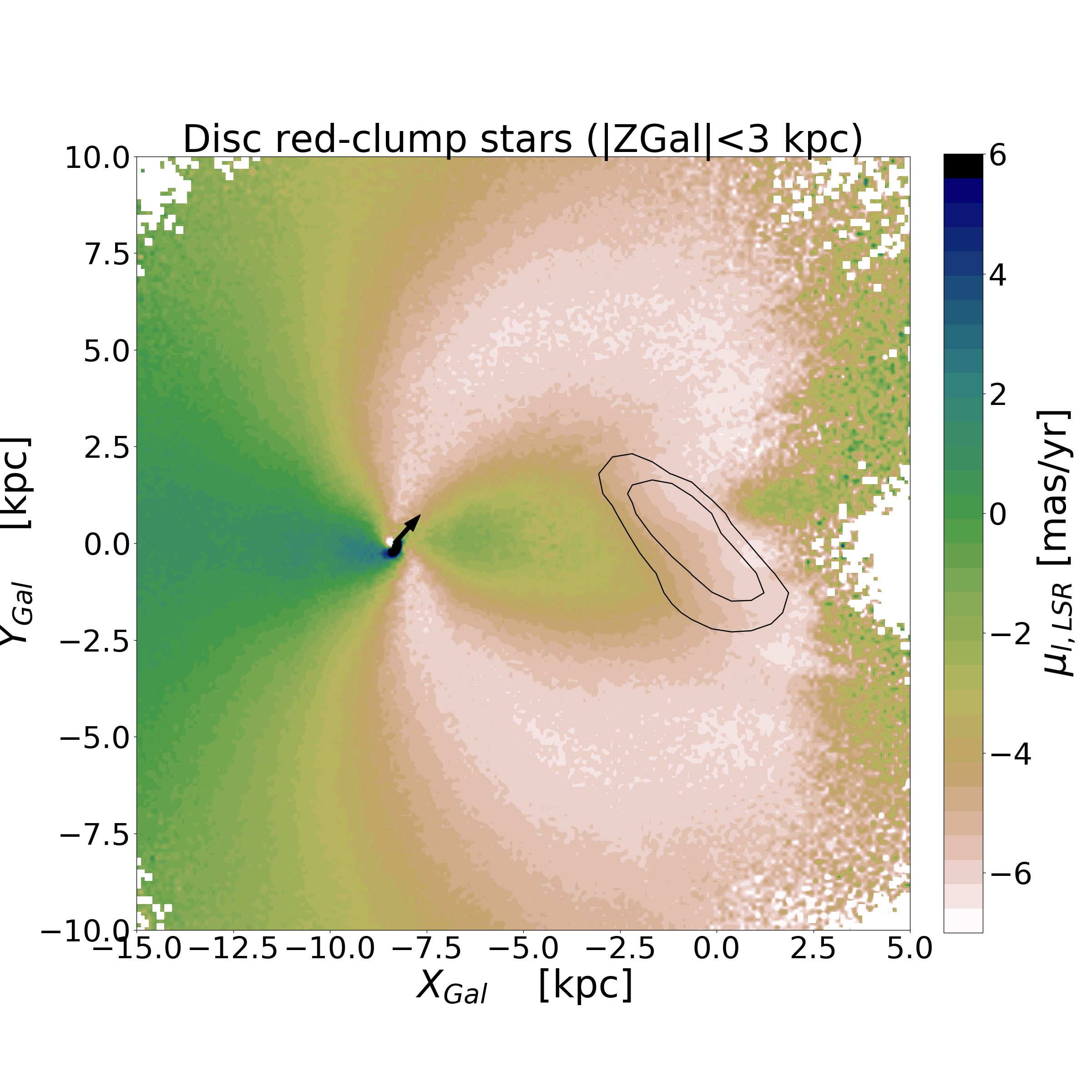}\\
 	\includegraphics[width=0.49\textwidth]{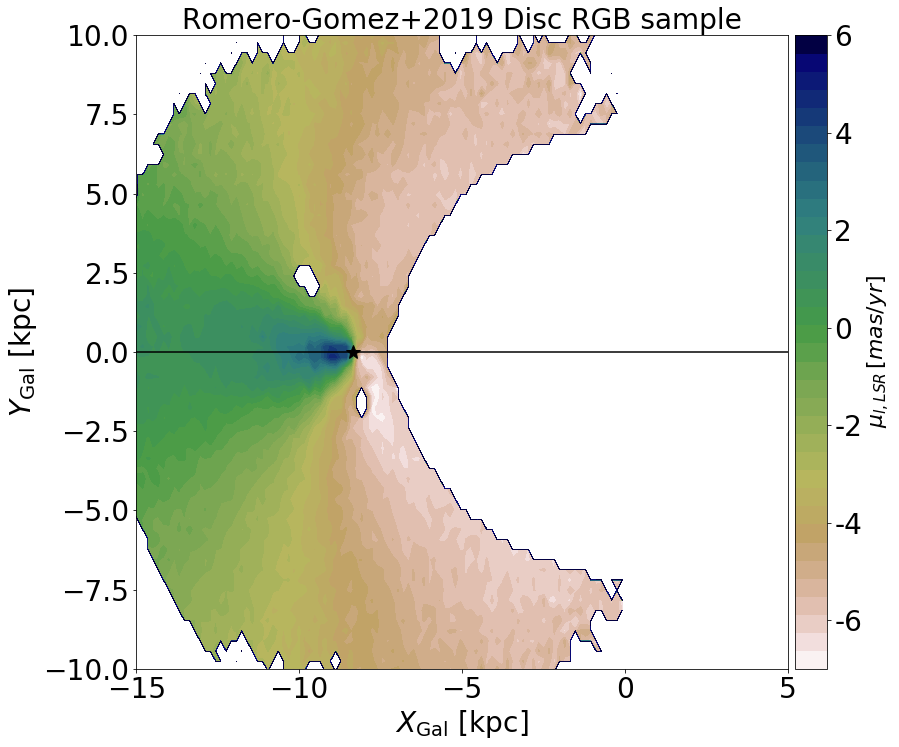}
 	\caption{Top panel: Median proper motion in Galactic longitude, $\mu_l, {\rm LSR}$, per pixel in Cartesian Galactic co-ordinates, for the disc RC sample shown in Fig. \ref{bar_rc_stars}. Overplotted are the highest density contours from Fig. \ref{bar_rc_stars}, highlighting the overdensity of the Galactic bar. The arrow highlights the direction of the solar motion used to correct the proper motion map. The large $\mu_l$ values close to the solar position point to a residual correction that may be necessary. Bottom panel: The same proper motion map for the disc red-giant sample used in \citet{Romero-Gomez2019}.}
 	\label{kin_rc}
\end{figure}

\begin{enumerate}
\item {\it Main {\tt StarHorse} reliability flag}: If this digit equals to 1, then the star has a very broad distance PDF:

\quad IF $0.5\cdot({\tt dist84}-{\tt dist16})/{\tt dist50} < \begin{cases}
0.35 & {\tt logg50}> 4.1  \\
1.0347-0.167\cdot {\tt logg50} & {\tt logg50}\leq4.1 
\end{cases}$ THEN 0 ELSE 1

We justify this definition, a cut in the posterior $\log g$ vs. distance plane, in Appendix \ref{outflag1}. The essence of this definition is that median statistics of the posterior parameters for stars where this digit equals to 1 should be treated with utmost care, as their combination often yields unphysical results. For instance, some stars fall in places of the extinction-corrected CMD that is inconsistent with any stellar model (due to complex multi-modal PDFs; see Appendix \ref{pdfappendix}), meaning that their median posterior absolute magnitude, distance, and extinction should not be used together (see for instance the unphysical 'nose' feature between the main sequence and the red-giant branch in Fig. \ref{cmds2}, bottom right panel). We verified that this effect only occurs for faint stars with very uncertain parallaxes ($\sigma_{\varpi}^{\rm cal} / \varpi^{\rm cal} \gtrsim 22\%$) - which is when the PDF of inverse parallax becomes very noisy and biased (see Fig. \ref{sigdist_piepi} and \citealt{Bailer-Jones2015a, Astraatmadja2016a, Luri2018}). This results in a poor discrimination between dwarfs and giants for these typically faint ($G\gtrsim 16.5$; see Fig. \ref{gmaghisto}) stars. Although their median effective temperatures and extinctions may still be useful, {\it their median 1D distances and other parameters should not be used}. We discuss the issue in more detail in Appendix \ref{outflag1}. In future {\tt StarHorse} runs we will resort to a more sophisticated treatment of multimodal posterior PDFs. 
\item {\it Large distance flag}: For some stars (especially extragalactic objects that are still bright enough to be in {\it Gaia} DR2, such as stars in the Magellanic Clouds or the Sagittarius dSph), {\tt StarHorse} delivers very large posterior distances, many of which are likely affected by significant biases due to the dominance of the Galactic prior used to infer them: IF ${\tt dist50}<20$ THEN 0 ELIF ${\tt dist50}<30$ THEN 1 ELSE 2
\item {\it Unreliable extinction flag}: Significantly negative extinctions, or $A_V$ values close to the prior boundary at $A_V=4$ should be treated with care: IF $({\tt AV95} > 0$ AND ${\tt AV95}<3.9$ THEN 0 ELIF ${\tt AV95}<0$ THEN 1 ELIF ${\tt AV84} < 3.9$ THEN 2 ELSE 3
\item {\it Large $A_V$ uncertainty flag}: Very large extinction uncertainties point to either incomplete or very uncertain input data: IF $0.5\cdot ({\tt AV84-AV16})<1$ THEN 0 ELSE 1
\item {\it Very small uncertainty flag}: Very small posterior uncertainties are most likely underestimated and indicate poor {\tt StarHorse} convergence (either due to inconsistent input data or too coarse model grid size). These results should therefore also be used with care. The definition is as follows: IF $0.5*{\tt (dist84-dist16)/dist50}<0.001$ OR $0.5*({\tt av84-av16})<0.01$ OR $0.5*({\tt teff84-teff16})<20.$ OR $0.5*({\tt logg84-logg16})<0.01$ OR $0.5*({\tt met84-met16})<0.01$ OR $0.5*{\tt (mass84-mass16)/mass50}<0.01$ THEN 1 ELSE 0.
\end{enumerate}

\begin{figure*}\centering
 	\includegraphics[width=0.85\textwidth]{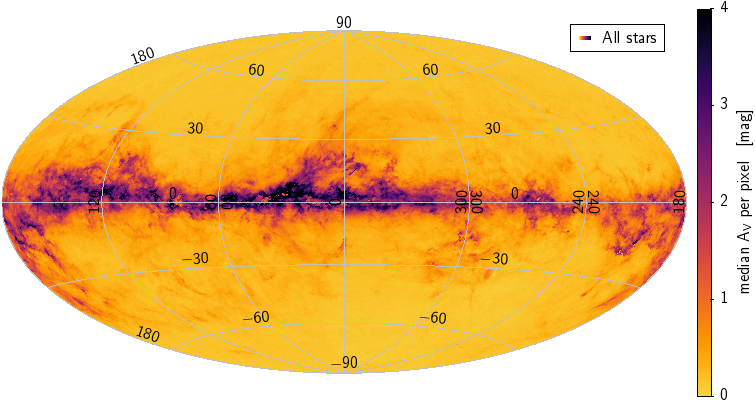}
 	\caption{All-sky median {\tt StarHorse} extinction map using all converged stars up to $G<18$. % (top panel), and divided into six wide distance bins up to 3 kpc, as indicated in each of the subplots.
    }
 	\label{avmaps}
\end{figure*}

\subsection{Data access}

{\tt StarHorse} delivered distances and extinctions for 265,637,087 objects, of which 151,506,183 pass the post-calculo quality flags included in {\tt SH\_OUTFLAG}, and 136,606,128 stars pass both the {\tt SH\_OUTFLAG} as well as the {\tt SH\_GAIAFLAG} that includes the recent recommendations of \citet{Lindegren2018}. For clarity, all our calibration choices are listed in Table \ref{calibtable}. The main statistics are summarised in Table \ref{summarytable}. Our results, together with documentation, can be queried via the AIP {\it Gaia} archive at \url{gaia.aip.de}. Example queries can be found in Appendix \ref{examplequeries}. In addition, the output files are available for download in HDF5 format at {\tt \url{data.aip.de}}. The digital object identifier for this dataset is {\tt doi:10.17876/gaia/dr.2/51}.

\begin{figure*}\centering
 	\includegraphics[width=0.49\textwidth]{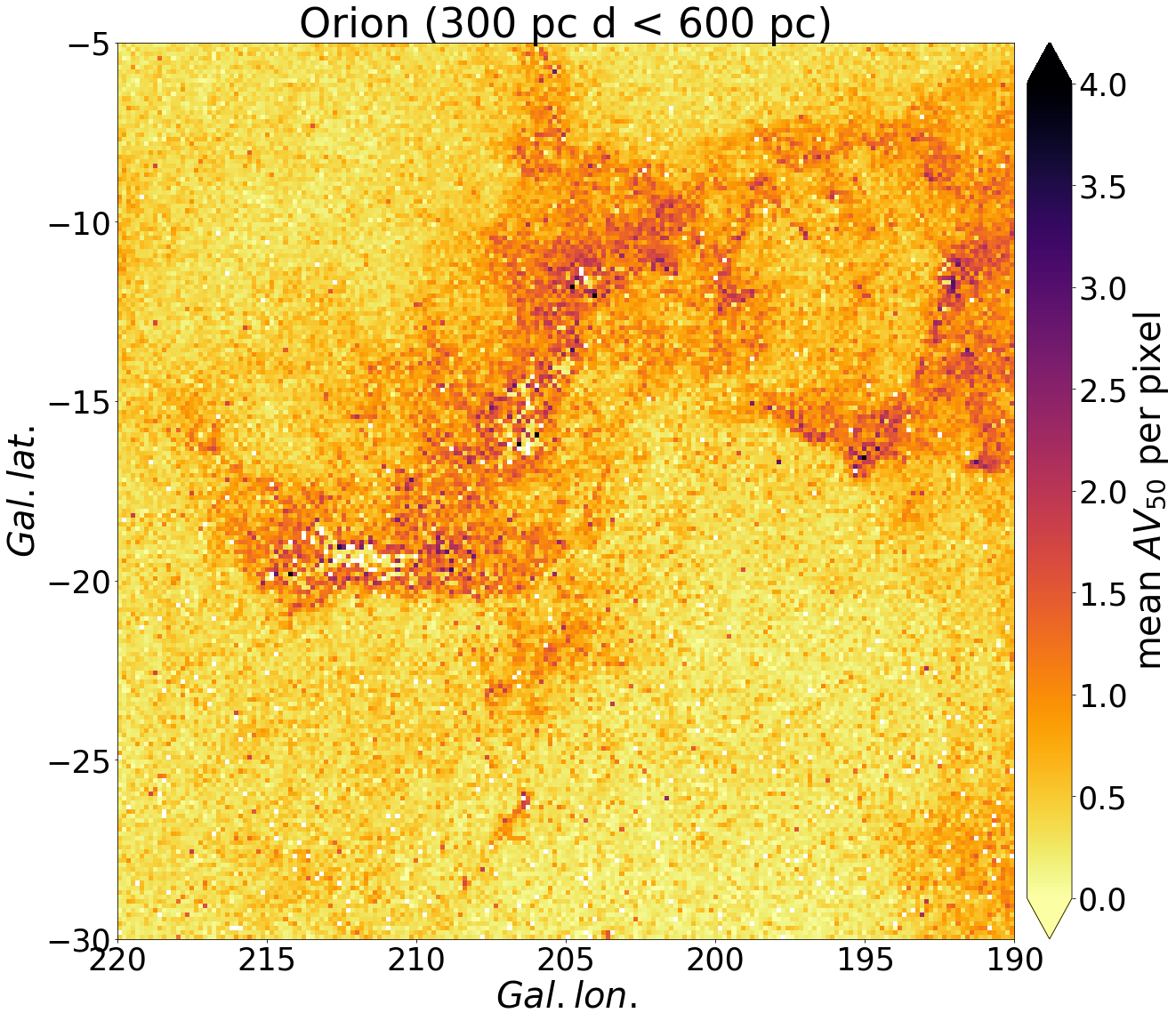}
 	\includegraphics[width=0.49\textwidth]{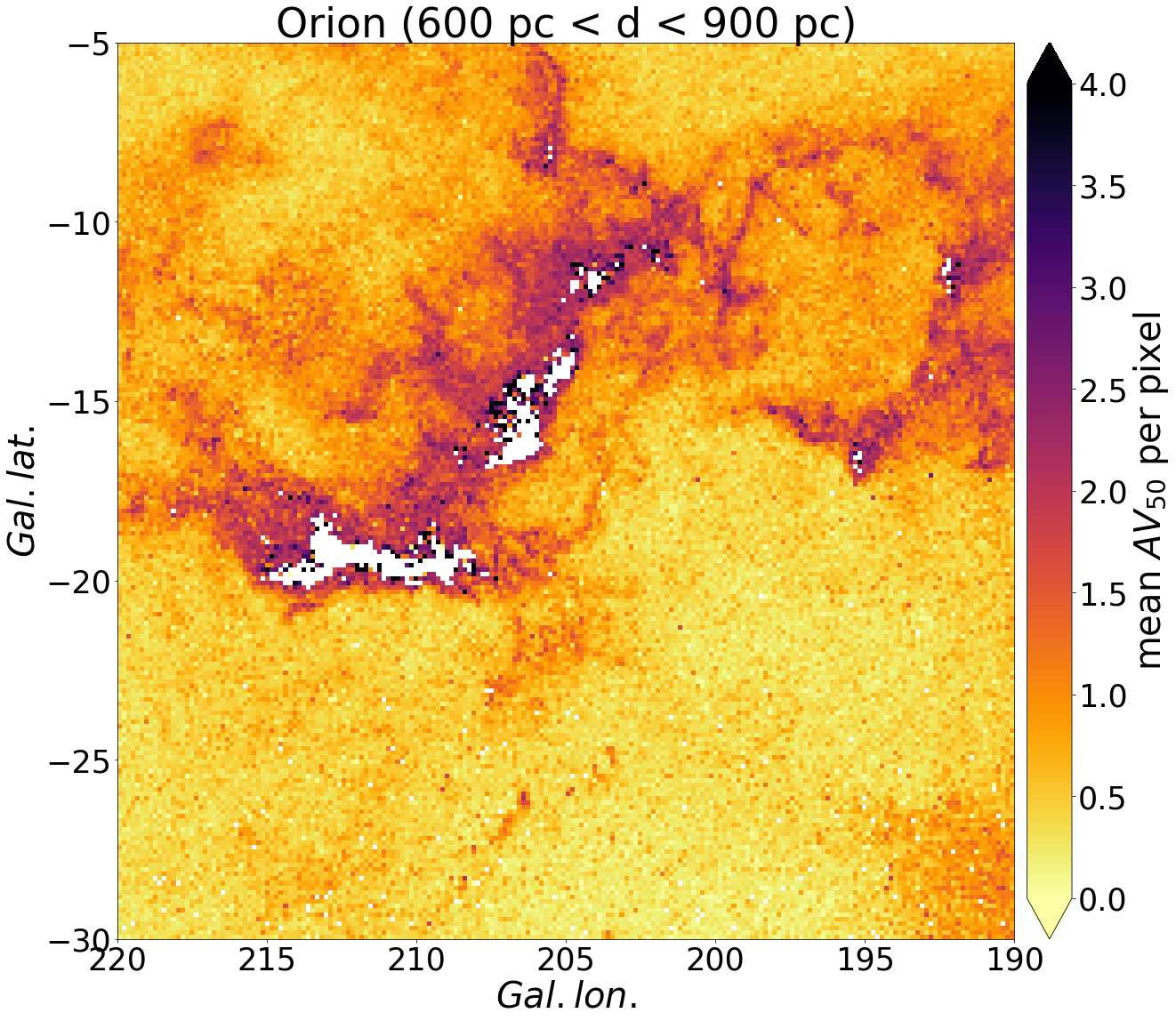}
 	\includegraphics[width=0.49\textwidth]{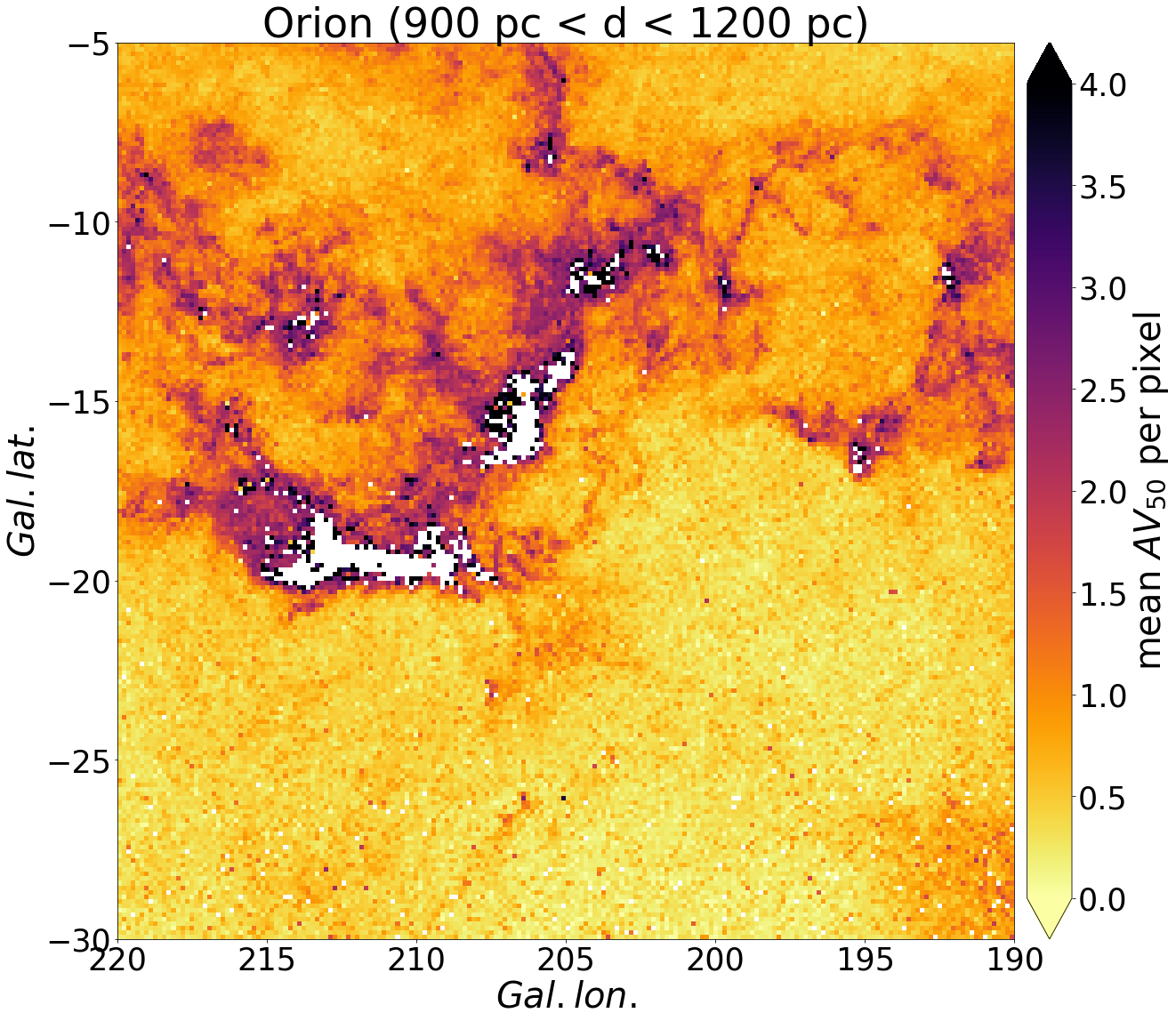}
 	\includegraphics[width=0.49\textwidth]{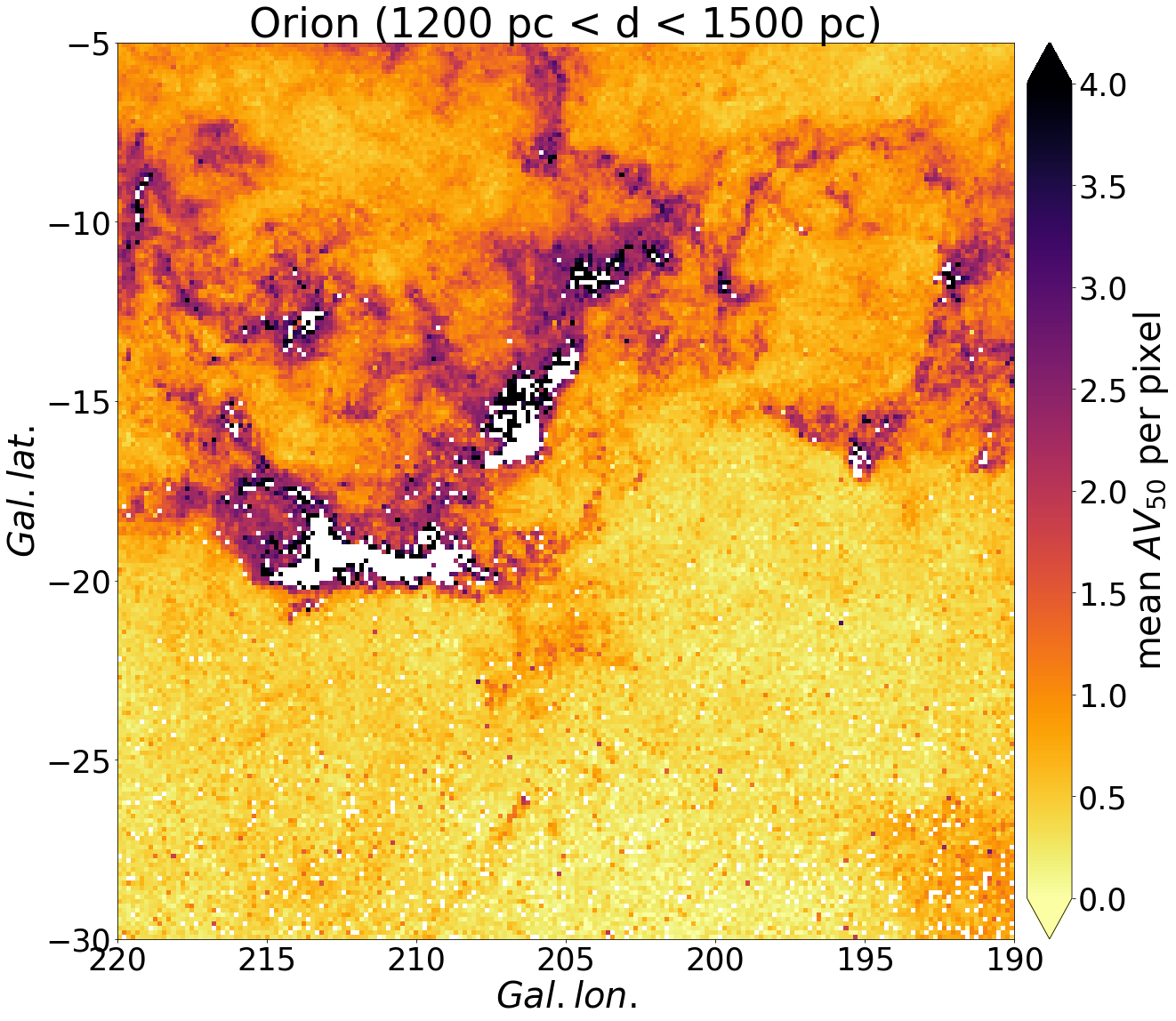}
    \caption{Distance-binned extinction maps for the Orion region, using the same dimensions as \citet{Zari2017}. The number of stars contained in each subplot is 228,808, 282,009, 297,862, and 246,266, respectively.}
 	\label{avmaps2}
\end{figure*}

\begin{figure*}\centering
 	\includegraphics[width=0.33\textwidth]{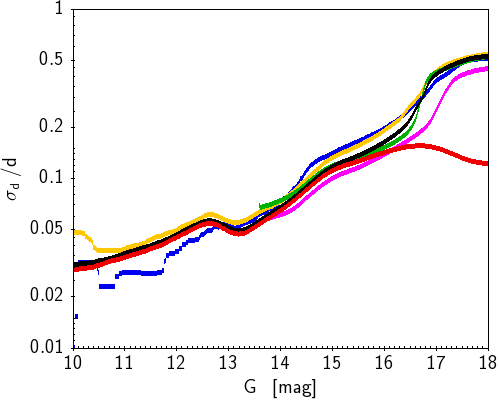}
    \includegraphics[width=0.33\textwidth]
{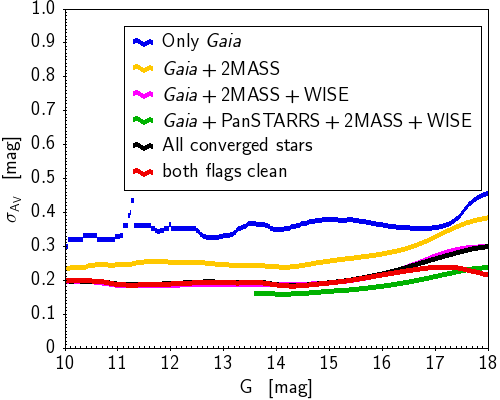}
    \includegraphics[width=0.33\textwidth]
{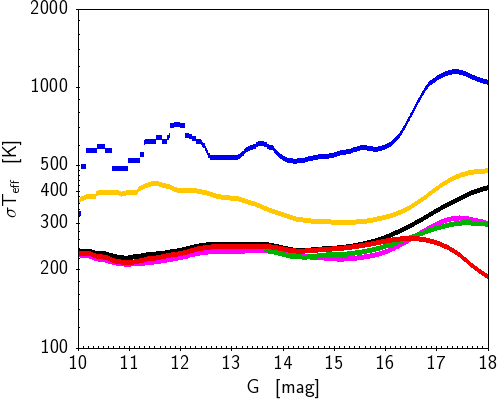}
 	\includegraphics[width=0.33\textwidth]{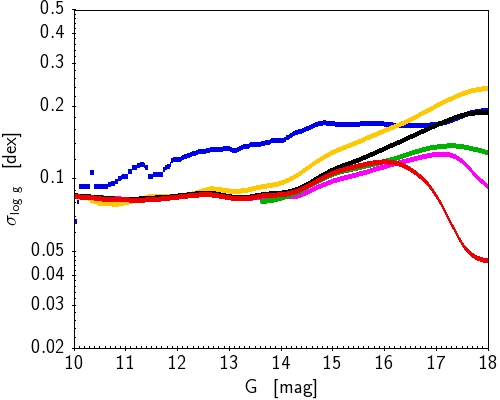}
    \includegraphics[width=0.33\textwidth]
{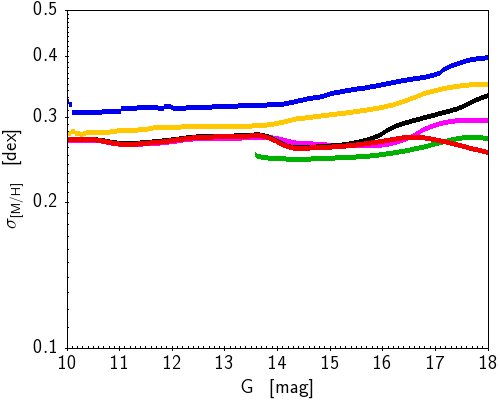}
    \includegraphics[width=0.33\textwidth]
{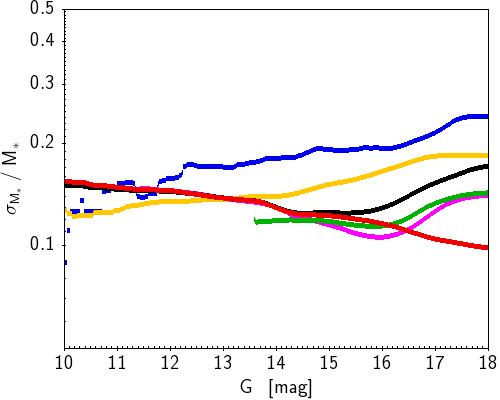}
 	\caption{{\tt StarHorse} posterior parameter precision as a function of {\it Gaia} DR2 $G$ magnitude, showing the median trends for different subsets of the data, and highlighting the improvement in precision when including additional photometry. Top row: Primary output parameter precision (from left to right: $\sigma_d/d, \sigma_{A_V}, \sigma_{T_{\rm eff}}$). Bottom row: Secondary output parameter precision (from left to right: $\sigma_{\log g}, \sigma_{[M/H]}, \sigma_{M_{\ast}}/M_{\ast}$). }
 	\label{precision_vs_Gmag}
 \end{figure*}

\begin{figure*}\centering
 	\includegraphics[width=0.33\textwidth]{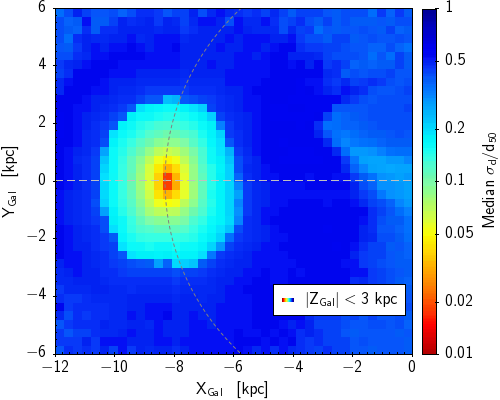}
 	\includegraphics[width=0.33\textwidth]{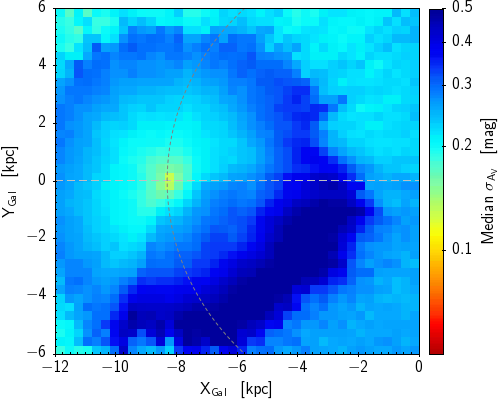}
 	\includegraphics[width=0.33\textwidth]{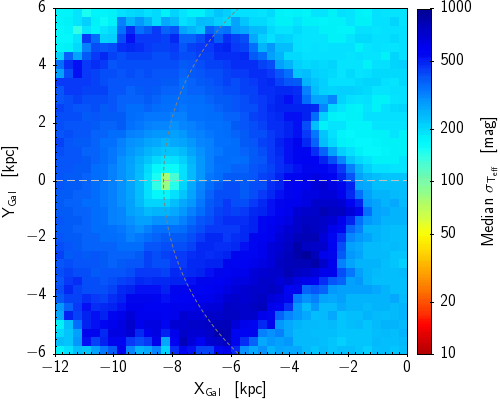}\\
 	\includegraphics[width=0.33\textwidth]{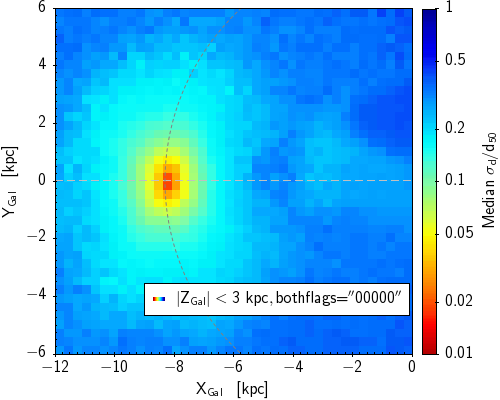}
 	\includegraphics[width=0.33\textwidth]{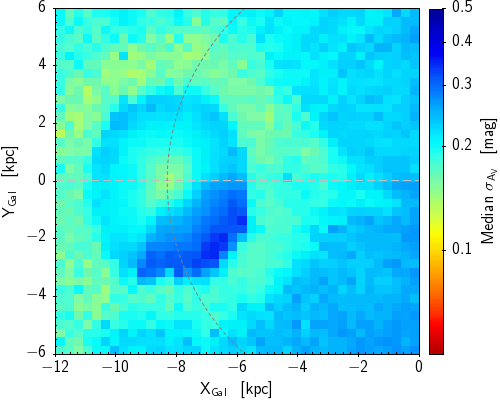}
 	\includegraphics[width=0.33\textwidth]{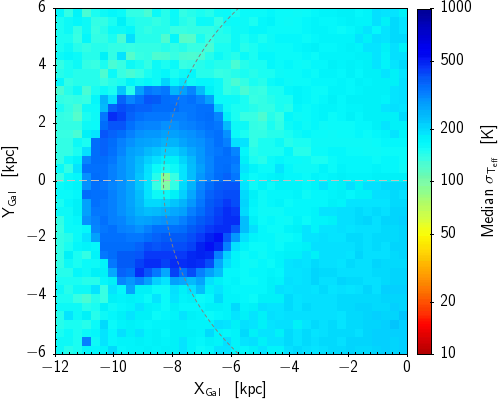}
 	\caption{Median uncertainty distributions $\sigma_d/d_{50}$ (left) $\sigma_{A_V}$ (middle), and $\sigma_{T_{\rm eff}}$ (right) as a function of position in Galactic co-ordinates. Top row: for all converged stars. Bottom row: for the {\bf {\tt SH\_GAIAFLAG}$="000"$ \& {\tt SH\_OUTFLAG}$="00000"$} sample.}
 	\label{xymaps_sigcoloured}
\end{figure*}

%%%%%%%%%%%%%%%%%%%%%%%%%%%%%%%%%%%%%%%%%%%%%%%
\section{{\tt StarHorse} {\it Gaia} DR2 results}\label{results}%%%%%%%%%%%%%%%%%%%%%%%%%
%%%%%%%%%%%%%%%%%%%%%%%%%%%%%%%%%%%%%%%%%%%%%%%

\subsection{Summary}\label{summary}

Table \ref{summarytable} summarises the results of the present {\tt StarHorse} run for {\it Gaia} DR2 and puts them in context with previous results available from the literature (three references for distances and extinctions for {\it Gaia} stars observed by spectroscopic surveys, as well as the two only studies that attempted to determine distances and extinctions, respectively, for the whole {\it Gaia} DR2 dataset). In particular, the table informs about sample sizes, magnitude ranges, and the typical precision in the primary output parameters $d, A_V,$ and $T_{\rm eff}$. In this table, we also define some useful sub-samples of the {\it Gaia} DR2 {\tt StarHorse} data (identified by colour in some of the subsequent plots) that are used throughout this paper. These are: 
\begin{enumerate}
\item stars with (recalibrated) parallaxes more precise than 20\% (blue colour; 39\% of the converged stars),
\item stars with {\tt SH\_GAIAFLAG} equal to "000" (cyan colour; 88\% of the converged stars),
\item stars with {\tt SH\_OUTFLAG} equal to "00000" (orange colour; 57\% of the converged stars),
\item stars with {\tt SH\_OUTFLAG} equal to "00000" and {\tt SH\_GAIAFLAG} equal to "000" (red colour; 52\% of the converged stars).
\end{enumerate}
The $G$ magnitude distribution for each of these sub-samples is shown in Fig. \ref{gmaghisto}. In this paper, we will mainly concentrate on the "both-flags"-cleaned sample.

Figure \ref{bigcorner} presents the output of the {\tt StarHorse} code for the {\it Gaia} DR2 sample in one plot. The figure displays the distributions and correlations of the median {\tt StarHorse} primary output parameters $T_{\rm eff}, d,$ and $A_V$, and their respective uncertainties, as well as $G$ magnitude and parallax signal-to-noise ratio. The grey contours in this plot refer to all converged stars, whereas the red contours emphasise the results for the stars with both {\tt SH\_GAIAFLAG} and {\tt SH\_OUTFLAG} equal to "00000". For a plot including also the secondary output parameters $\log g,$ [M/H], and $M_{\ast}$, we refer to Fig. \ref{hugecorner}. 

The panels in the diagonal row of Fig. \ref{bigcorner} provide the one-dimensional distributions in $G$ magnitude, parallax signal-to-noise, the median output parameters, and the distributions of the corresponding uncertainties (in logarithmic scaling) as area-normalised histograms. Each of the panels also illustrates the effect of applying the recommended flags: the red uncertainty distributions are typically confined to smaller values than the faint grey ones.

The off-diagonal plots of Fig. \ref{bigcorner} show the correlations between the output parameters. We observe complex structures in many of these panels, most of which are due to physical correlations stemming from stellar evolution or selection effects. For example, the strong bimodality between giants and dwarfs in the $\log g$ vs. $T_{\rm eff}$ diagram (see third column, second panel from top in Fig. \ref{hugecorner}) is reflected in many of the panels, most notably the distance distribution (fourth column). In addition, we note that some of the complex structure disappears when the flag cleaning is applied to the data. Different behaviour of the red and grey distributions in some panels should warn the user about potentially spurious correlations that may appear when using the full {\tt StarHorse} sample.

\subsection{Extinction-cleaned CMDs}\label{cmds}

As a first sanity check, in Fig. \ref{cmds2} we present {\tt StarHorse}-derived {\it Gaia} DR2 colour-magnitude diagrams (CMDs) for the full converged sample (i.e. excluding mostly white dwarfs and galaxies) in four magnitude bins. Focussing first on the top left panel ($G<14$), we note very well-defined features of stellar evolution the CMD: a thin main sequence (broadening in the very blue and very red regimes), a well-populated sub-giant branch, as well as a very thin red clump, the red giant branch, and the asymptotic giant branch. We also notice more subtle features such as the red-giant bump or the secondary red clump.

As we move to fainter magnitude bins, the number of objects grows, but also the typical uncertainty in the main input parameter parallax, resulting in a gradual broadening of the sharp stellar-evolution features observed in the top left panel of Fig. \ref{cmds2}. In the lower left panel, for example, we begin to note some additional features that are not directly related to stellar evolution. For example, the almost vertical arm at $(BP-RP)_0$ below the main sequence is related to problematic astrometry (large {\tt ruwe} values). Furthermore, the discrete stripes in the red main sequence are related to the finite mass, age, and metallicity resolution of our stellar model grid used. Some other unphysical structures, such as the nose between the main sequence and the giant branch, are induced by poor convergence of {\tt StarHorse} (see Sect. \ref{outflag}). The higher relative number of giant stars in the fainter magnitude bins with respect to the $G<14$ sample is an effect of stellar population sampling.

Figure \ref{cmds1} shows another collection of {\tt StarHorse} CMDs, now highlighting the sub-samples defined in Table \ref{summarytable}. As discussed in Sect. \ref{outflag}, the full {\tt StarHorse} sample occupies a larger volume in the CMD, including an unphysical region in-between the main sequence and the red-giant branch that is due to stars with poorly determined parallaxes. These stars disappear when applying the {\tt SH\_OUTFLAG} (orange dots in lower middle panel), and a further cleaning using the {\tt SH\_GAIAFLAG} results in a nice physical CMD for 129 million stars (lower right panel of Fig. \ref{cmds1}). 

Comparing the upper right and lower right panel of Fig. \ref{cmds1}, we see that the number of red giants in the latter is much higher, leading to a slight broadening of the RC locus and a substantially higher number of AGB stars. This is due to the fact that {\tt StarHorse} is able to determine still surprisingly precise ($\sim30\%$) photo-astrometric distances for giants with poor parallax measurements.

\begin{figure*}\centering
 	\includegraphics[width=0.33\textwidth]{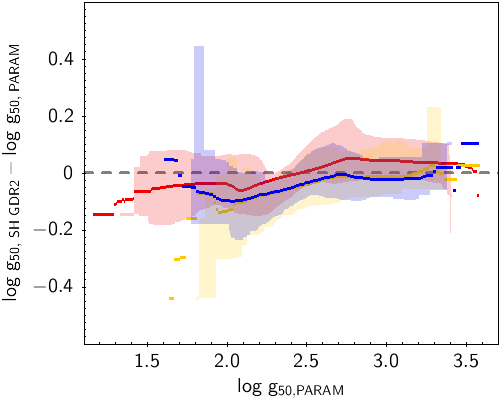}
 	\includegraphics[width=0.33\textwidth]{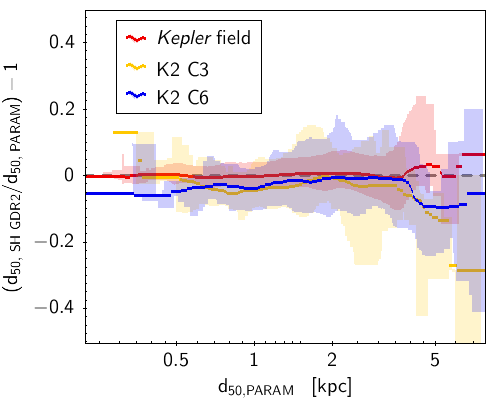}
 	\includegraphics[width=0.33\textwidth]{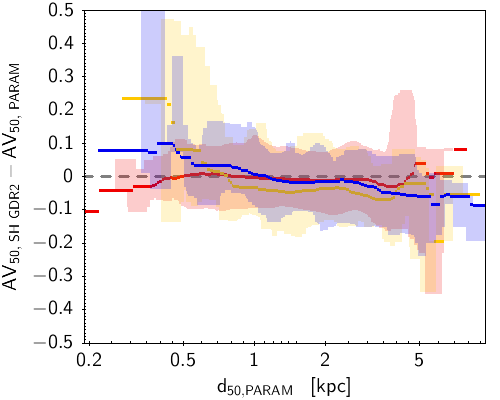}\\
 	\includegraphics[width=0.33\textwidth]{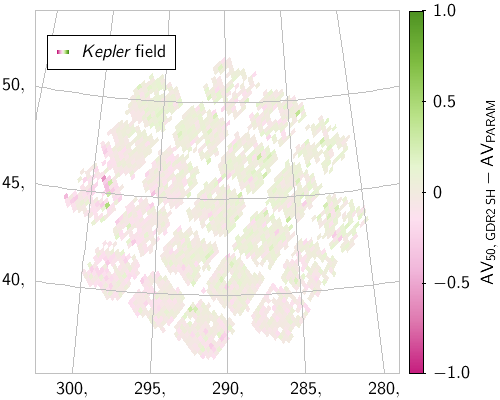}
 	\includegraphics[width=0.33\textwidth]{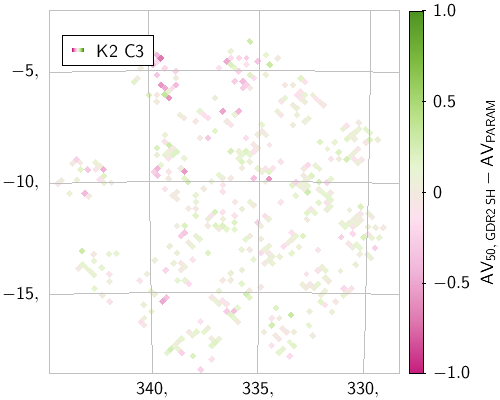}
 	\includegraphics[width=0.33\textwidth]{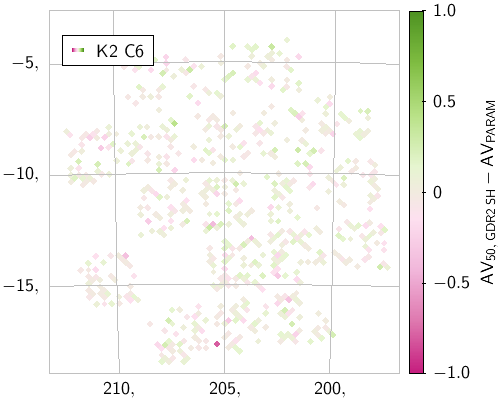}
 	\caption{Comparison of the results of this work with the results obtained by \citet{Khan2019} for the solar-like oscillating giants in the {\it Kepler} field and the K2 C3 and C6 fields. The surface gravities were obtained from the $\nu_{\rm max}$ scaling relation, while the distances and extinctions were computed with PARAM \citep{Rodrigues2017}, using the seismic parameters $\Delta\nu, \nu_{\rm max}$, as well as spectroscopic ({\it Kepler}) or photometric ({\it K2} effective temperatures and metallicities as an input.}
 	\label{seismocomparison}
\end{figure*}

\subsection{{\it Kiel} diagrams}

Figure \ref{kieldiagrams1} shows {\it Kiel} diagrams ($\log g$ vs. $T_{\rm eff}$) using the median posterior {\tt StarHorse} results, for the full sample of converged stars and for the flag-cleaned sample defined in Table \ref{summarytable}. The middle column of that figure show the the median distance and median $A_V$ extinction in each pixel of the {\it Kiel} diagram, respectively. The right column show their respective uncertainties in each pixel. The complex dependence of the uncertainties on the stellar parameters reflects the abrupt decrease in precision below $\varpi^{\rm cal}/\sigma_{\varpi}^{\rm cal}=5$ seen in Fig. \ref{sigdist_piepi}.

\begin{figure*}\centering
 	\includegraphics[width=0.33\textwidth]{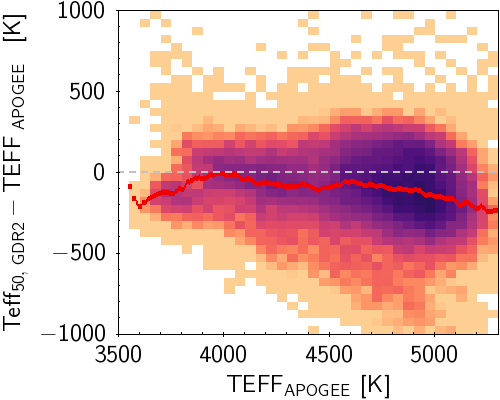}
 	\includegraphics[width=0.33\textwidth]{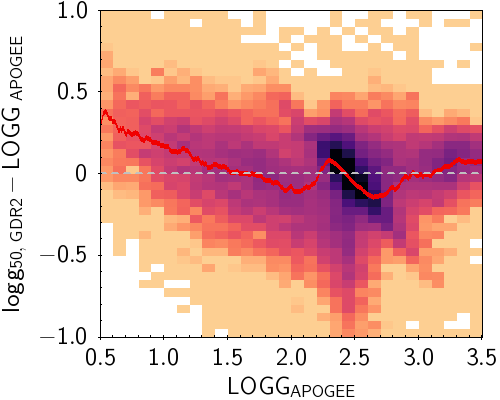}
 	\includegraphics[width=0.33\textwidth]{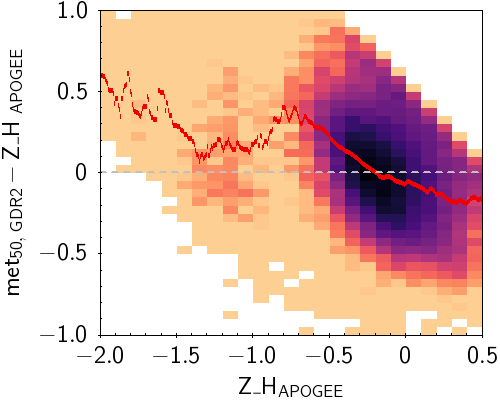}\\
 	\includegraphics[width=0.39\textwidth]{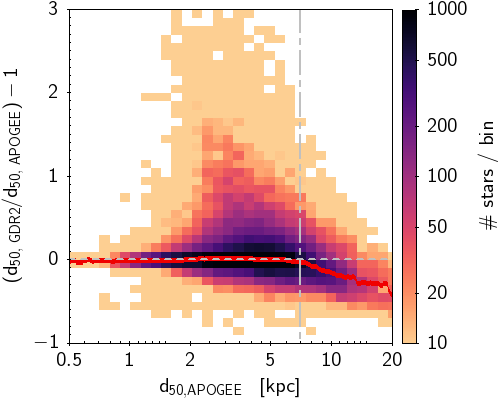}
 	\includegraphics[width=0.6\textwidth]{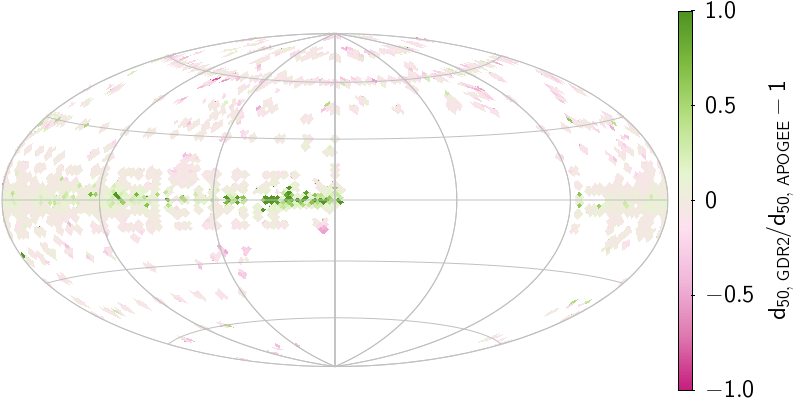}\\
 	\includegraphics[width=0.39\textwidth]{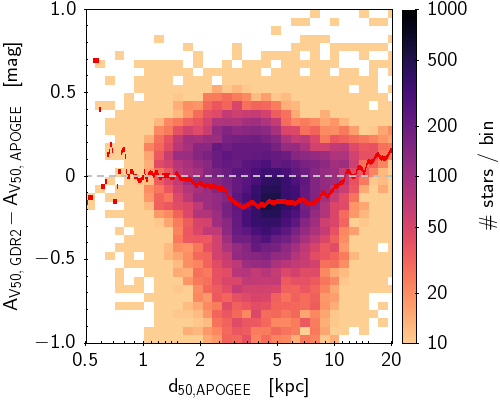}
 	\includegraphics[width=0.6\textwidth]{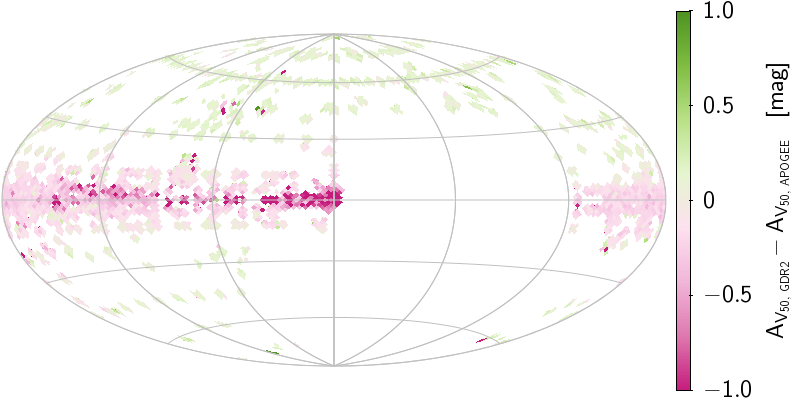}
 	\caption{Comparison of the results of this work with APOGEE results for the stars contained in SDSS DR14. Top row: comparison with the ASPCAP spectroscopic pipeline results. Middle and bottom rows: comparison with the {\tt StarHorse} distances and extinctions, respectively, obtained from combining the ASPCAP results with {\it Gaia} DR2 and additional photometry (Santiago et al., in prep.).}
 	\label{apogeecomparison}
\end{figure*}

\begin{figure}\centering
 	\includegraphics[width=0.5\textwidth]{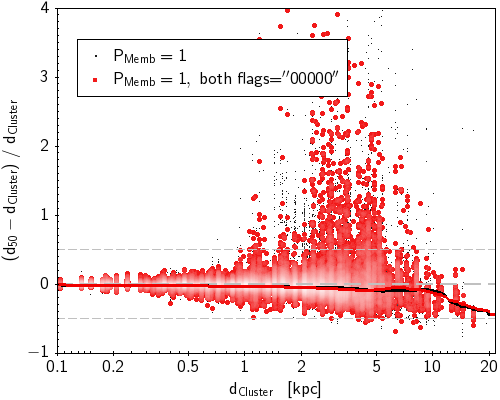}
 	\caption{Star-to-star comparison to the distance scale of the open cluster sample of \citet{Cantat-Gaudin2018}: relative distance difference as a function of cluster distance (for 60,284 stars with membership probabilities $P_{\rm memb}=1$). Only 1\% of the flag-cleaned stars in this sample show distance deviations greater than 50\% (defined by the short-dashed grey lines).}
 	\label{clustercomparison1}
 \end{figure}

\begin{figure*}\centering
 	\includegraphics[width=0.32\textwidth]{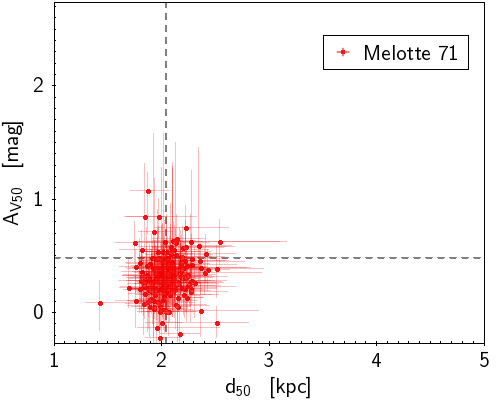}
 	\includegraphics[width=0.32\textwidth]{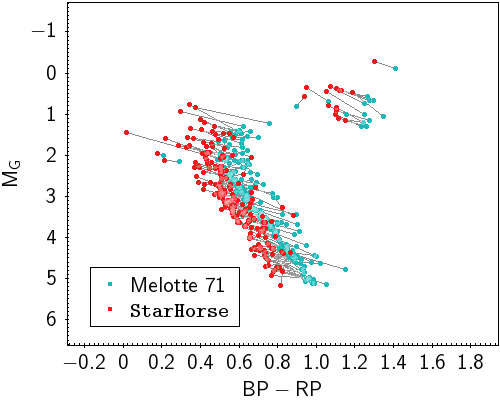}
 	\includegraphics[width=0.32\textwidth]{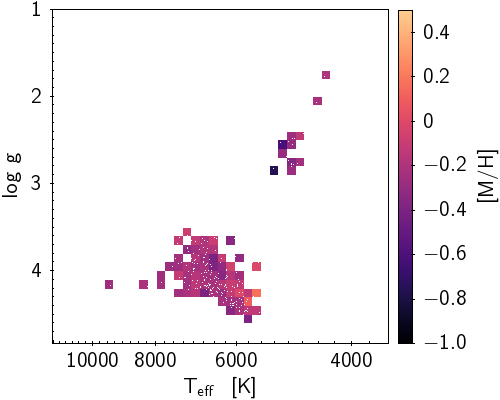}
 	\includegraphics[width=0.32\textwidth]{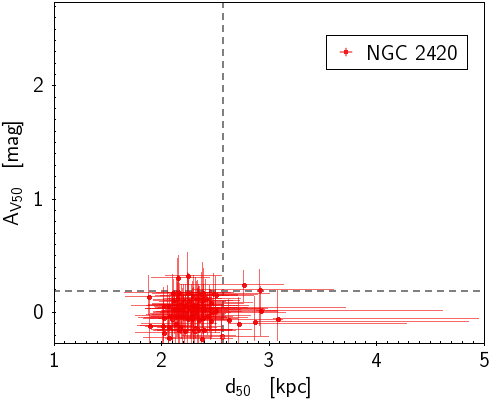}
 	\includegraphics[width=0.32\textwidth]{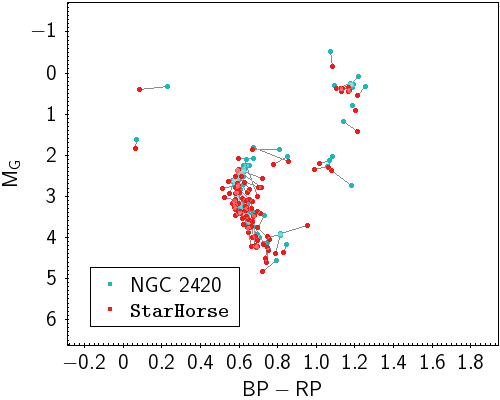}
 	\includegraphics[width=0.32\textwidth]{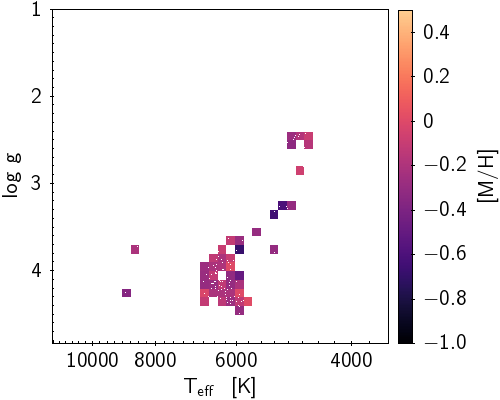}
    \includegraphics[width=0.32\textwidth]{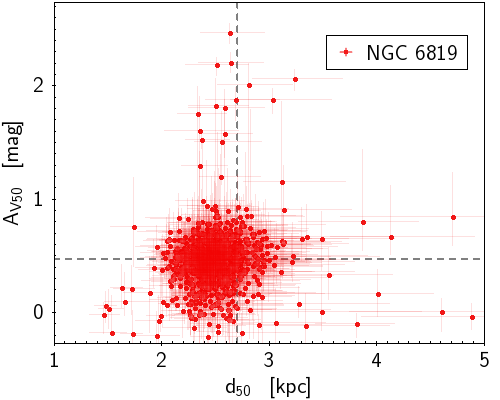}
 	\includegraphics[width=0.32\textwidth]{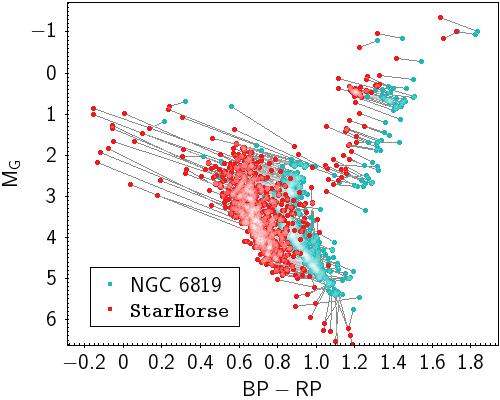}
 	\includegraphics[width=0.32\textwidth]{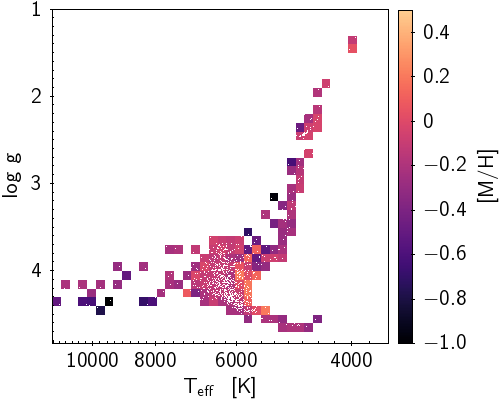}
 	\includegraphics[width=0.32\textwidth]{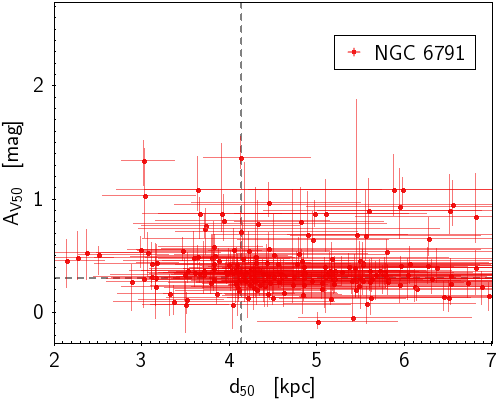}
 	\includegraphics[width=0.32\textwidth]{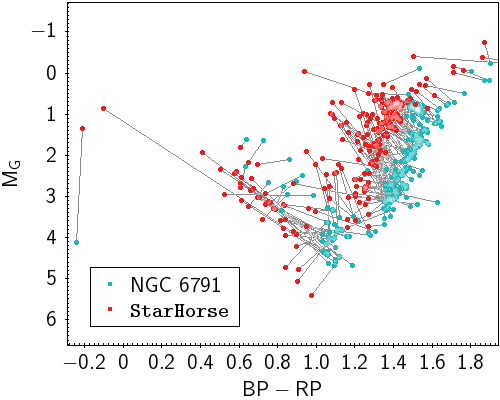}
 	\includegraphics[width=0.32\textwidth]{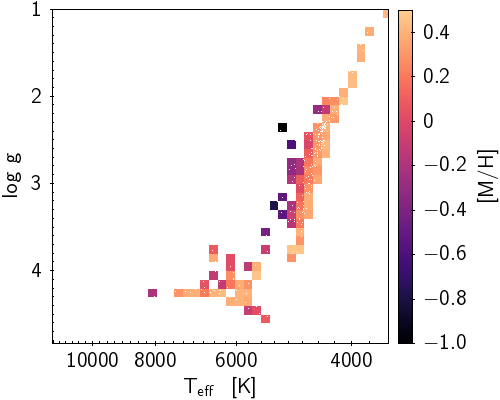}
 	\caption{Comparison to the distance scale of the open cluster sample of \citet{Bossini2019}: Detailed comparison to the four distant ($d>2$ kpc) clusters studied with high-resolution spectroscopy (Melotte 71, NGC 2420, NGC 6819, and NGC 6791). In each panel, we show only stars with $P_{\rm memb}=1$ according to \citet{Cantat-Gaudin2018}. The first panel in each row shows the distance-$A_V$ plane (grey-dashed lines refer to the values determined by \citealt{Bossini2019}), while the second panel displays the colour-magnitude diagram (only parallax-corrected in cyan, {\tt StarHorse}-corrected in red). The third panel shows the metallicity-coloured {\it Kiel} diagram for each cluster. The metallicities derived from high-resolution spectroscopy quoted by \citet{Bossini2019} for these clusters are [Fe/H]$=-0.27$ for Melotte 71, $-0.05$ for NGC 2420, $0.0$ for NGC 6819, and $+0.4$ for NGC 6791. The fact that we see on average much more metal-rich stars in NGC 6791 is encouraging: there is at least some metallicity sensitivity in the photometric data used in this work.}
 	\label{clustercomparison2}
 \end{figure*}

\subsection{Stellar density maps and the emergence of the Galactic bar}\label{maps}

Figure \ref{rzmaps} presents four projections of the stellar density distribution in Galactocentric co-ordinates for the flag-cleaned sample. The solar position (in kpc) is at $(X_{\rm Gal}, Y_{\rm Gal}, Z_{\rm Gal})=(8.2, 0, 0.025)$. The figure emphasises the loss of stars near the Galactic midplane towards the inner Galaxy, which is due to both the high dust extinction affecting the {\it Gaia} selection function, and the low number of stars that pass the flag quality criteria in these regions. 
Several conclusions can be drawn from Fig. \ref{rzmaps}, as we describe next.

As pointed outed by \citet{Bailer-Jones2015a, Luri2018}, for example, the naive $1/\varpi$ estimator provides biased distances, especially in the case of low parallax precision, extending the observed volume to unplausibly large distances. On the other hand, the exponentially decreasing density prior recently used by \citet{Bailer-Jones2018} is more apt for main-sequence stars and tends to underestimate the distances to distant luminous giant stars. The {\tt StarHorse} results for those stars, taking into account photometric information as well as more complex priors, show for the first time that {\it Gaia} DR2 already allows us to probe stellar populations in the bulge and beyond. A detailed comparison with \citet{Bailer-Jones2018} is presented in Section \ref{bj18}. 

The clearest novel feature of the {\tt StarHorse} density map shown in Fig. \ref{rzmaps} is the presence of a stellar overdensity coinciding with the expected position of the Galactic bar, inclined by about 40 degrees with respect to the solar azimuth, and with a semi-major axis of about 4 kpc. This almost direct detection of the Galactic bar is confirmed with {\tt StarHorse} distances for APOGEE stars and discussed in detail in a separate paper (Queiroz et al., in prep.). The significance of the result lies in the fact that although we are using a prior for the Galactic bulge-bar \citep{Robin2012a}, its shape and inclination angle are quite different from our bar prior (see Fig. \ref{bar_rc_stars}), even when invoking an interplay with possible observational biases.

The presence of the Galactic bar in the {\it Gaia} DR2 data is even more prominent when we focus only on the red-clump stars. Fig. \ref{bar_rc_stars} shows the resulting density map when selecting flag-cleaned RC stars close to the Galactic plane ($|Z_{\rm Gal}|<3$ kpc) from the {\tt StarHorse} {\it Kiel} diagram (4500 K$< {\tt teff50} < 5000$ K, $2.35<{\tt logg50} < 2.55, -0.6<{\tt met50} < 0.4$). The density contrast of the RC bar with respect to the RC population in front of the bar amounts to almost 50. This could in fact be a physical feature of the Galactic disc: the RC is a tracer of the young-to-intermediate age population ($\sim1-4$ Gyr; e.g. \citealt{Girardi2016}), and the star-formation history in the inner disc outside the bar region is still poorly constrained. It is more likely, however, that the observed shape of the bar (and especially the density drop in front of it) in Fig. \ref{bar_rc_stars} is a combined effect of the {\it Gaia} DR2 selection function, the stellar density profile of the inner disc, our adopted bulge prior, and the quality flag cuts used to produce Fig. \ref{bar_rc_stars}. At the present stage, we therefore caution the reader not to take the star count numbers in this map at face value, and refer to Queiroz et al. (in prep.) for a more in-depth discussion.

The lower right panel of Fig. \ref{rzmaps} shows the density map in Galactocentric cylindrical co-ordinates $R_{\rm Gal}$ vs $Z_{\rm Gal}$. Especially in this panel we note two overdensities in the direction of the Magellanic Clouds. These are mostly composed of stars belonging to the Clouds that have been forced to smaller distances by our Milky Way prior (which does not contain any extragalactic stellar population, only a smooth halo with a power-law density). The results for these stars have not been excluded from our analysis, but should be used with caution. The same is true for other nearby galaxies with resolved stellar populations, such as the Sagittarius dSph, Fornax, etc.

\subsection{Kinematic maps}\label{kinematics}

Several studies have already used our distances for the {\it Gaia} DR2 sub-sample of stars with radial velocity measurements \citep{Katz2019} in kinematic analyses of the Galactic disc. \citet{Quillen2018} used our results to study the arches and ridges in velocity space found by \citet{GaiaCollaboration2018c, Antoja2018a} and \citet{Kawata2018a}, attributing some of them to stellar orbit crossings with spiral arms. \citet{Monari2018} used our distances to counter-rotating stars in the Galactic halo to measure the escape speed curve and the mass of the Milky Way. Recently, \citet{Carrillo2019} used {\tt StarHorse} distances together with {\it Gaia} DR2 positions, proper motions, and line-of-sight velocities, to study the 3D velocity distribution in the Milky Way disc. They confirmed the bulk vertical motions see in earlier data, consistent with a combination of breathing and bending modes, and identified a strong radial
$V_R$ gradient in the Galactic inner disc, transitioning smoothly from 15 km/s/kpc at Galactic azimuth $\Phi_{\rm Gal}\sim50\deg$ to -15 km/s/kpc at Galactic azimuth $\Phi_{\rm Gal}\sim-50\deg$. Our {\tt StarHorse} results were essential for this type of work, since they enabled the authors to probe much farther heliocentric distances.

To further illustrate the accuracy of our distances for distant red-giant stars, in Fig. \ref{kin_rc} we %reproduce one of the kinematic maps presented by \citet{Romero-Gomez2018} - their Fig. 7 - 
show a proper-motion map of the disc red-clump sample used in Fig. \ref{bar_rc_stars} and Sect. \ref{maps}. Both panels of Fig. \ref{kin_rc} show proper motion in Galactic longitude corrected for the solar motion, $\mu_{l, \rm LSR}$, as a function of Galactic position. The top panel shows the {\tt StarHorse} red-clump stars, while the bottom panel shows the red-giant sample studied by \citet{Romero-Gomez2019}. For comparability, we assume the same values for the solar motion \citep[$U_{\odot}=11.1$ km/s, $V_{\odot}=12.24$ km/s;][]{Schoenrich2010} and the distance to the Galactic Centre \citep[8.34 kpc;][]{Reid2014} as in \citet{Romero-Gomez2019}, although the residual small-scale dipole variations close to the solar position suggest that the solar motion correction may have to be slightly revised. 

The study of \citet{Romero-Gomez2019} concerned the morphology and kinematics of the Galactic warp, so it mainly focussed on the motions perpendicular to the Galactic disc, $\mu_b$. Here we show that the $\mu_l$ map of the RGB sample in the bottom panel of Fig. \ref{bar_rc_stars} compares quite well to our red-clump sample shown in the top panel. 
Since here we are more interested in the possible kinematic effects of the Galactic bar, %the bottom panel of Fig. \ref{kin_rc} shows the same data as the upper panel, but now with a different colour map extending to slightly lower values of $\mu_l$. This plot 
we can now study the bulk motions in the Galactic plane out to larger distances from the Sun, using a cleaner sample of RC stars. We highlight several dynamical features present in this sample. 

The prominent symmetric arc features around the solar position towards the outer and inner disc are produced by the Galactic rotation curve, and follow the overall expected trends (see e.g. Fig. 3 in \citealt{Brunetti2010} for a prediction of the $\mu_l$ map for an axisymmetric disc). It is interesting to see that the proper motion contours in the inner disc coincide with the angle of the Galactic bar (defined by stellar density). Qualitatively this coherent motion seen in the region of the bar agrees with earlier predictions by \citet[][their Fig. 8]{Brunetti2010} and the disc red-clump test particle simulations of \citet{Romero-Gomez2015}. %The deviation from axisymmetry towards lower bulk proper motion seen towards longitudes of $240\deg \lesssim l \lesssim 300\deg$ is also seen in other datasets (e.g. \citealt{Kharchenko2013}, Fig. 3, \citealt{Zari2018}, Fig. 11), although possibly not as pronounced as in our Fig. \ref{kin_rc}. We suggest that this feature, along with the coherent motion seen in the region of the bulge, are kinematic imprints of the Galactic bar, as also predicted by \citet[][their Fig. 8]{Brunetti2010} and confirmed by disc RC test particle simulations (\citealt{Romero-Gomez2015}).

A more quantitative comparison to kinematic Galactic models including effects of the Galactic bar is left to future studies.

\subsection{Extinction maps}\label{extmaps}

Figures \ref{avmaps} and \ref{avmaps2} show {\tt StarHorse}-derived two-dimensional (2D) median extinction maps. Figure \ref{avmaps} shows the all-sky $A_V$ map in Aitoff projection. %The top panel shows all stars, while the smaller panels show the median extinction per HealPix cell in six consecutive distance bins up to 3 kpc. 
The overall appearance of this figure compares very well to the expected 2D extinction map (e.g. \citealt{Andrae2018}, Fig. 21; \citealt{Lallement2018}, Fig. 6). 

In Fig. \ref{avmaps2}, we show median extinction maps in four distance bins between 300 and 1500 pc, for the Orion region. The four panels show how with increasing distance extinction from molecular clouds gradually fills the Galactic plane. In principle, our results can thus be used to construct 3D extinction maps (e.g. \citealt{Schlafly2011, Green2015}) and infer the three-dimensional dust distribution in the extended solar vicinity (e.g. \citealt{Capitanio2017, RezaeiKh.2018, Lallement2019, Zucker2019}). 

\begin{figure*}\centering
 	\includegraphics[width=0.4\textwidth]{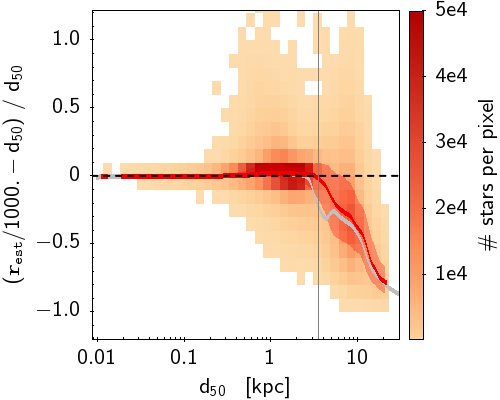}
 	\includegraphics[width=0.4\textwidth]{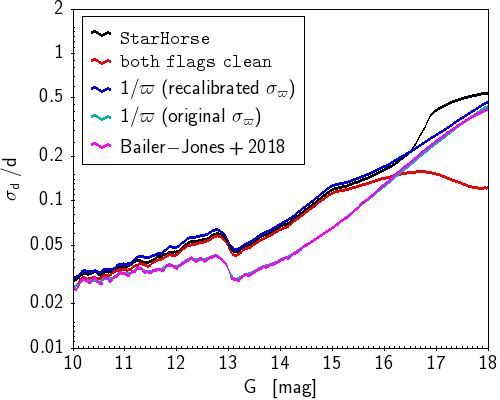}
 	\includegraphics[width=0.4\textwidth]{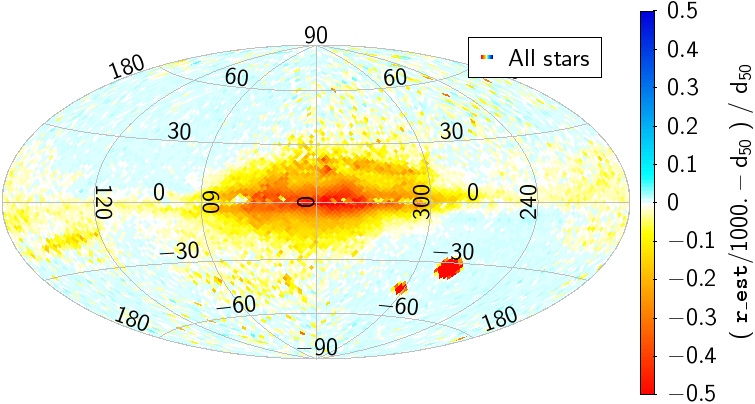}
 	\includegraphics[width=0.4\textwidth]{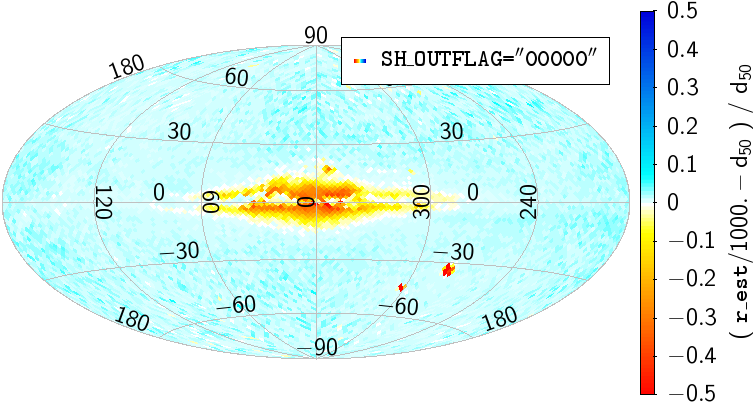}
 	\caption{Comparison of {\tt StarHorse} results with the {\it Gaia} DR2-only-derived distances of \citet{Bailer-Jones2018}. The top left and the bottom panels show the median relative differences with respect to our results. Top left panel: Median difference as a function of distance. The red density distribution and the running median correspond to the flag-cleaned results, while the grey running median corresponds to the full converged sample. Bottom panels: dependence on sky position for the full sample (left) and filtering on the {\tt SH\_OUTFLAG} (right). Top right: comparison of statistical uncertainties. See Sec. \ref{bj18} for details.}
 	\label{accuracy_bj18}
 \end{figure*}

\begin{figure*}\centering
 	\includegraphics[width=0.49\textwidth]{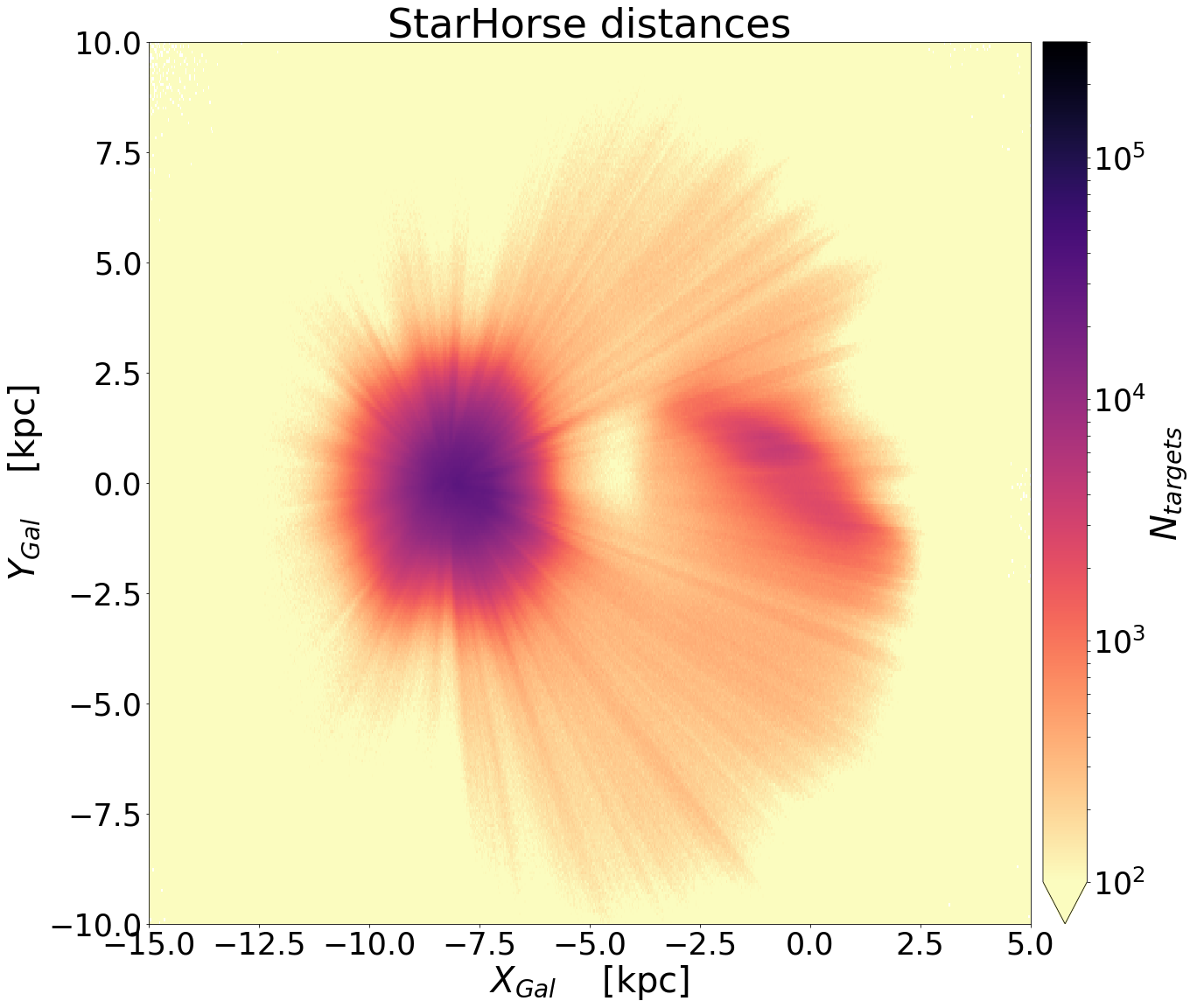}
 	\includegraphics[width=0.49\textwidth]{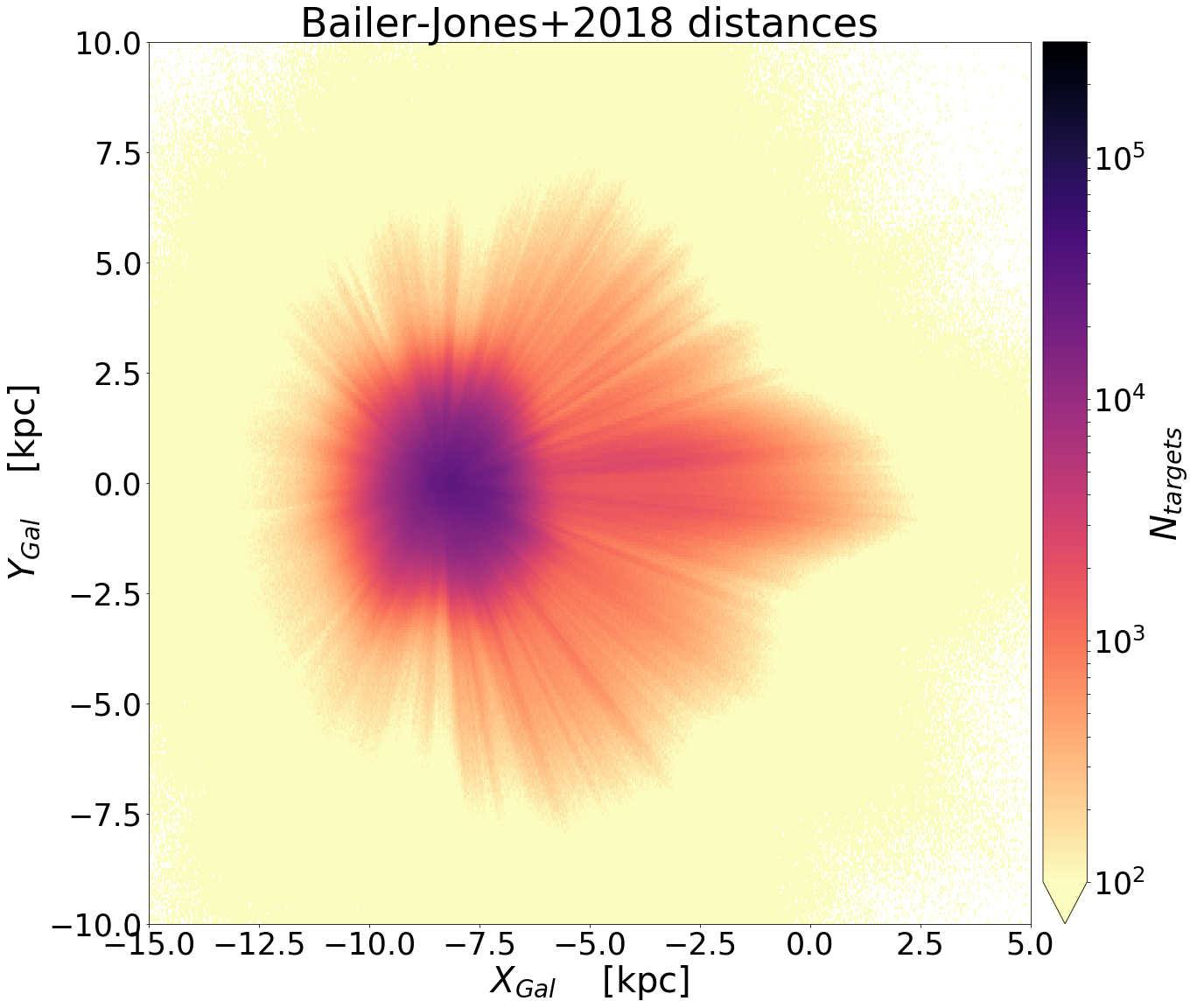}\\
 	\caption{Comparison of Galactic Cartesian density maps for the {\tt SH\_GAIAFLAG}$=="000", ${\tt SH\_OUTFLAG}$="00000"$ sample, resulting from \citet{Bailer-Jones2018} geometric distances (right) and {\tt StarHorse} (left; same as Fig. \ref{rzmaps}, top left panel). In both panels, the Sun is located at $(X_{\rm Gal}, Y_{\rm Gal})=(-8.2 {\rm kpc},0)$. }
 	\label{xymaps}
\end{figure*}

\begin{figure*}\centering
 	\includegraphics[width=0.4\textwidth]{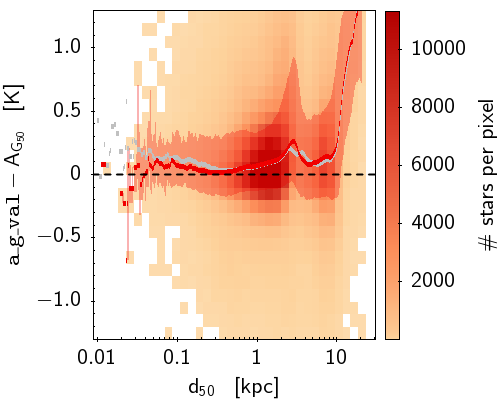}
 	\includegraphics[width=0.4\textwidth]{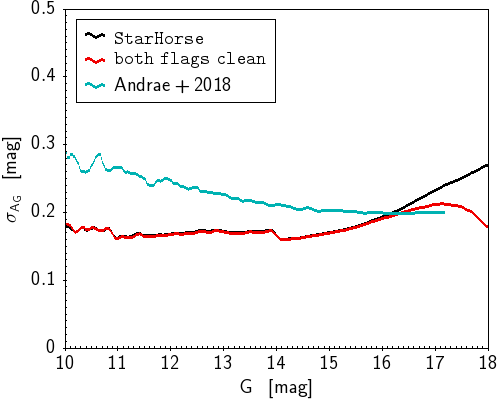}
 	\includegraphics[width=0.4\textwidth]{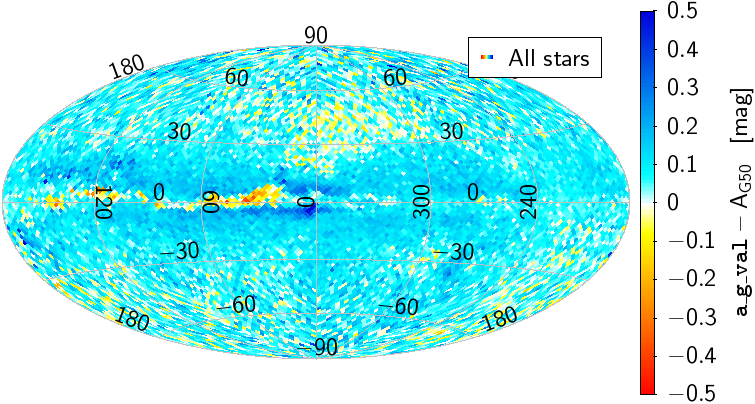}
 	\includegraphics[width=0.4\textwidth]{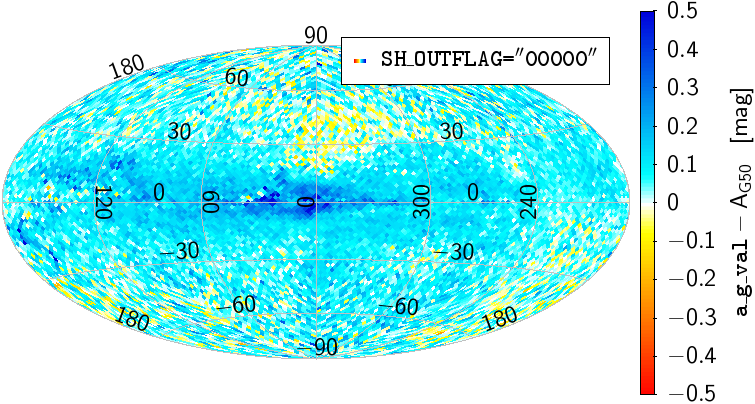}
 	\caption{Comparison of {\tt StarHorse} results with the {\it Gaia} DR2-only-derived $A_G$ extinction estimates of \citet{Andrae2018}, in the same style as the comparison to the \citet{Bailer-Jones2018} distances shown in Fig. \ref{accuracy_bj18}. See Sec. \ref{andrae} for details.}
 	\label{accuracy_apsis1}
 \end{figure*}

\begin{figure*}\centering
 	\includegraphics[width=0.4\textwidth]{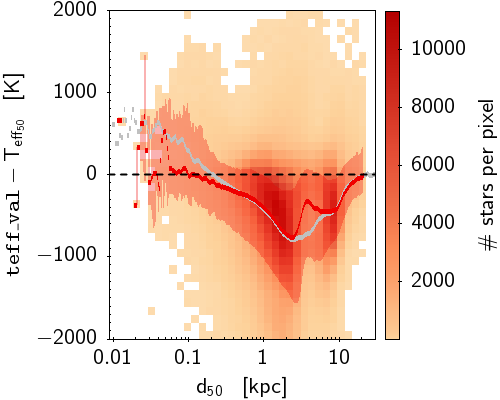}
 	\includegraphics[width=0.4\textwidth]{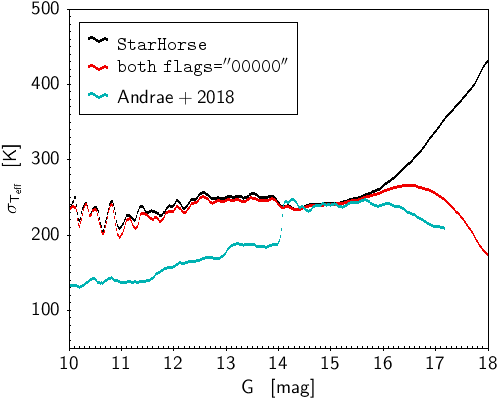}
 	\includegraphics[width=0.4\textwidth]{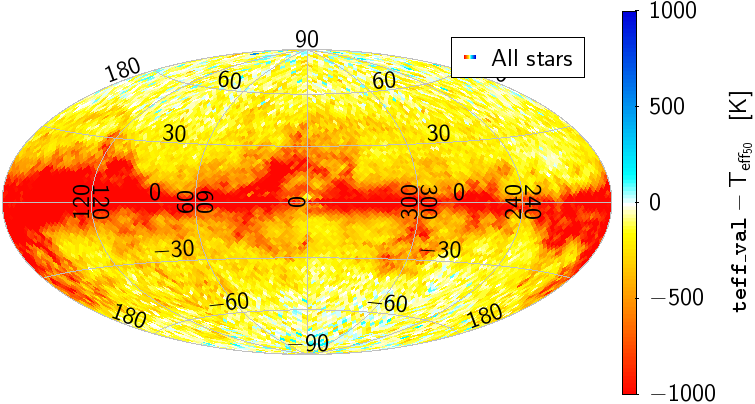}
 	\includegraphics[width=0.4\textwidth]{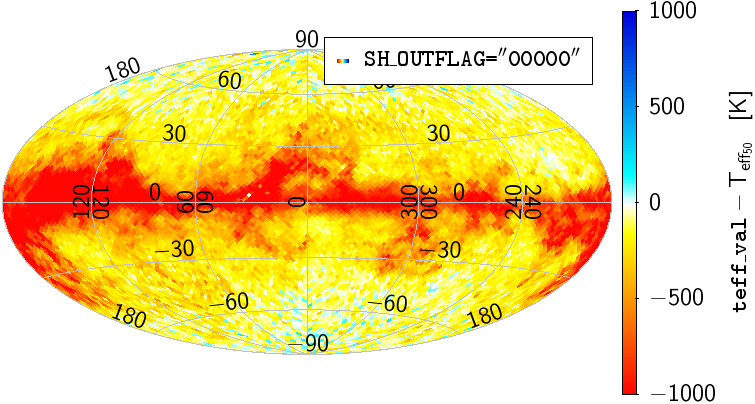}
 	\caption{Comparison of {\tt StarHorse} results with the {\it Gaia} DR2-only-derived $T_{\rm eff}$ estimates of \citet{Andrae2018}, in the same style as the comparison to the \citet{Bailer-Jones2018} distances shown in Fig. \ref{accuracy_bj18}. See Sec. \ref{andrae} for details.}
 	\label{accuracy_apsis2}
 \end{figure*}

\begin{figure*}\centering
 	\includegraphics[width=0.49\textwidth]{im/CMD_bothflags.png}
 	\includegraphics[width=0.49\textwidth]{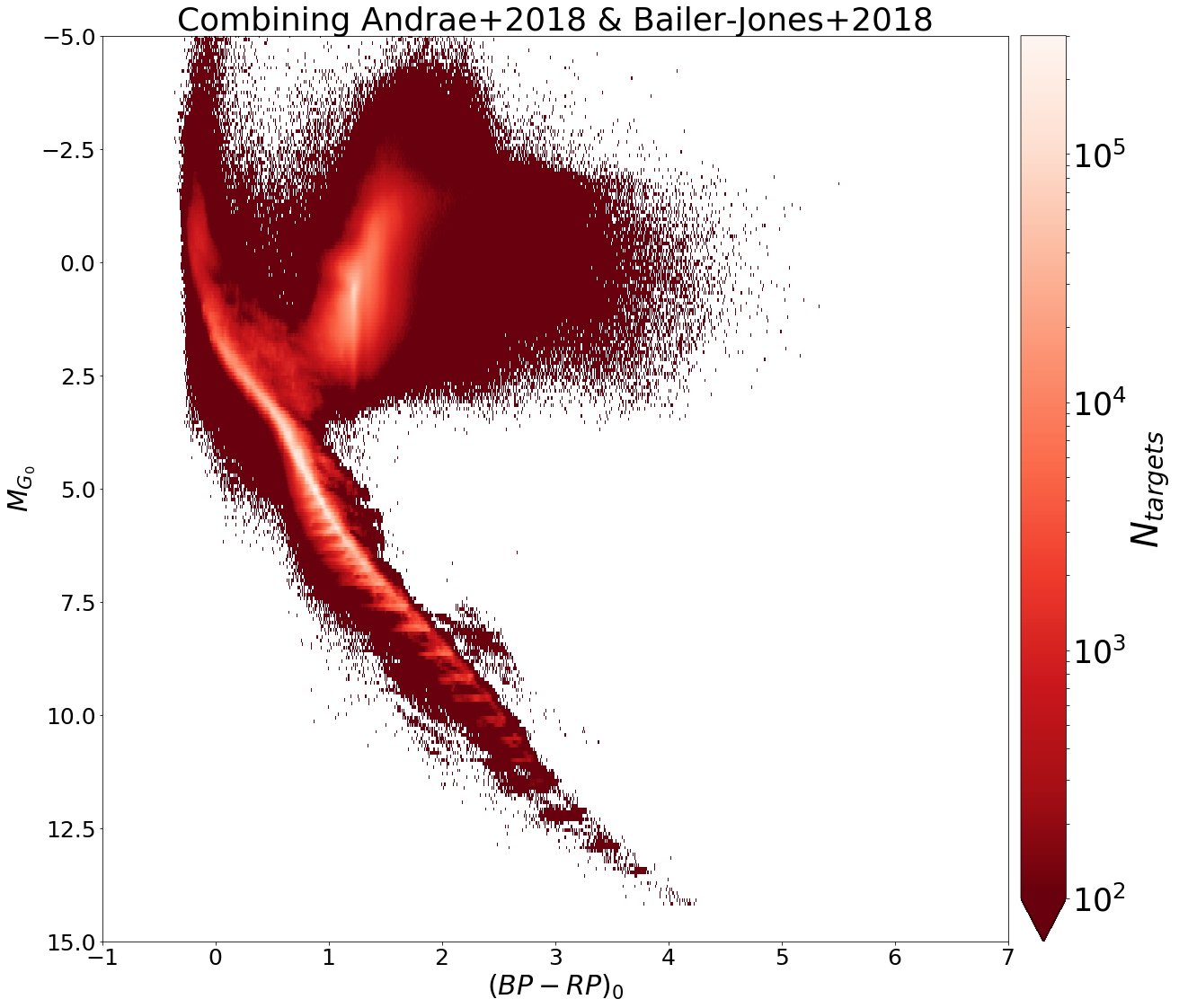}
 	\caption{Comparison of the {\tt StarHorse} posterior {\it Gaia} DR2 flag-cleaned colour-magnitude diagram (136,606,128 stars) to the one obtained by \citet{Andrae2018} (criterion: {\tt a\_g\_val!$=$NULL}; 84,498,216 stars).
    }
 	\label{cmds4}
 \end{figure*}

\begin{figure*}\centering
 	\includegraphics[width=0.49\textwidth]{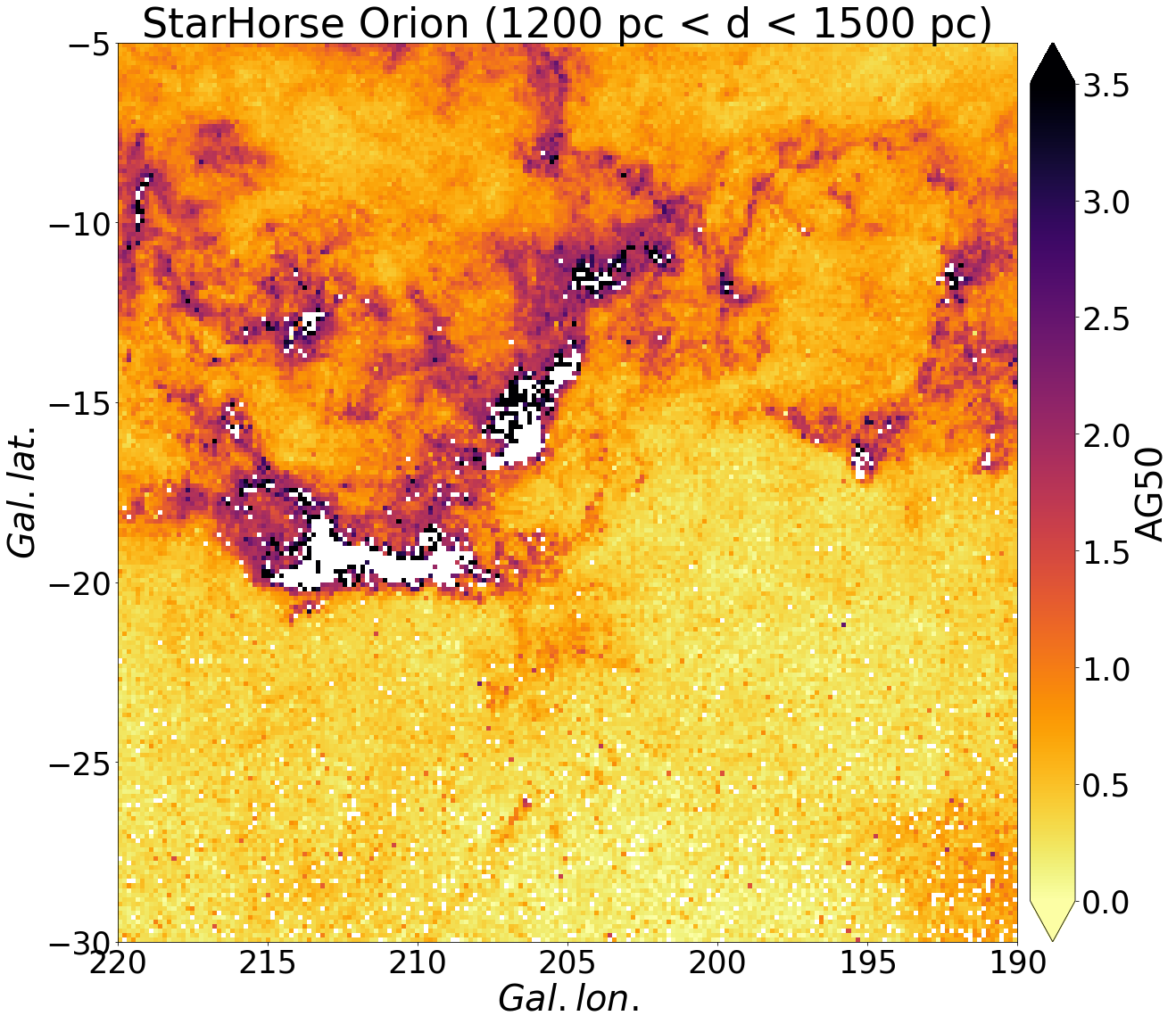}
 	\includegraphics[width=0.49\textwidth]{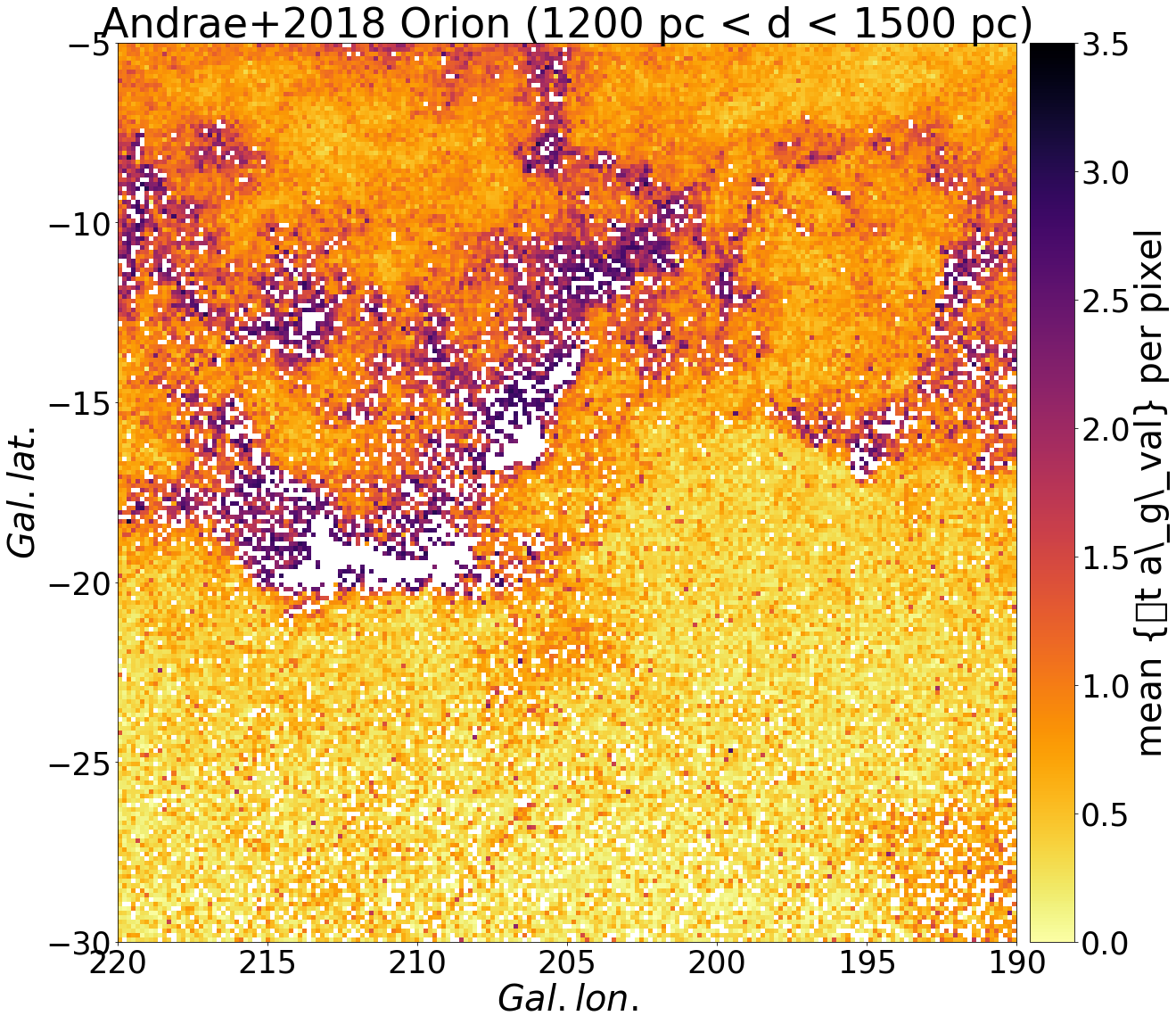}
    \caption{Comparison of the $A_G$ extinction maps obtained from {\tt StarHorse} (with the minimal flag {\tt SH\_OUTFLAG[0]}="0") and {\it Apsis} \citep{Andrae2018} for a distance slice in the Orion region. The number of stars contained in the left plot is 246,266, while in the right plot there are 129,549 stars.}
 	\label{avmaps3}
\end{figure*}

\section{Precision and accuracy}\label{uncertainties}

\subsection{Overall precision}\label{precision}

Figure \ref{precision_vs_Gmag} shows the median relative uncertainties in the {\tt StarHorse} output parameters as a function of Gaia DR2 G magnitudes. 
In all panels, we again show the results for all converged stars (in black, as before), and for the flag-cleaned sample (in red, as before). The other coloured lines shown in Fig. \ref{precision_vs_Gmag} demonstrate the precision improvement obtained by adding more photometric data to the {\it Gaia} DR2 data. The blue curves denote the running median uncertainty for stars with only {\it Gaia} DR2 photometry, while the other coloured lines refer to stars for which other data are available (as indicated in the legend in the middle panels). The green curve stands for the stars with complete photometric information ({\tt SH\_PHOTOFLAG}=="GBPRPgrizyJHKsW1W2").

Uncertainties in most quantities increase with $G$, as expected, due to the increasing uncertainties in the astrometry and photometry (we note the logarithmic y-axis in all panels of Fig. \ref{precision_vs_Gmag}, except the top middle). The fundamental determinant for the distance precision (as well as for most of the other {\tt StarHorse} output parameters) is of course not the magnitude itself, but the parallax signal-to-noise ratio \citep[e.g.][see also Fig. \ref{sigdist_piepi}]{Bailer-Jones2018}. The complex correlations between the output parameters and their uncertainties are shown in Fig. \ref{bigcorner} for the primary output parameters. For a global picture of the parameter and uncertainty trends including the secondary output parameters, we refer to Fig. \ref{hugecorner}. For the sake of brevity, however, here we focus our discussion mainly on the median uncertainty trends with $G$ magnitude (which is correlated with $\varpi/\sigma_{\varpi}$) shown in Fig. \ref{precision_vs_Gmag}. 

The distance precision plot (top left panel of Fig. \ref{precision_vs_Gmag}) deserves some further discussion. To begin with, in the bright regime ($G_{DR2} < 14$, including the radial-velocity sub-sample of $7\cdot10^6$ stars), the vast majority of stars have uncertainties of 8\% or less in distance, as expected from the exquisite parallax quality of {\it Gaia} DR2 \citep{Lindegren2018, Arenou2018}. Focussing on the orange and blue lines of this plot, it may be surprising that the addition of 2MASS magnitudes to the input data seems to worsen the distance precision. In fact, the most precise distances for stars with $G<12.5$ are obtained when only using {\it Gaia} DR2 data. This observation points to a tension between the 2MASS and the {\it Gaia} DR2 data: The range of acceptable distances for these stars is precisely determined by their measured parallax (we assume the parallax offset to be fixed and only a function of magnitude; see Table \ref{calibtable}); so that the three {\it Gaia} DR2 passbands alone already constrain the space of possible stellar parameters and extinctions. The three 2MASS magnitudes alone also constrain effective temperature and extinction, so if these two independent constraints are in tension with each other (most likely due to an underestimated - systematic - parallax uncertainty), the uncertainty on the output distance increases. 

For a similar, but not identical reason, the addition of Pan-STARRS1 magnitudes to the set of {\it Gaia} DR2+2MASS+AllWISE photometry does not improve the distance precision, but has a slight effect in the opposite direction (compare magenta and green lines in Fig. \ref{precision_vs_Gmag}). We suggest that this points to an inconsistency between the {\it Gaia} DR2 photometry with the Pan-STARRS1 one. Since the {\it Gaia} DR2 photometry is of unprecendented precision, and the transmission curves and zeropoints are well-characterised (at least for not too red stars, $G_{\rm BP}-G_{\rm BP}\lesssim1.5$) by \citet{MaizApellaniz2018}, we tentatively suggest that this indicates a need for additional corrections of the PS1 zeropoints. However, we decided to keep the PS1 photometry as input where possible, since the five optical passbands considerably help in increasing the precision of extinction and metallicity (see top middle and bottom middle panel of Fig. \ref{precision_vs_Gmag}).

The wiggle in the median uncertainty at $G\sim13$ is due to the decrease in parallax uncertainty at that magnitude transition (\citealt{Lindegren2018}). The sharp increase in median distance uncertainty at $G\simeq16.5$ is due to the transition into the low-signal-to-noise parallax regime. In particular, the distance uncertainties are much larger for faint main-sequence stars, which fill the locus of $G_{DR2} >$ 16.5 and $\sigma_d / d > 0.5$, whereas the (predominantly photometric) distances to distant red giants remain more precise (see Fig. \ref{sigdist_piepi}). 

The flag-cleaned results, by construction, yield much more precise results also in the faint regime. The drop in median uncertainty for those stars is due to the distance precision cut embedded in the definition of the {\tt SH\_OUTFLAG} (see Appendix \ref{outflag1}).

For the precision in $A_V$ extinction (top middle panel of Fig. \ref{precision_vs_Gmag}), we note a flat trend as a function of $G$, with the uncertainty increasing significantly only in the regime where parallaxes and distances become much more uncertain ($G\simeq16.5$). We also note the expected increase in precision when including more photometric passbands (see also Table \ref{summarytable}). 

Similar observations hold for the median uncertainties in effective temperature as well as the uncertainties in the secondary output parameters $\log g$, [M/H], and stellar mass $M_{\ast}$. For the latter we also note an (at first sight puzzling) decreasing trend of the overall median uncertainty (black line in bottom right panel of Fig. \ref{precision_vs_Gmag}) up to $G\sim14$, which is an effect of the different sampling of stellar populations at different magnitudes. \\

Figure \ref{xymaps_sigcoloured} shows the median uncertainties in the primary output parameters $d, A_V,$ and $T_{\rm eff}$ for stars in the Galactic disc as a function of their position. The top row shows the precision in each pixel in the $X$ vs. $Y$ plane for all converged stars, while the bottom row shows the corresponding results for the flag-cleaned sample. 

The top left panel demonstrates the sharp transition into the low-signal-to-noise parallax regime at heliocentric distances of $\sim2.5$ kpc (the {\it Gaia} DR2 "parallax sphere"). In the bottom left panel, this effect is much less severe, because of many distant giant stars passing the quality criteria of the {\tt StarHorse} flags. Even in the Galactic bar region, the typical uncertainties for the flag-cleaned sample only amount to $\sim30\%$.

The middle panels of Fig. \ref{xymaps_sigcoloured} especially highlight the decrease in $A_V$ precision in the quarter of the sky for which no Pan-STARRS1 photometry is available. We also note that outside the solar vicinity our extinction estimates are more precise in regions dominated by giant stars, resulting in a ring around the Sun for which the uncertainties are higher. The same is true for the effective temperatures (right panels), because the two posterior quantities are correlated (see  Appendix \ref{pdfappendix} for a short discussion of the correlations of correlations in the estimated parameters).

\subsection{Accuracy: Comparison to asteroseismology}\label{seismo}

It is difficult to find true benchmark tests for the distance, extinction, and stellar parameter scales of large surveys that are not themselves affected by significant systematic uncertainties. One of the most precise and widely used anchors in the context of spectroscopic surveys is the asteroseismic surface gravity scale defined by the seismic scaling relations for red giant stars \citep[e.g.][]{Holtzman2015,Valentini2016,Valentini2017}. 

In Fig. \ref{seismocomparison}, we show a comparison to the precise surface gravities, distances, and extinctions determined from asteroseismic data from the {\it Kepler} and {\it K2} missions \citep{Khan2019}. The surface gravities have been computed using the $\nu_{\rm max}$ scaling relation \citep{Brown1991, Kjeldsen1995} and can be considered accurate to 0.05 dex \citep[e.g.][]{Noels2016,Hekker2017}. 
Distances and extinctions were derived with the Bayesian stellar parameter estimation code PARAM \citep{daSilva2006, Rodrigues2014, Rodrigues2017}, using the global seismic oscillation parameters $\Delta\nu$ and $\nu_{\rm max}$ as well as effective temperatures and metallicities determined from APOGEE DR14 \citep{Abolfathi2018} and SkyMapper \citep{Casagrande2019} for the {\it Kepler} and {\it K2} fields, respectively. 

The $\log g$ comparison shown in the top left panel of Fig. \ref{seismocomparison} shows that our posterior gravity values perform unexpectedly well, with median biases below the 0.1 dex level. Since the $\log g$ information is mostly driven by the parallax measurements, the fact that our posterior $\log g$ values agree so well with the values obtained by using {\it Kepler} and {\it K2} data underlines the unprecedented quality of the {\it Gaia} DR parallaxes. It also means that our global zero-point correction inspired by \citet{Zinn2019} performs very well - slightly better in the {\it Kepler} field, as expected, but still acceptably in the K2 fields, although the the parallax zero-point in these fields is different \citep[$\sim-0.006$ mas in C3 and $\sim-0.017$ mas for C6, compared to $\sim -0.05$ mas for the {\it Kepler} field][]{Khan2019}. Another encouraging fact is that for all three fields we get the lowest biases for stars around the red clump ($2.3 \lesssim\log g \lesssim 2.7$). Although there are are few comparison stars in the upper RGB for the C3 field, it seems that these tend to have slightly more biased posterior $\log g$ values, in concordance with the different parallax zero-point for that field. In summary, the comparison to the asteroseismically detrmined surface gravities shows that our posterior $\log g$ estimates perform better than expected, with biases and precisions at the level of medium-to-high-resolution spectroscopy (at least for luminous red-giant stars out to $\sim5$ kpc).

Regarding the primary output parameters distance and extinction, the comparison with \citet{Khan2019} shows that our distances to red-giant stars seem to be accurate at the 10\% level with respect to the asteroseismic scale at least up to distances of around 5 kpc, with most accurate results achieved for the {\it Kepler} field (biases $<1\%$ up to $d\lesssim3.8$ kpc; see middle panel panel of Fig. \ref{seismocomparison}), for which our parallax zeropoint correction is most accurate. For C3 and C6, the parallax correction most likely overestimates the true parallaxes, which is why the {\tt StarHorse} distances are systematically lower than those from PARAM on the entire range of distances. As for the extinction comparison (right panel and bottom row of Fig. \ref{seismocomparison}), the picture is similar, with some systematics seen for the most nearby and the most distant stars in the {\it K2} fields, further corroborating the position-dependent parallax zero-point shift reported by \citet{Arenou2018} and \citet{Khan2019}.

\subsection{Accuracy: Comparison to APOGEE}\label{apogee}

To further test the accuracy of the {\tt StarHorse} {\it Gaia} DR2 results, we cross-matched them with the stars contained in the fourteenth data release of the Sloan Digital Sky Survey (SDSS DR14; \citealt{Abolfathi2018}), resulting in 210,545 stars, out of which 179,272 pass all {\it Gaia} DR2 {\tt StarHorse} flags. Furthermore, for this comparison we only consider stars with valid calibrated APOGEE stellar parameters, resulting in a total overlap sample of 59,351 giant stars. In Fig. \ref{apogeecomparison}, we show the differences with respect to the spectroscopically derived stellar parameters derived by the APOGEE Collaboration (which are of course of much higher precision), as well as {\tt StarHorse} distances and extinctions derived from those parameters together with {\it Gaia} DR2 parallaxes and additional photometry (same version of the code; Santiago et al., in prep.). We note that no {\it Gaia} DR2 photometry was used for the APOGEE {\tt StarHorse} run.

In the top row of Fig. \ref{apogeecomparison}, we show the differences between our photometric estimates and the values derived by the APOGEE Stellar Parameter and Chemical Abundances Pipeline (ASPCAP; \citealt{GarciaPerez2016}) as a function of the ASPCAP values, for effective temperature, surface gravity, and metallicity, respectively. Since the APOGEE stellar parameters are completely independent from ours, these comparisons possibly reveal the most important systematics of the results presented in this work. It is worth noting, however, that even the APOGEE sample cannot be considered a gold standard, since the photometric and spectroscopic effective temperature scales depend on the wavelength range and resolution used, and may still be subject to shifts of up to 100 K \citep[e.g.][]{Casagrande2014, Joensson2018}. In fact, to remove systematics with respect to the temperature scale defined by the infra-red flux method \citep{GonzalezHernandez2009}, for DR14 and following releases a metallicity- and temperature dependent calibration was applied to the raw ASPCAP results (\citealt{Holtzman2018, Joensson2018}).

The effective temperature comparison shown in the top left panel of Fig. \ref{apogeecomparison} shows very few systematics for the overlap sample inside the ASPCAP calibration range. Our effective temperature scale (defined by the PARSEC 1.2S isochrones) is offset by $-65$ K on average (median: $-46$ K) with respect to the APOGEE sample, with the difference being zero around 4500 K and increasing for both cooler and warmer stars. Since this difference is at the level of the systematics expected for the APOGEE $T_{\rm eff}$ scale, we can consider it insignificant. Perhaps more interesting is the overall spread of the temperature difference, which amounts to 197 K and is very similar to the formal uncertainties that {\tt StarHorse} delivers for $T_{\rm eff}$.

The second and third panel in the top row of Fig. \ref{apogeecomparison} show an analogous comparison for our secondary output parameters $\log g$ and [M/H]. As for $T_{\rm eff}$, also spectroscopic surface gravity values suffer from some level of systematics \citep{Holtzman2015, Valentini2016}. In DR14 and subsequent releases, however, the raw ASPCAP values have been carefully calibrated using precise $\log g$ values delivered by the CoRoT and {\it Kepler} asteroseismic missions (see \citealt{Holtzman2018} for details). The systematics seen in the $\log g $ comparison are therefore likely to be mainly due to our analysis (i.e. our priors), or intrinsic differences of the gravity scale of the PARSEC models with respect to the asteroseismic scale.

For the comparison to the (calibrated) ASPCAP metallicity scale, which is both much more precise and accurate than ours ($\sim 0.05$ dex; \citealt{Holtzman2018}), we see that {\tt StarHorse} tends to determine solar metallicities for the bulk of the APOGEE stars. This behaviour shows that the metallicity sensitivity of the broad-band photometric filters used in this work is very small for moderate metallicities ([M/H]$\gtrsim -1$), and the posterior metallicity estimates are in most cases dominated by the (broad) metallicity priors. However, for moderately metal-poor objects ($-2\lesssim$[M/H]$\lesssim -1$), our code seems to deliver somewhat more reliable metallicity estimates, enabling the construction of a candidate list of metal-poor stars, which may be followed up with spectroscopy (Chiappini et al., in prep.).

The middle row of Fig. \ref{apogeecomparison} displays the comparison with the distances to APOGEE stars obtained with the same version of the {\tt StarHorse} code, but including also the spectroscopic stellar parameters as input quantities, thereby yielding much more precise results \citep{Queiroz2018}. The left panel shows that we achieve remarkable overall concordance with the astro-spectro-photometric distance scale up to distances of $\sim 7$ kpc, with our photo-astrometric distances being increasingly too small towards more distant (especially extragalactic) stars, as could be expected (we note that this trend is comparable to the trend observed by comparing to open-cluster distances discussed in Sect. \ref{clusters}, and contrary to the trend observed when comparing to the distances of \citealt{Bailer-Jones2018}; see Sect. \ref{bj18}). The right panel also shows that the distance differences do not show any dependence on sky position, except for the very extincted regions close to the Galactic plane, where we see a tendency to overestimate distances compared to APOGEE (mostly a consequence of our $A_V=4$ boundary). Overall, however, since also the APOGEE-derived distances are model- and prior-dependent, the meaning of the median trend is limited and can be considered rather an internal consistency check.

The same is true for the extinction comparison (extinction difference as a function of sky position) shown in the bottom row of Fig. \ref{apogeecomparison}, although here we see more significant systematic trends. The bottom left panel shows that the median systematic differences as a function of distance are moderate ($\lesssim0.2$ mag), although certainly significant. In the bottom right panel, however, we observe a slight ($\lesssim 0.1$ mag) overestimation of the median extinction in the low-extinction regime at high latitudes with respect to the APOGEE-derived values, while at low latitudes the {\it Gaia} DR2+photometric extinction estimates are on average lower than the APOGEE ones by $\sim0.2$ mag for most of the parts of the Galactic disc, and severely underestimated in the most extinct regions, thus compensating the larger distances observed in the same sky regions (again mostly due to our $A_V$ prior, which is too restrictive for very extincted regions). These caveats should be kept in mind when using our catalogue.

\subsection{Accuracy: Open cluster comparison}\label{clusters}

By virtue of the precise {\it Gaia} DR2 astrometry, \citet{Cantat-Gaudin2018} were able to establish new membership probabilities and physical parameters for 1229 Galactic open clusters. In Fig. \ref{clustercomparison1}, we compare our results obtained for the most certain cluster members of \citet{Cantat-Gaudin2018} to the cluster distances reported in that work. These cluster distances were obtained by a maximum-likelihood analysis taking into account the systematic parallax uncertainties and the global parallax zero-point offset of $-0.029$ mas \citep{Lindegren2018}. 

Overall, Fig. \ref{clustercomparison1} shows similar trends as the comparison to the APOGEE-derived distances (left panel in the middle row of Fig. \ref{apogeecomparison}). We see slight negative median differences up for distances between $\sim2$ and 10 kpc, which, considering the systematic parallax uncertainties, is very much beyond the accuracy limits of the open cluster distance scale (we note the different global zeropoint applied in the bright regime). 
The obvious advantage of the cluster distances is the suppression of the statistical uncertainty with the square root of the number of members. However, the cluster members are affected by the same varying parallax zero-point as a function of sky position, magnitude, parallax, and/or colour. Therefore, also this comparison is not a fundamental distance comparison for the most distant clusters, since their distance uncertainty is dominated by systematics. In addition, the $P_{\rm memb}==1$ criterion used in Fig. \ref{clustercomparison1} only refers to astrometric membership; no photometry was used in the construction of the membership list of \citet{Cantat-Gaudin2018}.

To further illustrate the performance of {\tt StarHorse} for the open cluster sample, we show in Fig. \ref{clustercomparison2} a detailed comparison for the four most distant open clusters (Melotte 71, NGC 2420, NGC 6819, and NGC 6791) studied at high spectral resolution in the compilation of \citet{Bossini2019}. These authors have recently published revised Bayesian cluster parameters based on {\it Gaia} DR2 data and the membership list of \citet{Cantat-Gaudin2018}, thus providing also cluster ages and line-of-sight extinctions. In the left column of Fig. \ref{clustercomparison2}, we show the distance-extinction plane for each of the clusters, indicating also the size of the {\tt StarHorse} uncertainties of the individual members. Except for a number of outliers in NGC 6819, and for the problematic cluster NGC 6791 (which is also a highly debated object in the open-cluster literature; see e.g. \citealt{Linden2017, Villanova2018, Martinez-Medina2018}), we find that our results for the bulk of individual member stars cluster very well around the median distances and extinctions of \citet{Bossini2019}, within the known systematics.

The middle column of Fig. \ref{clustercomparison2} shows the effect of applying a {\tt StarHorse} extinction- and distance correction on the cluster colour-magnitude diagram, displaying the amount of noise added to the CMD when using our results (which, we recall, were obtained under the assumption that these stars are field stars). We find that the resulting diagrams are not much more noisy than the original cluster sequences, even for these distant populations, giving further confidence in our results. 

Finally, in the right column of Fig. \ref{clustercomparison2} we show the posterior {\it Kiel} diagrams for each of the four clusters, colour-coded by the median metallicity in each pixel, demonstrating that there is at least some metallicity sensitivity in the photometric data used in this work.

\subsection{Caveats}\label{caveats}

As can be expected from a data-intensive endeavour such as the one undertaken in this paper, there are several known caveats that should be taken into account when using our results. Some of them were discussed in the previous sections, but we list some additional considerations here. Specifically, for this work we did not attempt to correct the following effects (ordered by decreasing relative importance) that may have potential impacts on further scientific analyses. 

\begin{enumerate}
\item {\it Colour- and sky-position-dependent parallax zero-point shifts:} As has been demonstrated by \citet{Arenou2018, Lindegren2018a}, and \citet{Khan2019}, the parallax zero-point offset of {\it Gaia} DR2 depends on the magnitude, colour, and position in the sky, in a non-trivial manner. In this work we only account for a magnitude-dependent zero-point offset, which may lead to biased results in parts of the sky where the parallax zero-point shift is very different from the global shift applied here.
\item {\it Unresolved binaries:} For most binaries the primary star by far dominates the light budget (especially on the red-giant branch), so that we expect that our results do not suffer from significant binarity-induced biases in that regime. As for the main sequence, the exquisite quality of the {\it Gaia} DR2 photometry has shown that many star clusters show a well-populated equal-mass binary sequence. For these cases, we do expect significantly biased parameters. However, currently the only computationally tractable way to account for equal-mass binaries in the data is to use data-driven models for the colour-magnitude diagram \citep{Coronado2018}, which was explicitly not the aim of this work.
\item {\it Simple Galactic priors:} Due to optimization of computational resources, our priors do not include some well-established features of the Galaxy: for example a warped stellar disc, extended structures in the outer disc such as the Monoceros ring or Triangulum-Andromeda, or the presence of the nearby Magellanic Clouds, the Sagittarius dSph, etc. (see Sec. \ref{maps}). Also, the current extinction limit of $A_V=4$ could be replaced by a more informative prior.
\item {\it Uncertainties in the extinction curve:} Our results rely on the validity of the assumed extinction curve, which is limited in several respects. Most importantly, we do not allow for variable $R_V$ values (or $x$, in the \citealt{Schlafly2016} notation). In addition, by using the $G_{\rm BP}$ magnitudes we extrapolate the \citet{Schlafly2016} extinction law slightly into the blue. In the near future, with the {\it Gaia} DR3 BP/RP low-resolution spectra, it may be possible to simultaneously solve for stellar parameters, distance, extinction, and the extinction curve.
\item {\it Gaia DR2 photometry in crowded fields:} The {\it Gaia} DR2 aperture photometry is known to be prone to systematic errors in crowded regions of the Galactic disc \citep{Evans2018, Arenou2018}. For many applications, it will be sufficient to filter out data affected by this problem using the ${\tt phot\_bp\_rp\_excess\_factor}$ (as implemented in {\tt SH\_GAIAFLAG[1]}). 
\item {\it Systematic photometric errors in $G_{\rm BP}$ in the faint regime:} Faint sources ($G_{\rm BP}\gtrsim19$; 2\% of the converged stars) have been shown to be affected by background under-estimation, which leads to magnitude errors greater than 0.02 mag \citep[][Fig. 34b]{Arenou2018}. This may imply slight biases in our derived parameters for these stars.
\item {\it Systematic photometric errors in supplementary photometry:} Systematic photometric errors will result in slightly biased results, especially for the more delicate secondary output parameters. This is the reason why we introduced an uncertainty floor for the photometric data. As an example, in Sect. \ref{precision} we saw that the inclusion of the Pan-STARRS1 photometry in the input data yields more precise extinction and metallicity estimates, but slightly worsens the distance and $\log g$ precision. This fact suggests some remaining tensions between the zero-points or the passband definitions of the {\it Gaia} DR2 and Pan-STARRS1 photometric systems. Another potential problem arises for missing photometric uncertainties in 2MASS and WISE (0.03\% of the converged sample): in these cases, the catalogue magnitude values refer to upper limits and our fiducial uncertainties of 0.3 mag may be too optimistic.
\item {\it Uncertainties in the Gaia DR2 passband definitions:} Although we have used the improved transmission curves and recalibrated photometry of \citet{MaizApellaniz2018}, remaining uncertainties in the passband definition may impact our results. \citet{MaizApellaniz2018} have convincingly shown that more flux-calibrated spectro-photometry is necessary to characterise the on-board transmission curves of the {\it Gaia} photometers, especially in the red regime.
\item {\it Contamination by extragalactic objects and potentially erroneous cross-matches:} In this work we have used the carefully computed crossmatch tables provided as part of {\it Gaia} DR2 \citep{Marrese2019}, so that the occurence of erroneous crossmatches should be minimal. Also the contamination of our catalogue by galaxies with observed colours similar to those of stars is possible, although very unlikely in our magnitude regime.
\end{enumerate}

\section{Comparison to {\it Gaia} DR2 results}\label{dpac_comp}

Figures \ref{accuracy_bj18} through \ref{avmaps3} present a comparison of the {\tt StarHorse} primary output parameters with the widely used {\it Gaia} DR2-based distance catalogue of \citet{Bailer-Jones2018}, and with the {\it Gaia} DR2 astrophysical parameters presented by \citet{Andrae2018}. In this section, we discuss these comparisons in detail. 

\subsection{Comparison to the \citet{Bailer-Jones2018} distances}\label{bj18}

Shortly after {\it Gaia} DR2, \citet{Bailer-Jones2018} released a catalogue of geometric Bayesian distances for 1.33 billion stars inferred from the {\it Gaia} DR2 parallaxes. Their goal was to provide homogeneous distance estimates for the entirety of {\it Gaia} DR2 stars, "independent of assumptions about the physical properties of, or interstellar extinction towards, individual stars." The authors rely on a solid theoretical background \citep{Bailer-Jones2015a, Astraatmadja2016}, and carefully calibrated their geometric distance prior as a function of galactic longitude and latitude using a {\it Gaia} DR2-like stellar density model \citep{Rybizki2018}, and their results have been shown to provide precise results also beyond the {\it Gaia} DR2 parallax sphere.

The main advantages of the approach taken by \citet{Bailer-Jones2018} are 1. a very clean selection function (they provide mode statistics for virtually all 1.33 billion stars with measured parallaxes, and 2. a smooth transition between the likelihood-dominated and the prior-dominated regime of the {\it Gaia} DR2 data. According to the authors, the main drawbacks are 1. lower precision than could be achieved by including more information, and 2. biased distances for certain subsets of objects (e.g. distant giants, extragalactic stars). With the {\tt StarHorse} results, we can now quantify these statements.

Figure \ref{accuracy_bj18} shows a comparison of {\tt StarHorse} and \citet{Bailer-Jones2018} distances. The top left panel shows the median relative distance difference as a function of {\tt StarHorse} distance. For the flag-cleaned sample, we observe an overall concordance between both distance scales up to distances of $\sim 3$ kpc, and then a continuously growing deviation, in the sense that the \citet{Bailer-Jones2018} distances are typically smaller than the {\tt StarHorse} ones. This behaviour is expected, since the exponentially decaying space density prior employed by \citet{Bailer-Jones2018} tends to confine stars to distances within $\sim 6$ kpc. 

In the top right panel of Fig. \ref{accuracy_bj18}, we compare the median precision obtained with {\tt StarHorse} as a function of $G$ magnitude with the formal uncertainties given by \citet{Bailer-Jones2018}. Surprisingly at first sight, the median \citet{Bailer-Jones2018} distance uncertainties are smaller than the corresponding {\tt StarHorse} uncertainties. However, it should be taken in mind that we increased the {\it Gaia} DR2 parallax uncertainties, in accordance with the recent analysis of \citet{Lindegren2018a}. In the same panel of Fig. \ref{accuracy_bj18} we also show the (ill-defined) approximations of the uncertainties of inverse parallax distances, with and without our parallax uncertainty recalibration (see Table \ref{calibtable}). The offset between these two lines is essentially the same as the one between the \citet{Bailer-Jones2018} distances and the {\tt StarHorse} results, indicating that the parallax uncertainty is the driving parameter also for our distance precision, and suggesting that the \citet{Bailer-Jones2018} distance uncertainties are slightly underestimated.

The bottom panels of Fig. \ref{accuracy_bj18} show the median relative distance deviation as a function of sky position. In the bottom left panel, we show the "all converged stars" sample, while the bottom right one only contains the results with {\tt SH\_OUTFLAG}$==$"00000". The concordance of our flag-cleaned results with the distances of \citet{Bailer-Jones2018} is remarkable over most of the sky (<5\% differences in more than 90\% of the HealPix cells), excluding only the Inner Galaxy ($|l|\lesssim30\deg, |b|\lesssim10\deg$) and the Magellanic Clouds. The different picture for the full sample again cautions against the blind use of our non-flag-cleaned median distances.

The remaining differences are most probably due to the fact that the prior used in \citet{Bailer-Jones2018} is more apt for nearby main-sequence stars and therefore tends to underestimate distances to far-away giant stars, especially in the Galactic bulge (see \citealt{Bailer-Jones2018}). An additional effect is that close to the inner Galactic plane the stellar density does not decrease exponentially, but in fact is a more complex function than described by the exponentially decreasing space density prior. Finally, it is possible that the dust model used by \citet{Rybizki2018} is substantially different from the actual dust distribution in our Galaxy in the Inner Galaxy, and therefore the values of the prior length scale in this region could be underestimated.

Finally, in Fig. \ref{xymaps} we compare the Galactic density maps in Galactic Cartesian co-ordinates. In the region of highest density close to {\it Gaia} DR2 parallax sphere, the maps are very similar, as expected. The {\tt StarHorse} distances reach higher values due to the less restrictive density prior used. The most striking feature, however, is of course the emergence of the Galactic bar as a clear overdensity in the $XY$ plane (see Sect. \ref{maps}). 

Summarising this comparison, we can say that due to the relatively low impact of the additional photometric measurements on the distances, our distance estimates are not more precise than the distances obtained by \citet{Bailer-Jones2018}, even when rescaling their input parallax uncertainties. However, we argue that at large distances, our values are more accurate due to the choice of more informative Galactic priors, which allows us to see more substructure, including the direct imprint of the Galactic bar in the density maps.

\subsection{Comparison to {\it Gaia} DR2 {\it Apsis} results}\label{andrae}

As part of {\it Gaia} DR2, \citet{Andrae2018} published a catalogue of astrophysical parameters ($T_{\rm eff}, A_G, E(G_{BP}-G_{RP})$, radius, luminosity). Due to the limited number of observables used (parallax + {\it Gaia} DR2 photometry), the output parameters $T_{\rm eff}$ and $A_G$ are strongly correlated, and therefore suffer from a number of caveats documented in {\it Gaia} DR2 (see \citealt{Andrae2018} for an extensive discussion). In this work, we set out to improve on these initial results by including more photometric data in our analysis, thus creating more leverage to break the degeneracy between effective temperature and extinction. Different from \citet{Andrae2018} who used a machine-learning algorithm to infer astrophysical parameters, we have chosen a more classical approach: Bayesian parameter inference over a grid of stellar models. In this sense, both our method and our input data are quite different from the work carried out by \citet{Andrae2018}. 

Figures \ref{accuracy_apsis1} and \ref{accuracy_apsis2} show a comparison of the {\tt StarHorse} results for $A_G$ extinction and effective temperature, respectively, in a similar fashion as for the comparison to the \citet{Bailer-Jones2018} distances in the Fig. \ref{accuracy_bj18}. The top panels show the median absolute differences between the two results as a function of distance. For the flag-cleaned sample (red density and red running median line), despite the large spread we observe a remarkable overall concordance between the extinction scales of {\it Apsis} and {\tt StarHorse} up to a distance of $\sim 1$ kpc, followed by growing deviations towards larger distances. For nearby stars, we also note the effect of the $A_G$ positivity requirement imposed by the DR2 {\it Apsis} pipeline. For the effective temperatures, the agreement is significantly worse (we note the large y-axis scale), most likely due to the {\it Apsis} assumption of zero extinction and the fact the no stellar population prior was applied to the stellar model training dataset, resulting in too small $T_{\rm eff}$ values for reddened stars (see \citealt{GaiaCollaboration2018, Andrae2018} for details).

This explanation is confirmed in the bottom rows of Fig. \ref{accuracy_apsis1} and \ref{accuracy_apsis2}, which show the median absolute $A_G$ and $T_{\rm eff}$ differences as a function of sky position, for the full sample and the {\tt SH\_OUTFLAG}-cleaned sample, respectively. We find that the median $A_G$ deviations vary over the sky, indicating larger systematic differences close to the Galactic Centre. We can also make out the footprint of the Pan-STARRS1 survey in the $A_G$ comparison maps, indicating a slightly different extinction scale when these data are not available. The $T_{\rm eff}$ differences, on the other hand, follow the dust distribution in the Galaxy, an unphysical feature that is not present in the {\tt StarHorse} data, and that can also be explained by the zero-extinction assumption imposed for the {\it Gaia} DR2 {\it Apsis} run. 

The top right panels of Figs. \ref{accuracy_apsis1} and \ref{accuracy_apsis2} compare the quoted statistical uncertainties of {\tt StarHorse} and {\it Apsis}. However, the meaning of the {\it Apsis} uncertainties is probably limited, since the uncertainties are certainly dominated by systematics, as we have seen above. We therefore suggest that {\tt StarHorse}, in addition to providing more accurate results, also provides more realistic uncertainty estimates.

As an additional check, Fig. \ref{cmds4} shows a comparison of the {\tt StarHorse} extinction-corrected CMD with the one obtained from combining the \citet{Bailer-Jones2018} distances with the extinction estimates by \citet{Andrae2018} for the flag-cleaned sample. Apart from the increase in number counts (137 million vs. 84 million), we observe that the {\tt StarHorse} results (left panel) produce a much more populated lower main sequence and a more well-defined red clump when compared to the \citet{Andrae2018} CMD (right panel). 

Finally, Fig. \ref{avmaps3} shows a comparison of two-dimensional $A_G$ extinction maps in a narrow distance slice in the Orion region, showing the increase in number of stars with extinction estimates from {\tt StarHorse}, especially in dense obscured regions, which allows for more substructure to be revealed. 

In summary, we can confidently state that our astrophysical parameters are more accurate and have more reliable statistical uncertainties than the initial {\it Apsis} parameters obtained as part of {\it Gaia} DR2. This was expected from the inclusion of more multi-wavelength observations for a large part of the {\it Gaia} DR2 stars \citep{Andrae2018}. With this work we have verified this expectation quantitatively and provided improved results. We expect that with a machine-learning approach similar to that of \citet{Andrae2018} or \citet{Das2019}, accompanied by a better training sample, our results can be further improved.

In the near future, {\it Gaia} eDR3 (envisioned for summer 2020)\footnote{\url{https://www.cosmos.esa.int/web/gaia/release}} will provide improved photometry and astrometry for a similar number of sources as contained in DR2. This will enable short-term improvements on the results presented in this paper, with {\tt StarHorse} or similar codes.
{\it Gaia} DR3 (scheduled for spring 2021), will then provide much more precise astrophysical parameters determined from the BP/RP and RVS spectra. We imagine, however, that there may still be room for further improvements by adding additional constraints (such as near- and mid-infrared photometry).

%%%%%%%%%%%%%%%%%%%%%%%%%%%%%%%%%%%%%%%%%%%%%%%
\section{Conclusions}\label{conclusions}%%%%%%%%%%%%%%%%%%%%%%%%%%%%%%
%%%%%%%%%%%%%%%%%%%%%%%%%%%%%%%%%%%%%%%%%%%%%%%

In this work we have derived Bayesian stellar parameters, distances, and extinctions for 265 million stars brighter than $G=18$ with the {\tt StarHorse} code. By combining the precise parallaxes and optical photometry delivered by {\it Gaia}'s second data release ({\it Gaia} DR2) with the photometric catalogues of Pan-
STARRS1, 2MASS, and AllWISE, and the use of informative Galactic priors, our results substantially improve the accuracy of the extinction and effective temperature estimates provided with {\it Gaia} DR2 \citep{Andrae2018}, and arguably also the distances for distant giant stars, when compared to \citet{Bailer-Jones2018}. When cleaning our results for both unreliable input and output data, we obtain a sample of 137 million stars for which we achieve a median precision of 5\% in distance, 0.20 mag in $V$-band extinction, and 245 K in effective temperature for $G\leq14$, degrading slightly towards fainter magnitudes (12\%, 0.20 mag, and 245 K at $G=16$; 16\%, 0.23 mag, and 260 K at $G=17$, respectively). 

To verify our results, we presented distance- and extinction-corrected colour-magnitude diagrams, extinction maps as a function of distance, extensive density maps, as well as comparisons to asteroseismology, star clusters, the high-resolution spectroscopic survey APOGEE, and the {\it Gaia} DR2 astrophysical parameters and distances themselves. Furthermore, our results have already been used to infer the Galactic escape speed curve \citep{Monari2018}, to study the kinematic structure of the Galactic disc \citep{Quillen2018, Carrillo2019}, and for the survey simulations of the 4MOST spectroscopic survey \citep{deJong2019, Chiappini2019}. 

In this paper we also report for the first time a clear manifestation of the Galactic bar directly in the stellar density distributions. Considering that we assumed a vastly different prior for the density in the Galactic bulge, this observation can almost be considered a direct imaging of the Galactic bar. We also find a kinematic imprint of the coherent motion in the Galactic bar in the proper motion maps presented in Sect. \ref{kinematics}. A more detailed study of the Galactic bulge will be presented in Queiroz et al. (in prep.). We are confident that our value-added dataset will be useful for various other Galactic science cases, such as mapping the three-dimensional dust distribution within the Milky Way, or hunting for metal-poor stars.

We make our results available through the ADQL query interface of the {\it Gaia} mirror at AIP ({\tt gaia.aip.de}), and, together with complementary {\it Gaia} DR2 information, as binary tables at {\tt data.aip.de}. The digital object identifier of this dataset is {\tt doi:10.17876/gaia/dr.2/51}. 

\bibliographystyle{aa}
\bibliography{FA_library}

\begin{acknowledgements}
The authors thank Beno\^{i}t Mosser (Paris), Lola Balaguer (Barcelona), Ralf-Dieter Scholz (Potsdam), and the careful referee for valuable comments on the manuscript. We also thank Eleonora Zari (Leiden) for testing a preliminary version of the database. During the preparation of the data for the {\tt StarHorse} run, we have frequently used the {\it Gaia} archive at ESAC \citep{Salgado2017} as well as its mirrors at AIP and ARI. FA warmly thanks Alcione Mora and Juan González-Núñez (ESAC) for support with the {\it Gaia} archive in a critical moment.\\

During the analysis, we have made extensive use of the astronomical {\tt java} software TOPCAT and STILTS \citep{Taylor2005}, as well as the {\tt python} packages {\tt numpy} and {\tt scipy} \citep{Oliphant2007}, {\tt astropy}{AstropyCollaboration2013}, {\tt dask} \citep{DDT2016}, {\tt HoloViews}, and {\tt matplotlib} \citep{Hunter2007}. Figures \ref{bigcorner} and \ref{hugecorner} through \ref{badpdfs} were created using the {\tt corner} package \citep{Foreman-Mackey2016}. For Fig. \ref{kin_rc} we used {\tt galpy.util.bovy\_coords} \citep{Bovy2015} to transform the proper motions to the Galactic frame. This research has made use of the SVO Filter Profile Service (\url{http://svo2.cab.inta-csic.es/theory/fps/}; \citealt{Rodrigo2012, Rodrigo2013}) supported from the Spanish MINECO through grant AYA2017-84089.
\\

This work has made use of data from the European Space Agency (ESA) mission {\it Gaia} (\url{http://www.cosmos.esa.int/gaia}), processed by the {\it Gaia} Data Processing and Analysis Consortium (DPAC, \url{http://www.cosmos.esa.int/web/gaia/dpac/consortium}). Funding for the DPAC has been provided by national institutions, in particular the institutions participating in the {\it Gaia} Multilateral Agreement. 
This project has received funding from the European Union's Horizon 2020 research and innovation programme under the Marie Sk\l{}odowska-Curie grant agreement No. 800502 H2020-MSCA-IF-EF-2017. This work was partially supported by the MINECO (Spanish Ministry of Economy) through grant ESP2016-80079-C2-1-R (MINECO/FEDER, UE) and MDM-2014-0369 of ICCUB (Unidad de Excelencia Mar\'{i}a de Maeztu).
\end{acknowledgements}

\appendix

\section{Justification for {\tt StarHorse} flag definitions}\label{outflag1}

\begin{figure}\centering
 	\includegraphics[width=0.49\textwidth]{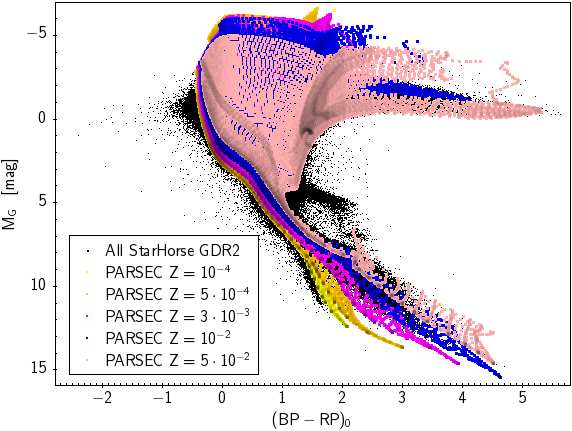}
 	\caption{{\tt StarHorse} posterior CMD for 1\% of the stars, comparing the results with the PARSEC 1.2S stellar models used to infer the results. Objects in unphysical regions of the posterior CMD result from poor median statistics of multi-modal posterior PDFs (see discussion in Sec. \ref{flags}) and are flagged in the first digit of {\tt SH\_OUTFLAG}. }
 	\label{cmdparsec}
 \end{figure}

\begin{figure*}\centering
 	\includegraphics[width=0.55\textwidth]{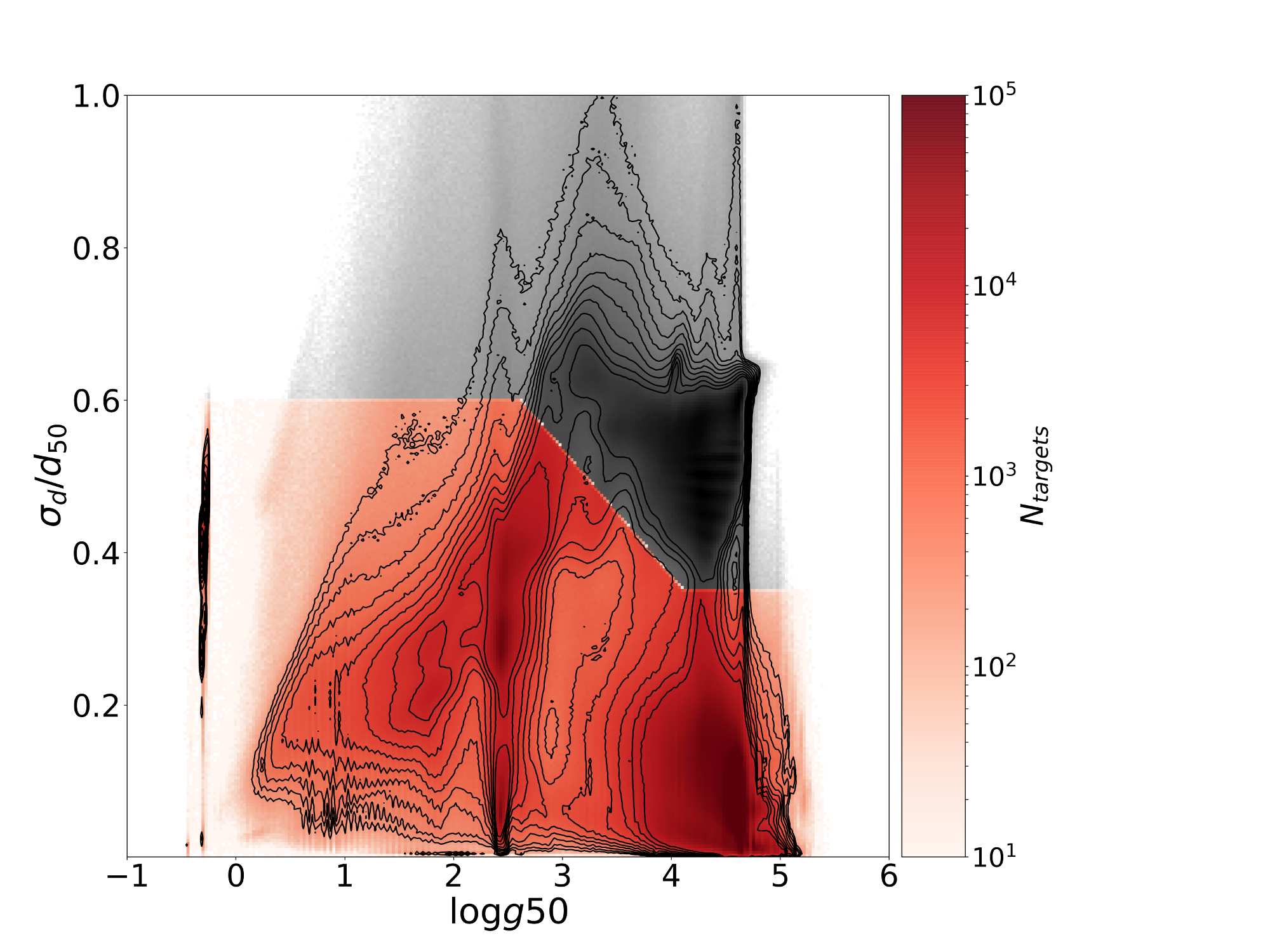}
 	\includegraphics[width=0.44\textwidth, trim={0.5cm 0cm 0 1cm},clip=True]{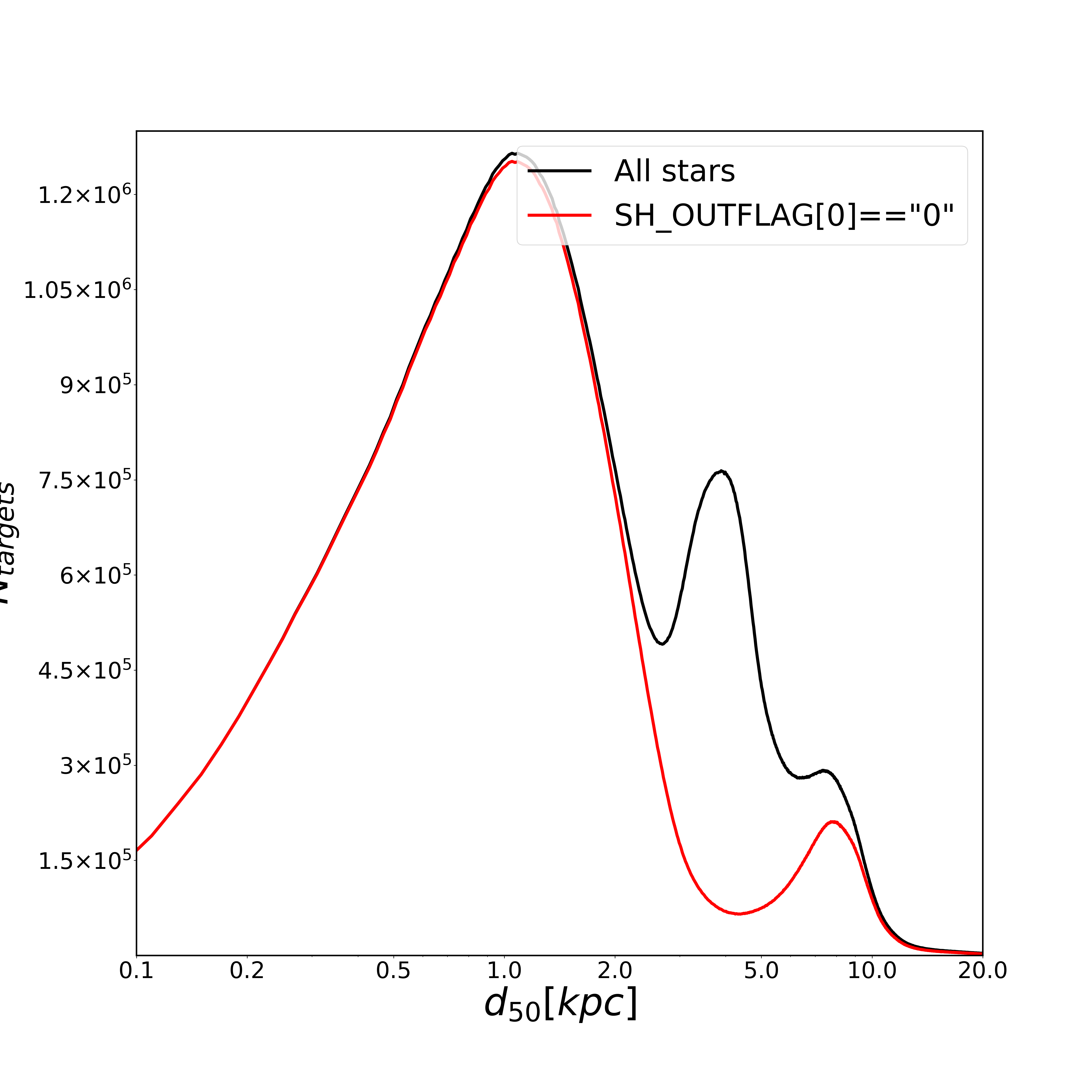}\\
 	\includegraphics[width=0.49\textwidth]{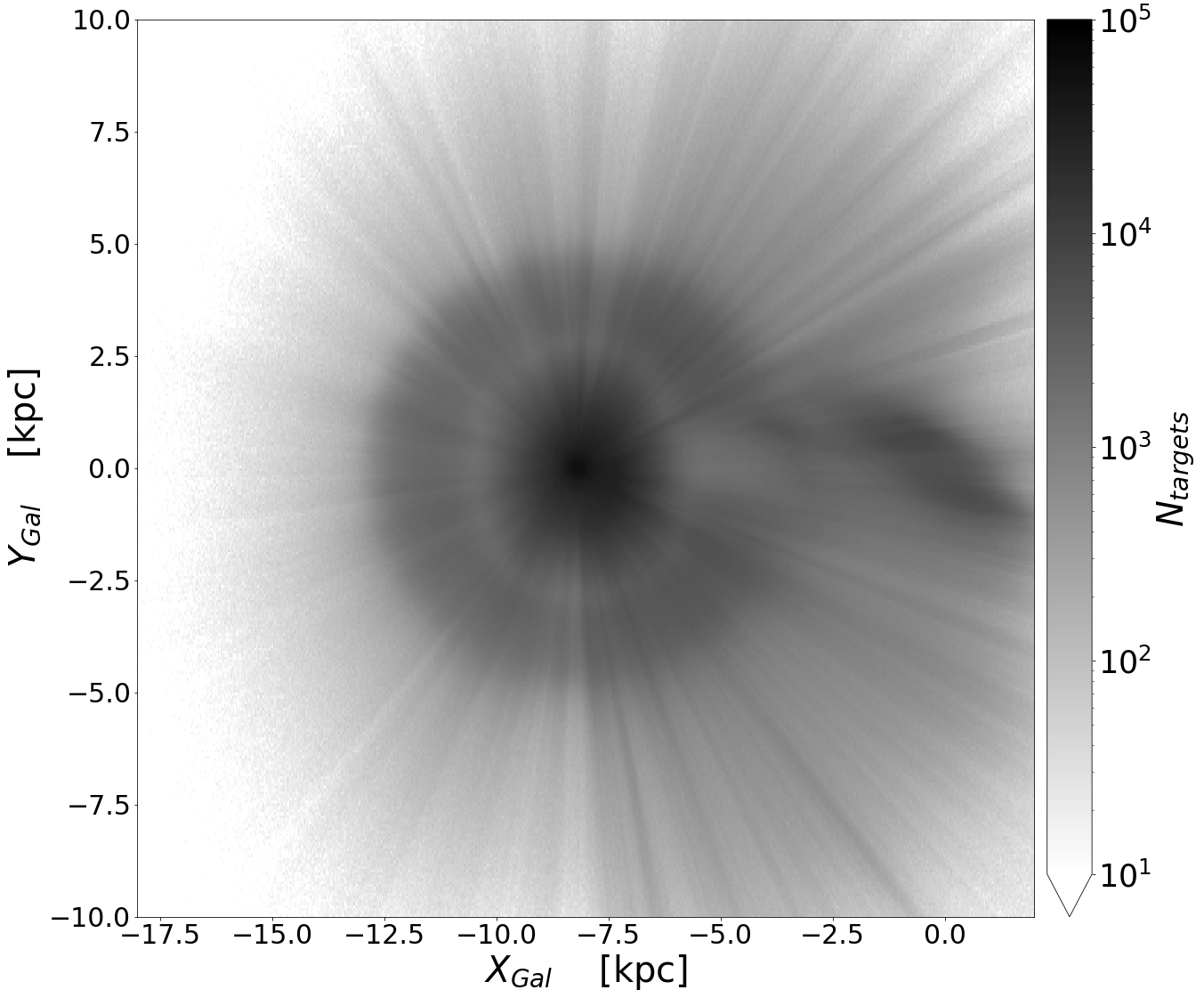}
 	\includegraphics[width=0.49\textwidth]{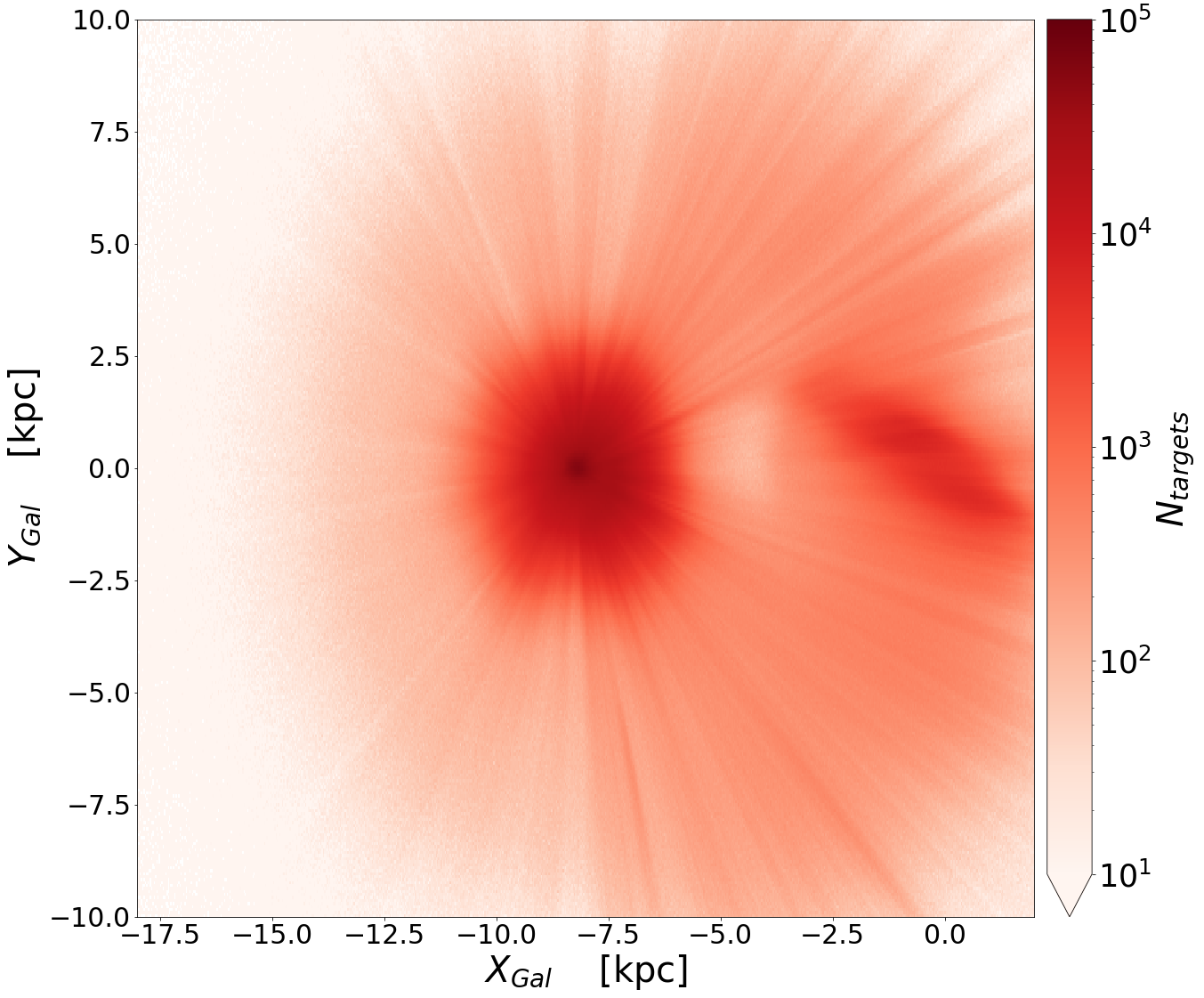}
 	\includegraphics[width=0.49\textwidth]{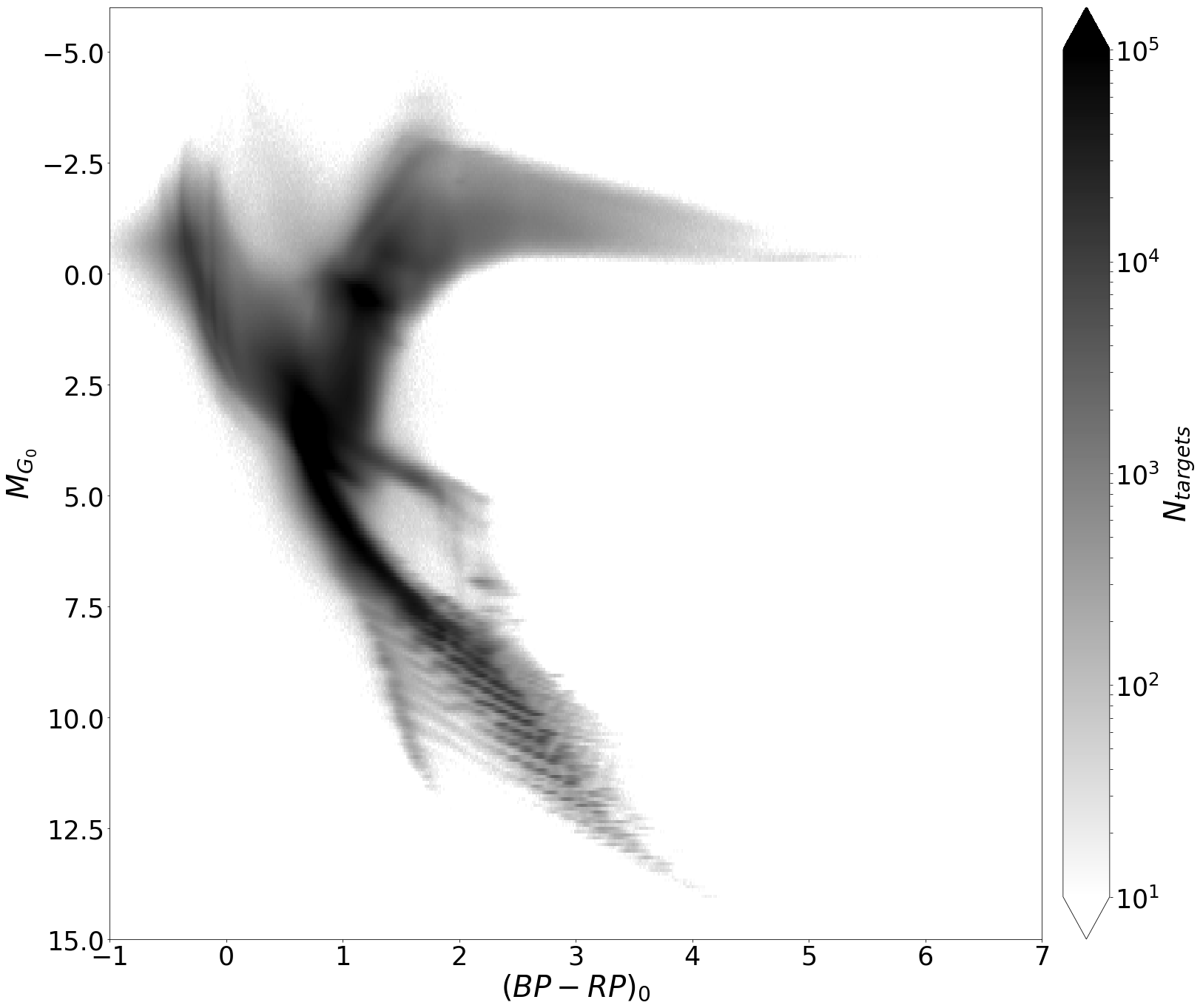}
 	\includegraphics[width=0.49\textwidth]{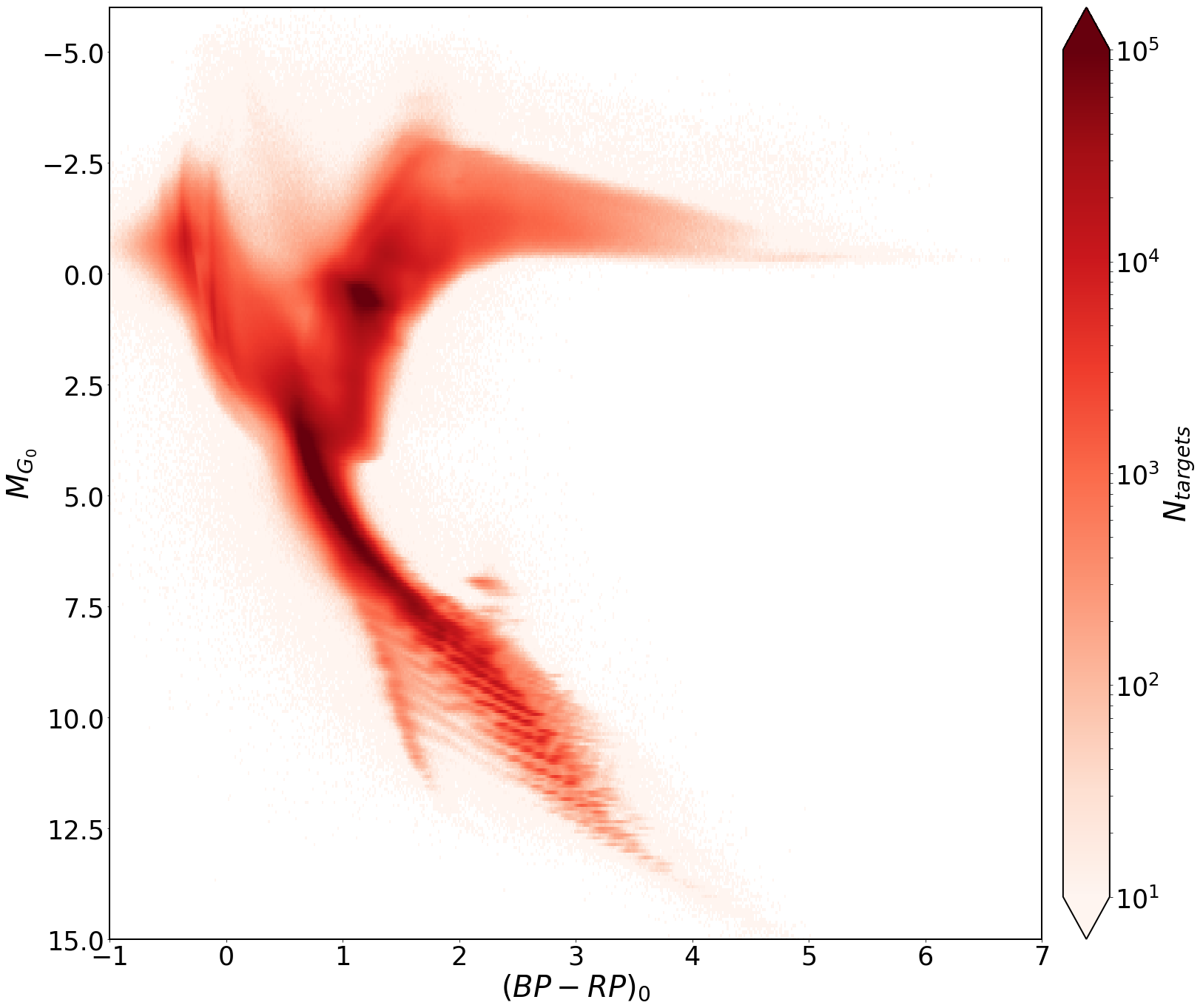}
    \caption{Illustration of the main {\tt StarHorse} reliability flag, {\tt SH\_OUTFLAG[0]}, corresponding to a cut in the $\sigma_d/d_{50}$ vs. $\log g_{50}$ diagram (top left panel). In each panel, the coloured distribution corresponds to converged stars with {\tt SH\_OUTFLAG[0]}$=="0"$, while the grey distribution corresponds to all stars. The top right panel shows the distance histograms for the two samples, and the middle panels show the density in Galactic $XY$ co-ordinates, highlighting the 'bloody-eye' effect of stars with poorly-constrained posterior distances that disappear when applying the reliability flag. The bottom panels show the CMDs of both samples, highlighting the disappearance of the nose feature when applying the reliability flag.}
 	\label{flagdefinition}
\end{figure*}

In the case of astro-photometric data with very uncertain parallaxes, Bayesian inference often results multimodal multidimensional posteriors. However, due to the huge data volume of {\it Gaia} DR2, the current {\tt StarHorse} version only saves the 1D median statistics for each posterior variable. This implies that in the case of a very bi- or even multimodal posterior, the median of one of the output parameters may lie in an unlikely, and sometimes unphysical, part of the model parameter space.

We explained this in Sec. \ref{flags}, and now illustrate the effect in some more detail using Figs. \ref{cmdparsec} and \ref{flagdefinition}. Figure \ref{cmdparsec} shows 1\% of the full {\tt StarHorse} output sample in an extinction-corrected {\it Gaia} colour-magnitude diagram. Overlaid are some of the PARSEC 1.2S models (for different metallicities) that were used to find the most likely combination of stellar parameters, distance, and extinction for each star. By construction, we expect that most of the stars fall in places compatible with at least one stellar model, and this is indeed the case for the vast majority of stars. 

Some stars, however, are situated outside the CMD space defined by the stellar models. As explained briefly in Sect. \ref{flags}, this means that the combination of their median posterior absolute magnitude, distance, and extinction should not be used together. The most prominent unphysical feature in the CMD is certainly the nose feature between the main sequence and the red-giant branch. 

We verified that this effect only occurs for faint stars with very uncertain parallaxes($\sigma_{\varpi}^{\rm cal} / \varpi^{\rm cal} \gtrsim 22\%$) - which is when the PDF of inverse parallax becomes seriously unbound (see Fig. \ref{sigdist_piepi} and \citealt{Bailer-Jones2015a, Astraatmadja2016a, Luri2018}). This results in a poor discrimination between dwarfs and giants for these typically faint ($G\gtrsim 16.5$; see Fig. \ref{gmaghisto}) stars. Although their median effective temperatures and extinctions may still be useful, their median 1D distances and other parameters should not be used. 

In any case, the flags provided together with the {\tt StarHorse} catalogue should only be regarded as a guidance. We encourage users of our data to apply their own quality cuts depending on their particular science case.

\subsection{The "bloody eye" effect for stars with poor parallaxes}\label{sauron}

In addition to the nose feature in the CMD , Figure \ref{flagdefinition} shows that the {\tt StarHorse} distances for the full $G<18$ sample (left panels) result in a very different appearance of the sampled space density in Galactic co-ordinates when compared to the flag-cleaned {\tt StarHorse} distances (right panels). This effect is a direct result of the poor data quality for faint stars and their consequently broad distance PDFs, as discussed above and in Sec. \ref{outflag}. Removing stars with such uncertain distances (typically $\sigma_d/d_{50}\gtrsim0.6$) leaves us with a much more meaningful density map.

\section{Parameter correlations and examples of {\tt StarHorse} joint posterior PDFs}\label{pdfappendix}

\begin{figure*}\centering
 	\includegraphics[width=0.99\textwidth]{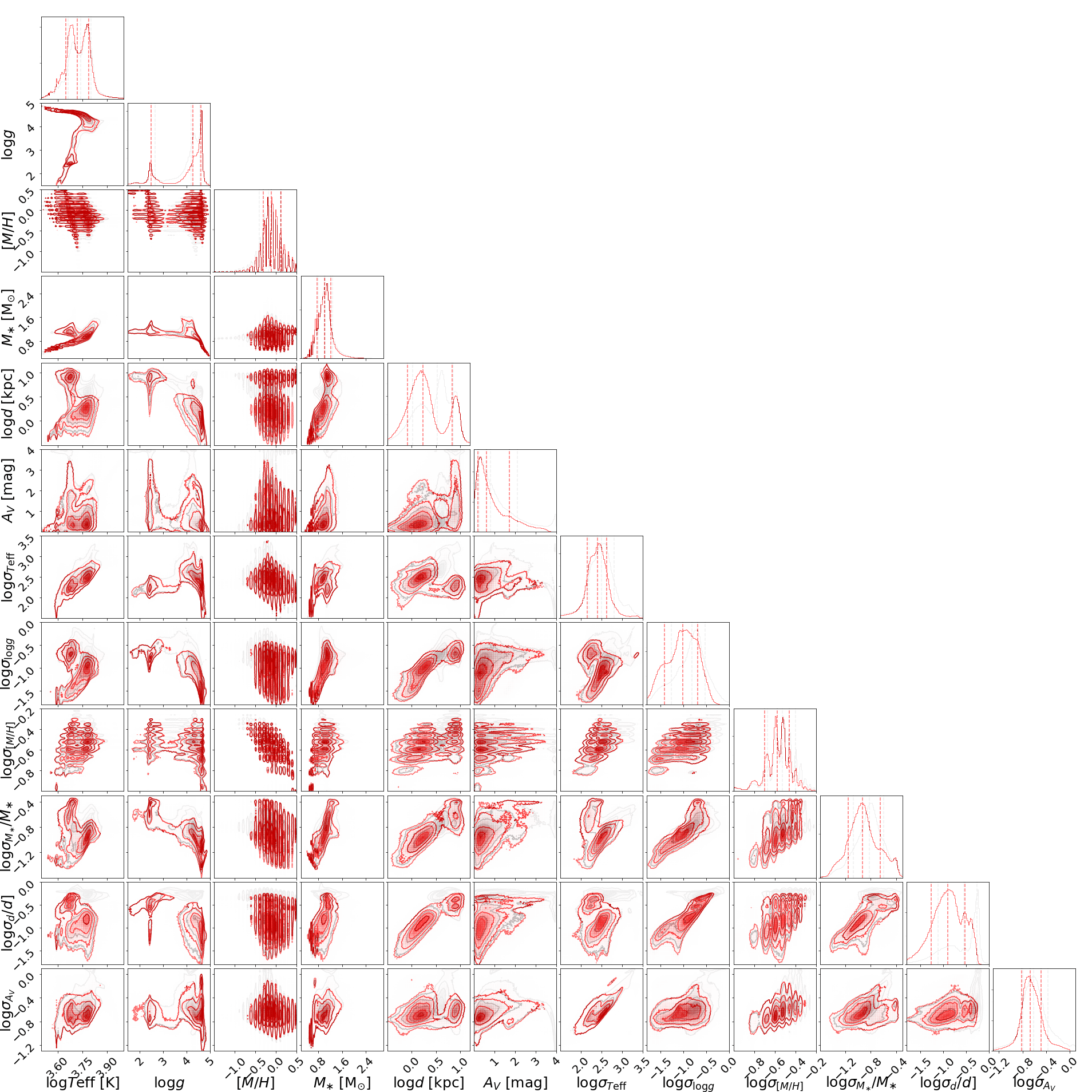}
 	\caption{{\tt corner} plot showing the correlations and distributions of {\tt StarHorse} median posterior output values $T_{\rm eff}, \log g, {\rm [M/H]}, M_{\ast}, d, A_V$, and their corresponding uncertainties. The grey contours show the distribution of the full sample, while the red contours show the distribution of all sources with {\tt SH\_OUTPUTFLAG}$==$"00000" and {\tt SH\_GAIAFLAG}$==$"000".}
 	\label{hugecorner}
 \end{figure*}

Figure \ref{bigcorner} presented the primary output of the {\tt StarHorse} code for the {it Gaia} DR2 sample in one plot, displaying the distributions and correlations of $T_{\rm eff}, d,$ and $A_V$, and their respective uncertainties, as well as $G$ magnitude and parallax signal-to-noise ratio. In Fig. \ref{hugecorner}, we now  display also the secondary output parameters $\log g,$ [M/H], and $M_{\ast}$, and their inter-correlations with the primary parameters. As in Fig. \ref{bigcorner}, the diagonal panels of Fig. \ref{hugecorner} provide the one-dimensional distributions for each parameter as area-normalised histograms, while the off-diagonal panels show the correlations between each of the output parameters. 

In addition to the observations described in Sect. \ref{summary}, we now also observe a griding effect in the [M/H] dimension, due to the finite metallicity resolution of our PARSEC model grid ($\sigma_{\rm [M/H]}$) chosen to optimise the computation cost. However, we recall that the quality of the output photometric metallicities is very diverse, and in many cases dubious (see Sects. \ref{apogee} and \ref{caveats}). We recall that the main objective of this paper is to deliver more precise distances, extinctions, and effective temperatures for a larger number of stars than provided in {it Gaia} DR2. The secondary output parameters ($\log g$, [M/H], $M_{\ast}$) mainly serve to attach stellar spectrum templates to {it Gaia} DR2 stars for the 4MOST Simulator \citep{deJong2019}, and to thereby assess the targeting strategy of the 4MOST low-resolution disc and bulge survey (4MIDABLE-LR; \citealt{Chiappini2019}).

To further gain insight into the correlations between the output parameters, we now discuss some randomly chosen full posterior PDFs. Ideally, in addition to the marginal median statistics for each output parameter, one would also like to report the full posterior. However, the large size of the posterior data files makes this completely unviable to even store this information (let alone publish it), even for the case of only thousands of stars. In this appendix, we therefore only show a few examples of {\tt StarHorse} joint posterior PDFs for the interested reader.

\begin{figure*}\centering
 	\includegraphics[width=0.49\textwidth]{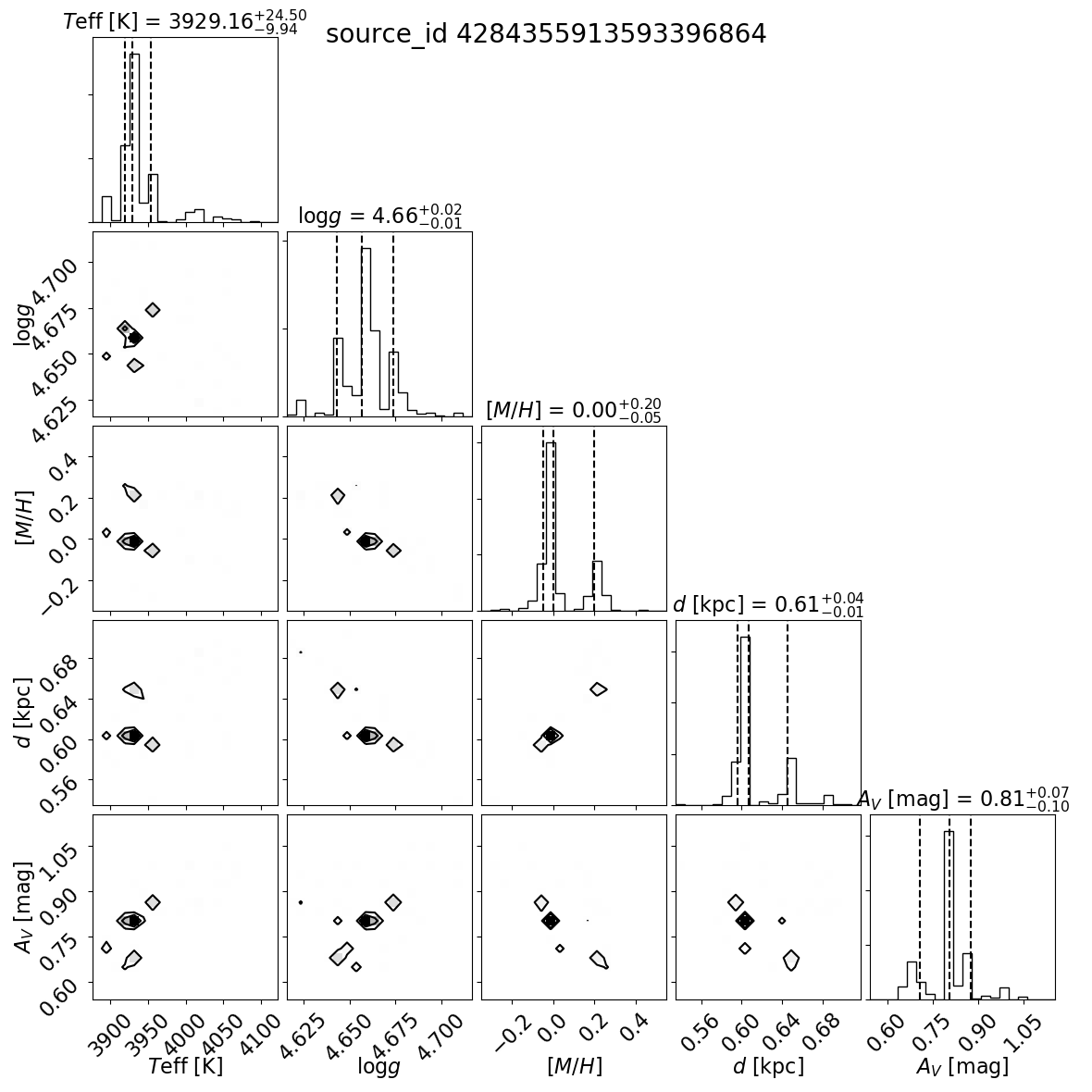}
 	\includegraphics[width=0.49\textwidth]{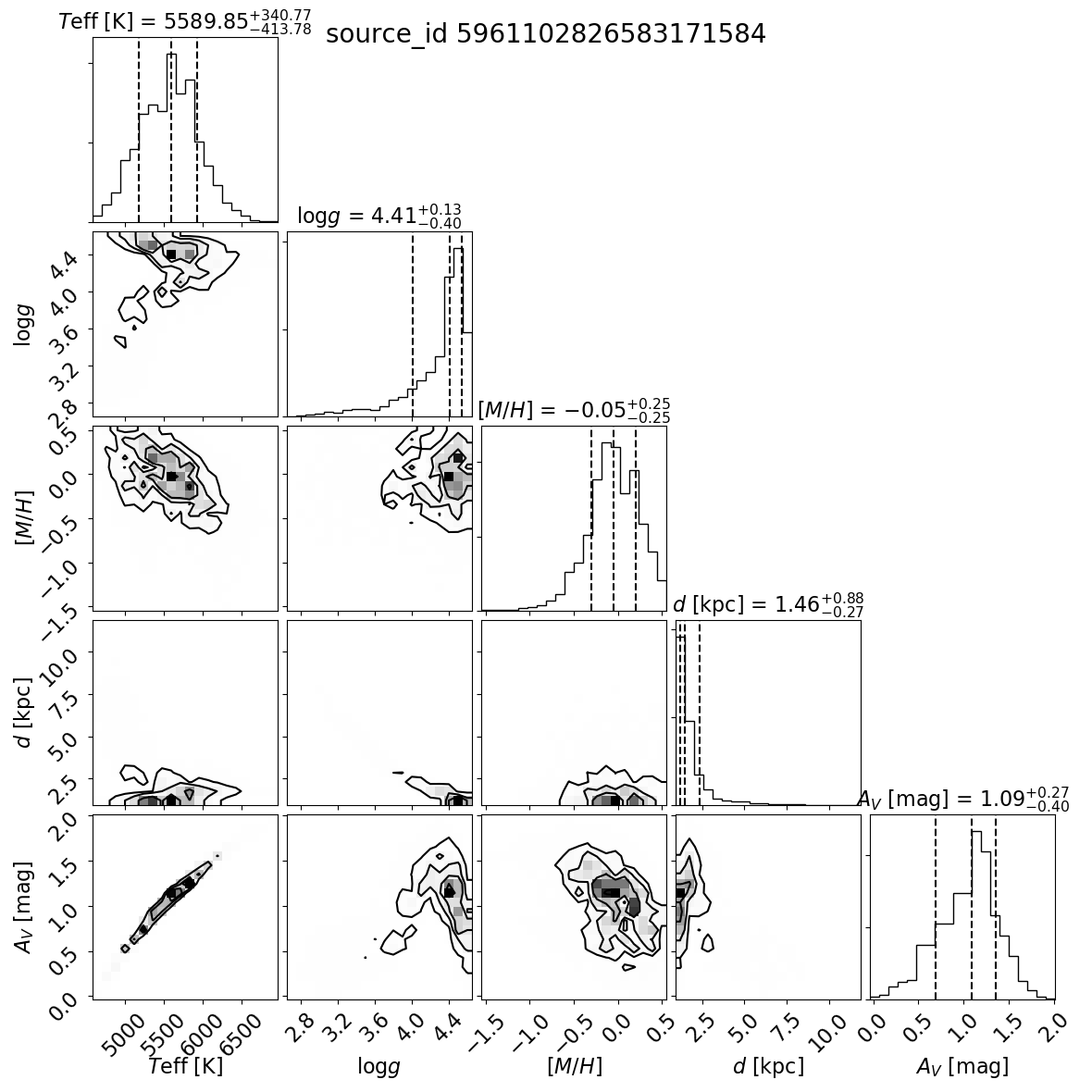}
 	\includegraphics[width=0.49\textwidth]{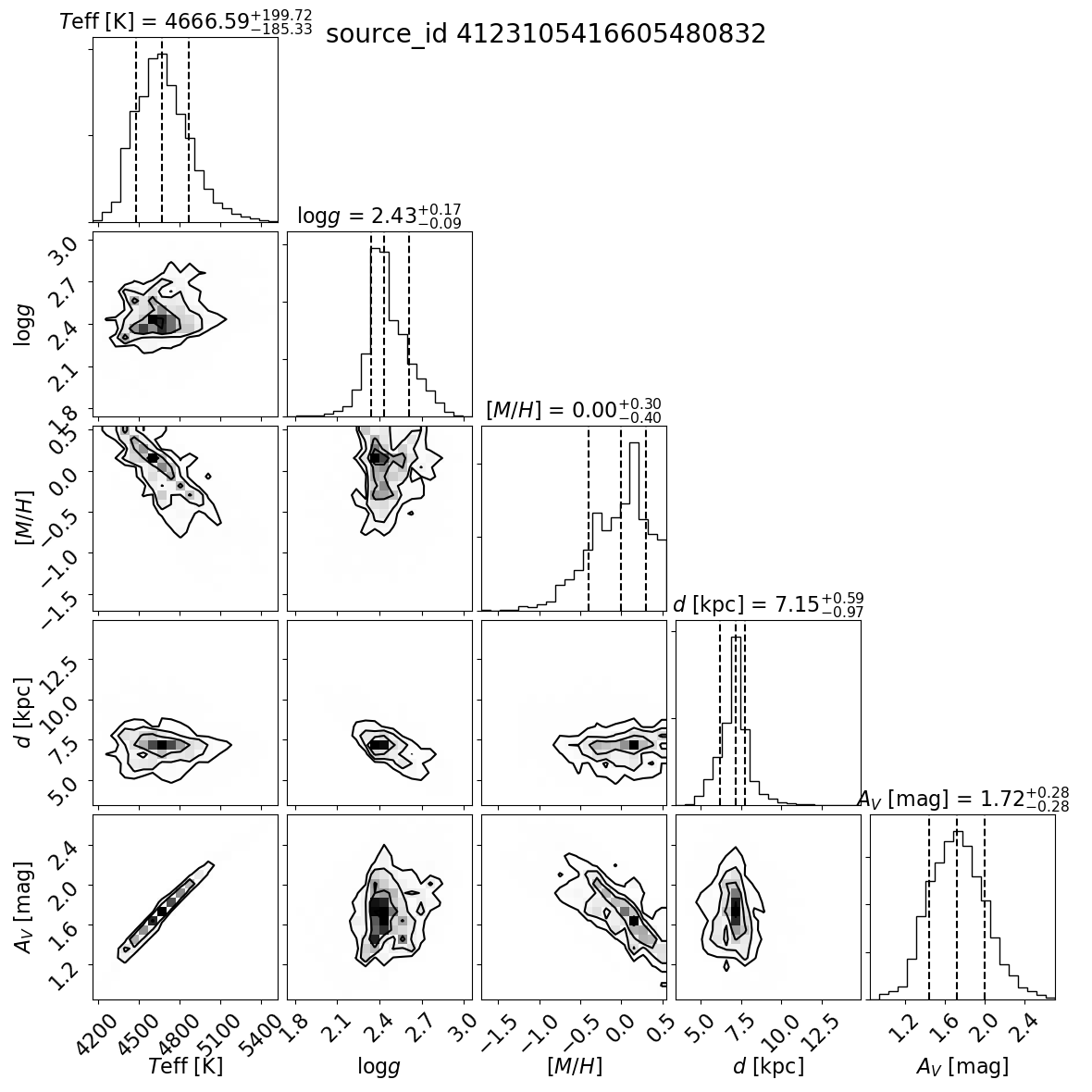}
 	\includegraphics[width=0.49\textwidth]{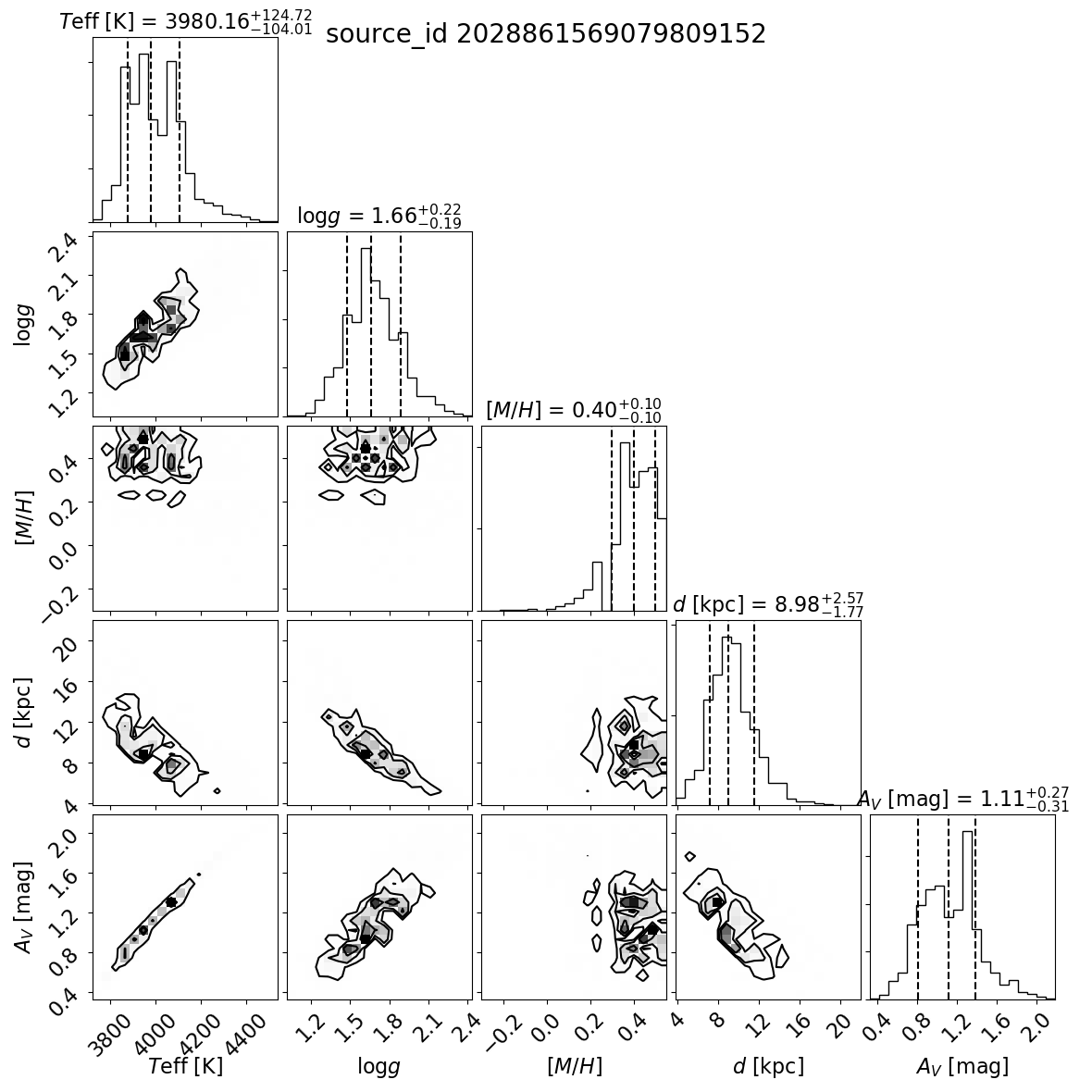}
    \caption{Example {\tt corner} plots of the posterior PDFs of four stars with well-determined parameters (${\tt SH\_OUTFLAG=="00000"} \& {\tt SH\_GAIAFLAG=="000"}$).}
 	\label{goodpdfs}
\end{figure*}

\begin{figure*}\centering
 	\includegraphics[width=0.49\textwidth]{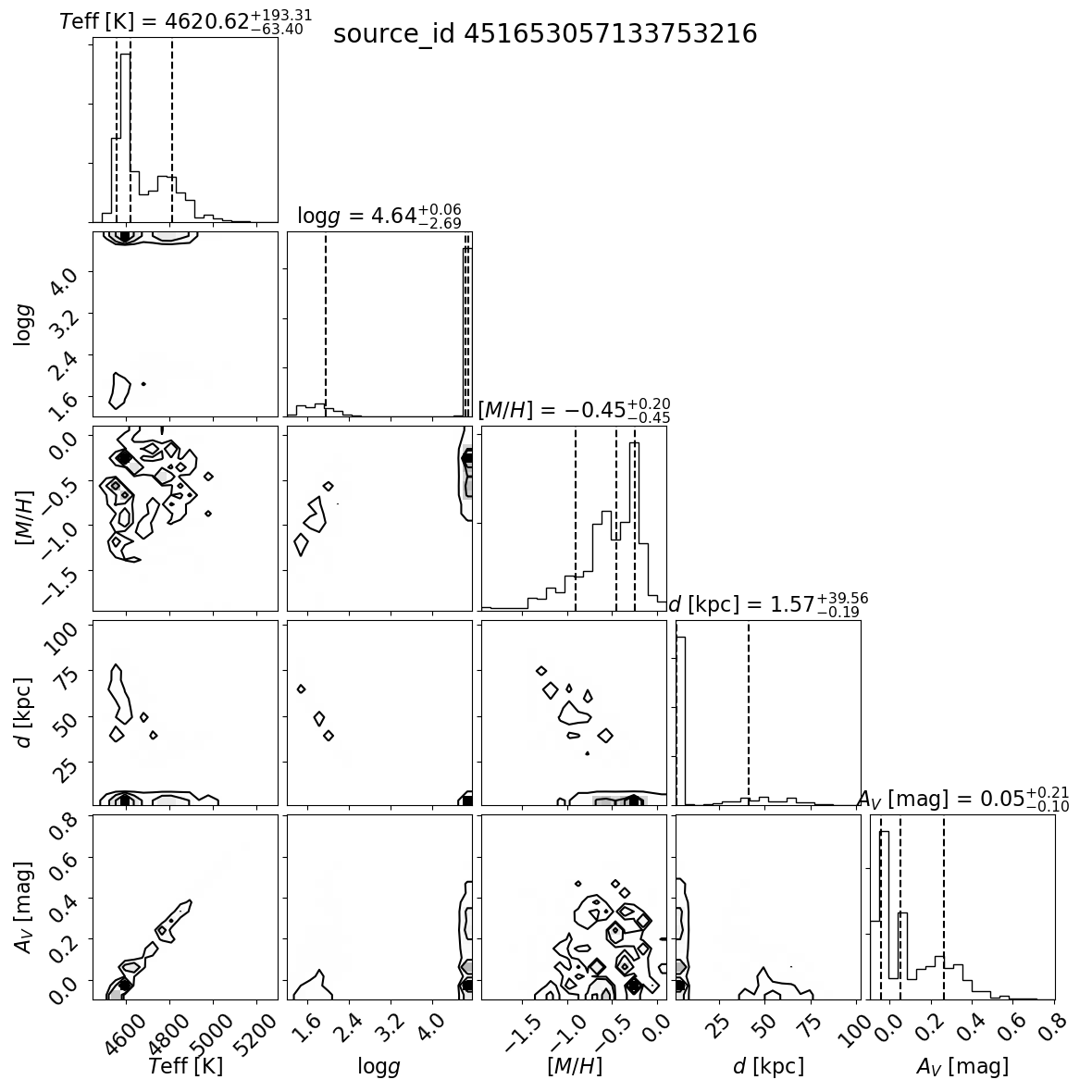}
 	\includegraphics[width=0.49\textwidth]{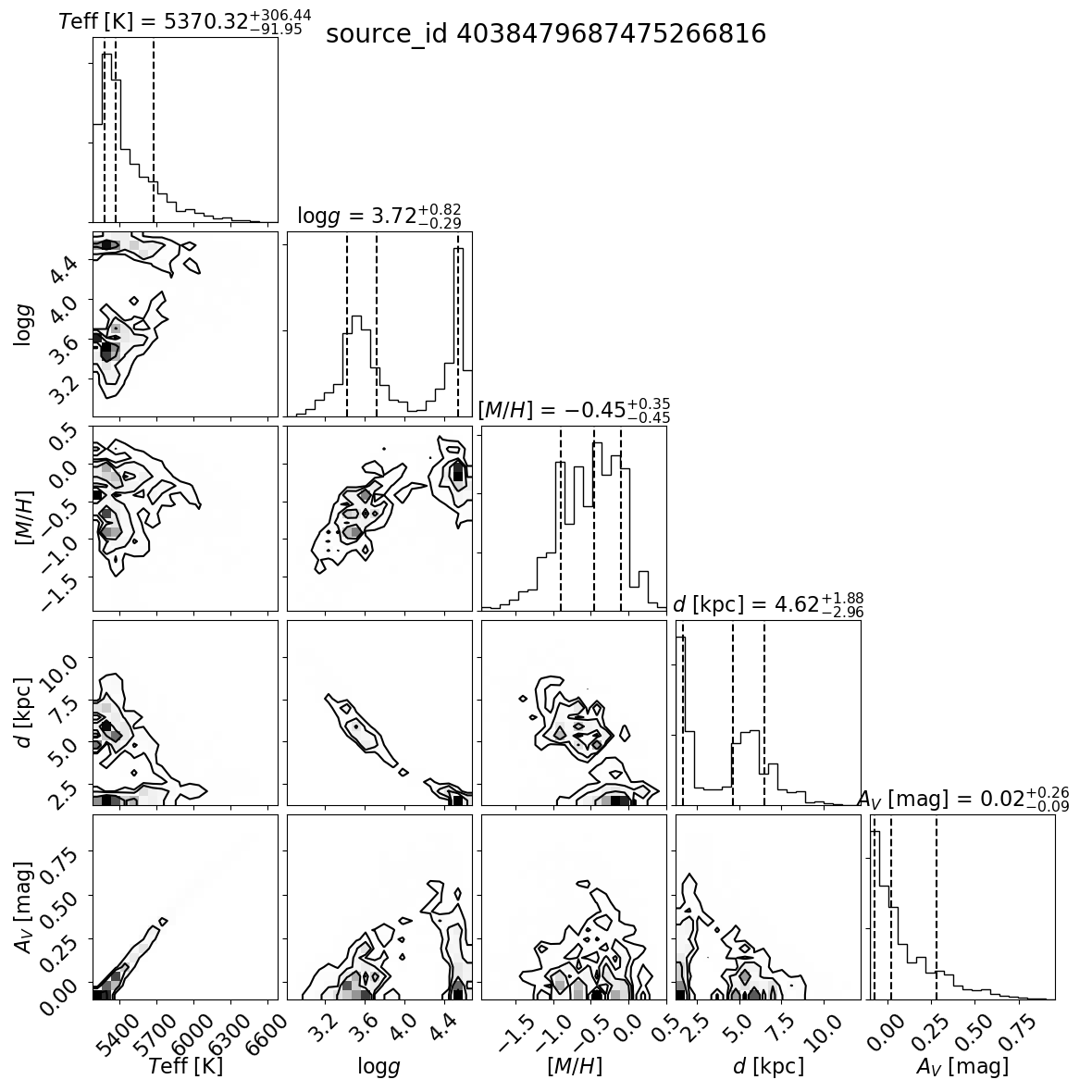}
 	\includegraphics[width=0.49\textwidth]{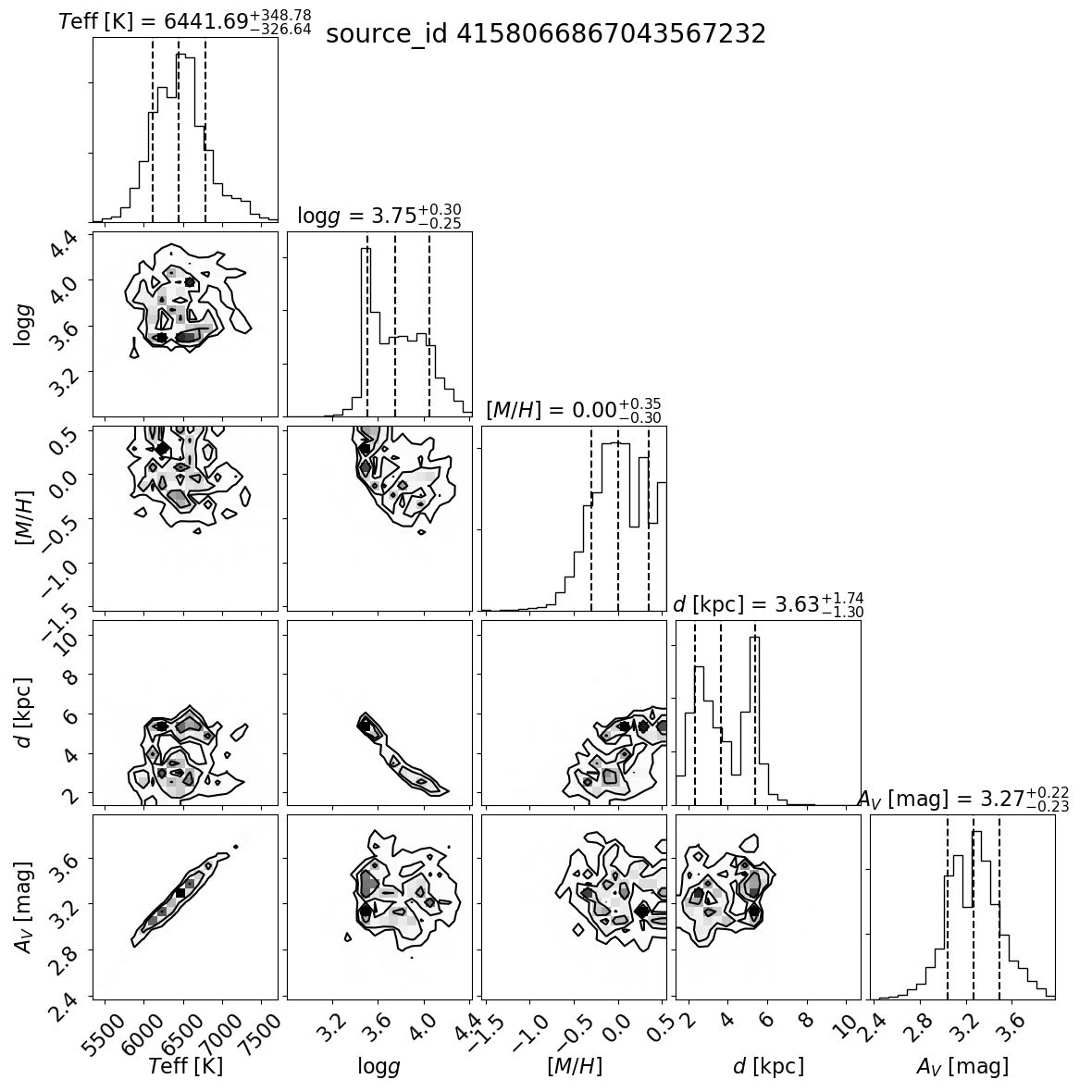}
 	\includegraphics[width=0.49\textwidth]{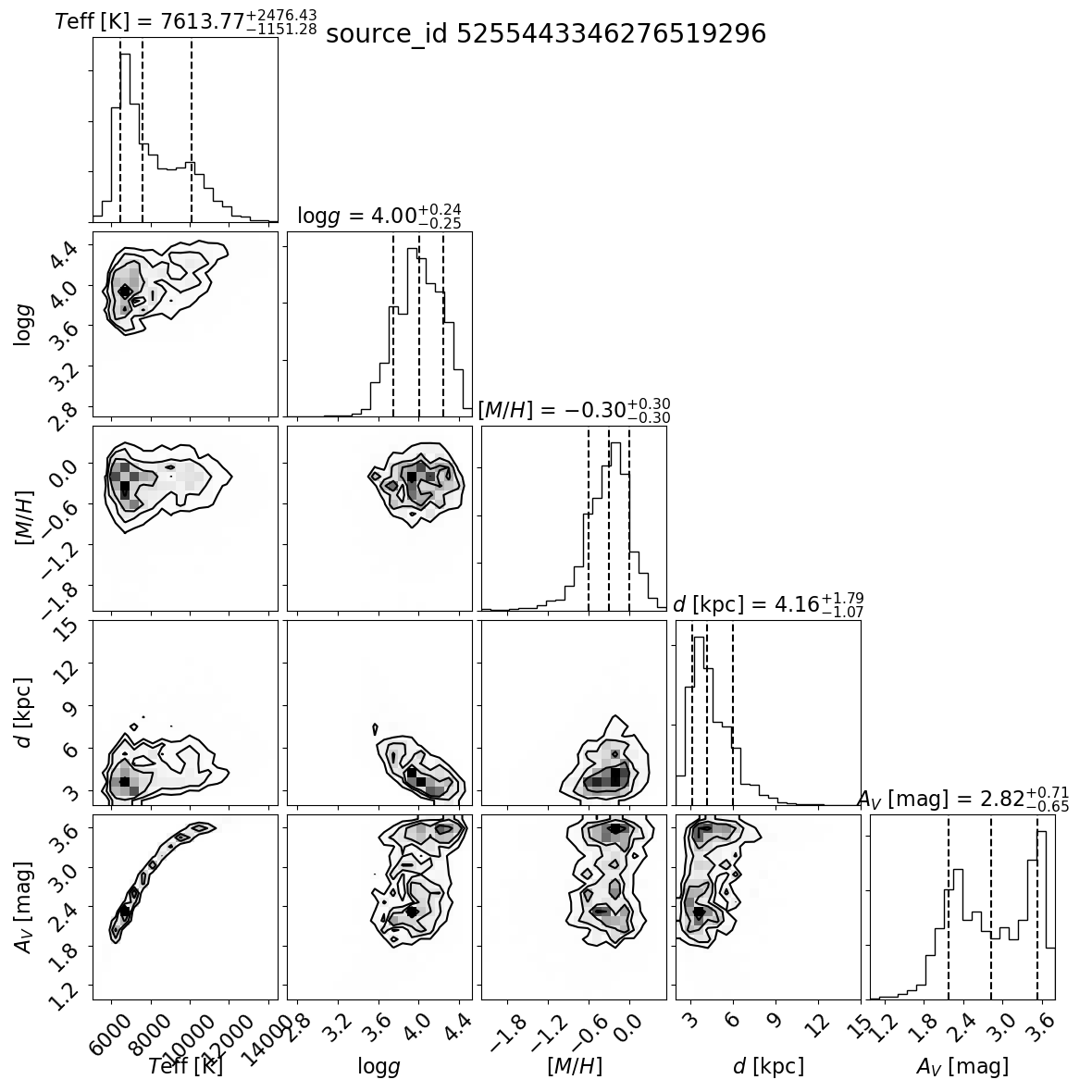}
    \caption{Example {\tt corner} plots of the posterior PDFs for stars with poorly determined parameters (${\tt SH\_OUTFLAG[0]=="1"}$).}
 	\label{badpdfs}
\end{figure*}

Figure \ref{goodpdfs} and \ref{badpdfs} show {\tt corner} plots \citep{Foreman-Mackey2016} of the full posterior PDF projected onto one- and two-dimensional subspaces. For visibility, we only show contours in each  of the plots. Figure \ref{goodpdfs} shows examples of stars with well-determined posterior parameters. As discussed in Sect. \ref{precision}, the precision in the output parameters mainly depends on the parallax signal-to-noise ratio, but the precision of effective temperature and extinction estimates is considerably increased when the full information is available. 

Finally, Fig. \ref{badpdfs} shows a few cases of stars for which {\tt StarHorse} was only able to determine very uncertain parameters (${\tt SH\_OUTFLAG[0]=="1"}$), due to the poor precision of the input parallaxes. These stars, some of them responsible for the unphysical nose in the CMD (see Sect. \ref{cmds2}) as well as the bloody-eye effect (see Sect. \ref{sauron}) display varying degrees of bimodality in their posterior PDFs, in many cases meaning that {\tt StarHorse} was unable to decide with certainty if these stars are dwarfs or giants. In consequence, as discussed in Sect. \ref{flags} and Appendix \ref{outflag1}, their median output parameters are not necessarily compatible with each other.

% 4 randomly selected good guys:
% 4284355913593396864 (nearby red dwarf, d~0.65 kpc, G~17.5)
% 2048736478663311872 (G dwarf, d~1.8 kpc, G~16.4)
% 452485872771888768 (RC star, d~3 kpc, G~15.2)
% 2028861569079809152 (AGB/upperRGB star, d~11+-3 kpc, G~15.3)

\section{Data model}\label{tables}

%In this appendix we provide a set of tables for the readership interested in the data products provided in this paper. In particular, Table \ref{resultstable} gives an overview over the binary output tables (we recall that all results are also available via the AIP {\it Gaia} data mirror at \url{gaia.aip.de}), while 
Table \ref{datamodel1} provides the data model for the provided {\tt StarHorse} output tables.

\begin{table*}
\centering
\caption{Data model of the {it Gaia} DR2 {\tt StarHorse} catalogue released via the {\it Gaia} mirror at {\tt gaia.aip.de}.}
\begin{tabular}{rlll}
ID & Column name & Unit & Description \\
\hline
0 & {\tt source\_id} &  & {\it Gaia} DR2 unique source identifier \\ 
%1 & random\_index &  & Random index used to select subsets \\ 
%2 & glon & deg & {\it Gaia} DR2 Galactic longitude \\ 
%3 & glat & deg & {\it Gaia} DR2 Galactic latitude \\ 
1 & {\tt SH\_PHOTOFLAG} &  & StarHorse photometry input flag \\ 
2 & {\tt SH\_PARALLAXFLAG} &  & StarHorse parallax input flag \\ 
3 & {\tt SH\_GAIAFLAG} &  & StarHorse {\it Gaia} DR2 quality flag \\ 
4 & {\tt SH\_OUTFLAG} &  & StarHorse output quality flag \\ 
5 & {\tt dist05} & kpc & StarHorse distance, 5th percentile \\ 
6 & {\tt dist16} & kpc & StarHorse distance, 16th percentile \\ 
7 & {\tt dist50} & kpc & StarHorse distance, 50th percentile \\ 
8 & {\tt dist84} & kpc & StarHorse distance, 84th percentile \\ 
9 & {\tt dist95} & kpc & StarHorse distance, 95th percentile \\ 
10 & {\tt AV05} & mag & StarHorse line-of-sight extinction at $\lambda=5420\ \AA$, $A_{\rm V}$, 5th percentile \\ 
11 & {\tt AV16} & mag & StarHorse line-of-sight extinction at $\lambda=5420\ \AA$, $A_{\rm V}$, 16th percentile \\ 
12 & {\tt AV50} & mag & StarHorse line-of-sight extinction at $\lambda=5420\ \AA$, $A_{\rm V}$, 50th percentile \\ 
13 & {\tt AV84} & mag & StarHorse line-of-sight extinction at $\lambda=5420\ \AA$, $A_{\rm V}$, 84th percentile \\ 
14 & {\tt AV95} & mag & StarHorse line-of-sight extinction at $\lambda=5420\ \AA$, $A_{\rm V}$, 95th percentile \\ 
15 & {\tt AG50} & mag & StarHorse line-of-sight extinction in the G band, $A_{\rm G}$, 50th percentile, derived from {\tt AV50} and {\tt teff50} \\ 
16 & {\tt teff16} & K & StarHorse effective temperature, 16th percentile \\ 
17 & {\tt teff50} & K & StarHorse effective temperature, 50th percentile \\ 
18 & {\tt teff84} & K & StarHorse effective temperature, 84th percentile \\ 
19 & {\tt logg16} & [dex] & StarHorse surface gravity, 16th percentile \\ 
20 & {\tt logg50} & [dex] & StarHorse surface gravity, 50th percentile \\ 
21 & {\tt logg84} & [dex] & StarHorse surface gravity, 84th percentile \\ 
22 & {\tt met16} & [dex] & StarHorse metallicity, 16th percentile \\ 
23 & {\tt met50} & [dex] & StarHorse metallicity, 50th percentile \\ 
24 & {\tt met84} & [dex] & StarHorse metallicity, 84th percentile \\ 
25 & {\tt mass16} & $M_{\odot}$ & StarHorse stellar mass, 16th percentile \\ 
26 & {\tt mass50} & $M_{\odot}$ & StarHorse stellar mass, 50th percentile \\ 
27 & {\tt mass84} & $M_{\odot}$ & StarHorse stellar mass, 84th percentile \\ 
28 & {\tt ABP50} & mag & StarHorse line-of-sight extinction in the $G_{\rm BP}$ band, $A_{\rm BP}$, 50th percentile, derived from {\tt AV50} and {\tt teff50} \\ 
%31 & ABPft50 & mag & StarHorse line-of-sight extinction in the G\_BP band, A\_BP, 50th percentile - valid for $G>$10.87 \\ 
%32 & ABPbr50 & mag & StarHorse line-of-sight extinction in the G\_BP band, A\_BP, 50th percentile - valid for G<10.87 \\ 
29 & {\tt ARP50} & mag & StarHorse line-of-sight extinction in the $G_{\rm RP}$ band, $A_{\rm RP}$, 50th percentile, derived from {\tt AV50} and {\tt teff50} \\ 
30 & {\tt BPRP0} & mag & StarHorse dereddened colour, $(G_{\rm BP}-G_{\rm RP})_0$, derived from {\tt phot\_bp\_mean\_mag, phot\_rp\_mean\_mag, ABP50}, and {\tt ARP50} \\ 
31 & {\tt MG0} & mag & StarHorse absolute magnitude, derived from {\tt phot\_g\_mean\_mag} (recalibrated), {\tt dist50}, and {\tt AG50} \\ 
32 & {\tt XGal} & kpc & StarHorse Galactocentric Cartesian X co-ordinate, derived from {\tt dist50} and assuming $R_0 = 8.2$ kpc \\ 
33 & {\tt YGal} & kpc & StarHorse Galactocentric Cartesian Y co-ordinate, derived from {\tt dist50} and assuming $R_0 = 8.2$ kpc \\ 
34 & {\tt ZGal} & kpc & StarHorse Galactocentric Cartesian Z co-ordinate, derived from {\tt dist50} and assuming $R_0 = 8.2$ kpc \\ 
35 & {\tt RGal} & kpc & StarHorse Galactocentric planar distance, derived from {\tt XGal} and {\tt YGal} \\ 
36 & {\tt ruwe} &  & {\it Gaia} DR2 renormalised unit-weight error (Lindegren 2018) \\ 
\end{tabular}
\label{datamodel1}
\end{table*}

\section{Example queries}\label{examplequeries}

In this appendix we show some examples of ADQL queries that can be used to access the {\tt StarHorse} {it Gaia} DR2 results via the {it Gaia} DR2 mirror archive at {\tt gaia.aip.de}.

The first example query shows how to extract the median distance and extinction for the first 50 objects of the flag-cleaned {\tt StarHorse} sample (using both the {\tt SH\_GAIAFLAG} and the {\tt SH\_OUTFLAG}; see Sect. \ref{flags}):

\begin{verbatim}
SELECT TOP 50 s.glon, s.glat, s.dist50, s.AV50, 
FROM gdr2_contrib.starhorse AS s 
WHERE s.SH_OUTFLAG LIKE '00000'
AND s.SH_GAIAFLAG LIKE '000'
\end{verbatim}

The second example shows how to access the first 50 rows of our results, cross-matched with the {\it Gaia} DR2 catalogue, cleaned only for the main {\tt StarHorse} output flag, {\tt SH\_OUTFLAG}[0]=="0":

\begin{verbatim}
SELECT TOP 50 s.*, g.ra, g.dec 
FROM gdr2.gaia_source AS g, 
     gdr2_contrib.starhorse AS s 
WHERE g.source_id = s.source_id 
AND s.SH_OUTFLAG LIKE '0%%%%'
\end{verbatim}

The next example shows how to access the first 50 rows of our catalogue cross-matched to the {\it Gaia} DR2 main source table, without cleaning for any {\tt StarHorse} flags, but selecting only red-clump stars with {\tt ruwe}$<1.3$:

\begin{verbatim}
SELECT TOP 50 s.*, g.ra, g.dec 
FROM gdr2.gaia_source AS g, 
     gdr2_contrib.starhorse AS s 
WHERE g.source_id = s.source_id
AND 4500 < s.teff50 AND s.teff50 < 5000
AND 2.35 < s.logg50 AND s.logg50 < 2.55
AND -0.6 < s.met50  AND s.met50 < 0.4
AND s.ruwe < 1.3
\end{verbatim}

The following example extracts an extinction-corrected, flag-cleaned CMD for stars in a 1 degree region around the centre of the open cluster NGC 6819 (see Fig. \ref{clustercomparison2}, third row):

\begin{verbatim}
SELECT g.source_id, s.MG0, s.BPRP0 
FROM gdr2.gaia_source AS g, 
     gdr2_contrib.starhorse AS s 
WHERE g.source_id = s.source_id
AND s.SH_OUTFLAG LIKE '00000'
AND s.SH_GAIAFLAG LIKE '000'
AND 1=CONTAINS(POINT('GALACTIC',g.l,g.b), 
                 CIRCLE('GALACTIC',74.,8.5, 1.))
\end{verbatim}

Finally, the last example shows how to retrieve all {\it Gaia} DR2 stars for which {\tt StarHorse} did not converge:

\begin{verbatim}
SELECT g.source_id, g.l, g.b, g.parallax, 
       g.parallax_error, g.phot_g_mean_mag, 
       g.phot_bp_mean_mag, g.phot_rp_mean_mag 
FROM gdr2.gaia_source AS g
LEFT OUTER JOIN gdr2_contrib.starhorse AS s 
ON (g.source_id=s.source_id)
WHERE g.phot_g_mean_mag <= 18.0
AND s.source_id IS NULL
\end{verbatim}

%-------------------------------------------------------------------
\end{document}